\documentclass[acmlarge]{acmart}
\makeatletter
\newcommand{\confshort}{\acmConference@shortname}
\newcommand{\conffull}{\acmConference@name}
\newcommand{\confdate}{\acmConference@date}
\newcommand{\confloc}{\acmConference@venue}
\AtBeginDocument{
  \fancypagestyle{firstpagestyle}{
    \fancyhead{}%
    \fancyfoot[C]{}%
  }
  \fancyhf{}
  \fancyhead[LO]{\@headfootfont\shorttitle}%
  \fancyhead[RE]{\@headfootfont\@shortauthors}%
  \fancyhead[LE]{\@headfootfont\footnotesize \confshort, \confdate, \confloc}%
  \fancyhead[RO]{\@headfootfont\footnotesize \confshort, \confdate, \confloc}%
  \fancyfoot[C]{}%
}
\makeatother
\acmBooktitle{\conffull\@ (\confshort), \confdate, \confloc}

\usepackage{multirow} 
\usepackage{enumitem} 
\usepackage[dvipsnames]{xcolor} 
\usepackage{makecell} 
\definecolor{resultBlue}{RGB}{0, 114, 178}    
\definecolor{resultOrange}{RGB}{213, 94, 0}   
\definecolor{specialBlue}{RGB}{0, 114, 178}    
\definecolor{specialOrange}{RGB}{213, 94, 0}   

\usepackage{colortbl}

\usepackage{tikz}
\usetikzlibrary{backgrounds}

\AtBeginDocument{%
  }

\copyrightyear{2026}
\acmYear{2026}
\setcopyright{cc}
\setcctype{by}
\acmConference[FAccT '26]{The 2026 ACM Conference on Fairness, Accountability, and Transparency}{June 25--28, 2026}{Montreal, QC, Canada}
\acmBooktitle{The 2026 ACM Conference on Fairness, Accountability, and Transparency (FAccT '26), June 25--28, 2026, Montreal, QC, Canada}
\acmDOI{10.1145/3805689.3812299}
\acmISBN{979-8-4007-2596-8/2026/06}

\newcommand\rev[1]{\textcolor{black}{#1}}

\begin{document}

\title{When and How AI Should Assist Brainstorming for AI Impact Assessment}

\author{Jarod Govers}
\orcid{0000-0002-7648-318X}
\affiliation{
\department{School of Computing and Information Systems}
  \institution{University of Melbourne}
  \city{Melbourne}
  \country{Australia}
  \postcode{3053}
}
\email{jgovers@student.unimelb.edu.au}
\author{Sanja Šćepanović}
\email{sanja.scepanovic@nokia-bell-labs.com}
\orcid{0000-0002-1534-8128}
\affiliation{%
  \institution{Nokia Bell Labs}
  \city{Cambridge}
  \country{United Kingdom}
}
\affiliation{%
  \institution{University of Oxford}
  \city{Oxford}
  \country{United Kingdom}
}
\author{Daniele Quercia}
\email{quercia@cantab.net} 
\orcid{0000-0001-9461-5804}
\affiliation{%
  \institution{Nokia Bell Labs}
  \city{Cambridge}
  \country{United Kingdom}
}
\affiliation{\institution{Politecnico di Torino}
\city{Turin}
\country{Italy}}

\renewcommand{\shortauthors}{Govers et al.}
\newcommand\TODO[1]{\textcolor{red}{(TODO: #1)}}
\newcommand\NOTE[1]{\textcolor{blue}{(NOTE: #1)}}


\begin{abstract}
A key task in AI practice is to assess potential impacts to prevent harm. Current AI tools assisting AI impact assessment have not been designed \rev{or evaluated} for collaborative team brainstorming, and they do not capture the range of views in diverse teams. We studied how AI can support team brainstorming during AI impact assessment and made three contributions. First, we adapted two structured methods from strategic foresight and co-designed AI interventions for them in five in-person workshops with $28$ participants in total. Second, we evaluated the interventions in ten in-person workshops with $54$ participants, finding that AI improved impact assessment quality and brainstorming perceptions for a general-purpose AI use (a chatbot companion) but not for a specialised one (a kidney allocation application). Third, our findings result in broader design guidance for AI assistance in brainstorming: AI should only offer hints and not solutions during early ideation, initiating interaction only when participants face fixation or saturation; it should facilitate structuring ideas during convergence; leverage expertise to refine ideas; and overall, it should serve more in support of tedious brainstorming process tasks, rather than ideation that teams value to do themselves.
\end{abstract}


\begin{teaserfigure}
  \centering
  \includegraphics[width=\textwidth]{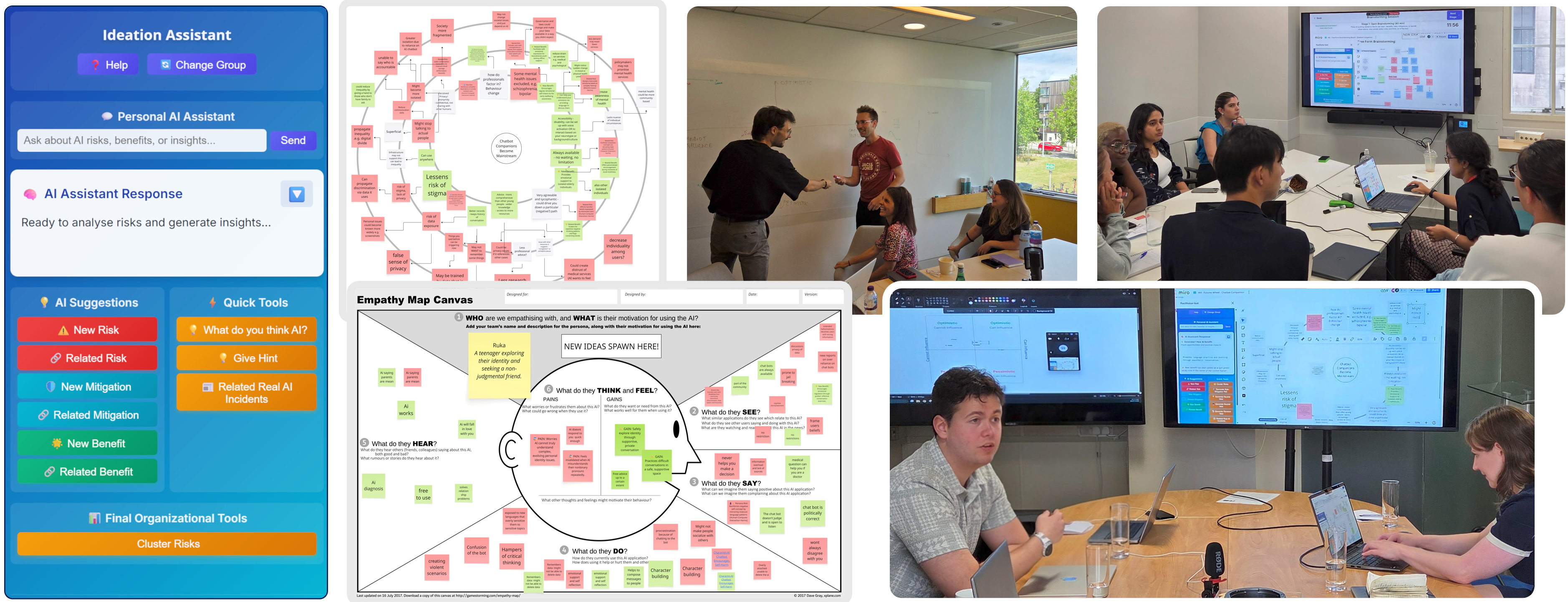}
  \caption{Structured team brainstorming of AI impacts with AI interventions; consented photos from workshops in three sites.}
  \Description{A composite image of five panels showing a user interface and workshop activities related to an AI Risk \& Mitigation Assistant. The first panel on the far left shows the user interface of the 'AI Risk \& Mitigation Assistant'. At the top, there is a persona selection dropdown and an input box for AI tasks. The main body of the interface is split into sections including 'AI Assistant Response', 'Empathy Mapping Mode', 'AI Suggestions' (with buttons like 'New Risk' and 'Related Harm'), 'Quick Tools', and 'Final Organisational Tasks'. To the right of the UI panel are two diagrams filled with sticky notes. The top one is a large Futures Wheel diagram with several rings of impact, populated with example impacts. Below this is a digital 'Empathy Map' canvas. It features a stylised head in the centre and is divided into quadrants labelled 'What do they THINK and FEEL?', 'What do they SEE?', 'What do they HEAR?', and 'What do they SAY \& DO?'. This map is also covered in red and green digital sticky notes. The right-hand side of the composite image consists of three photographs depicting workshop sessions. The top-middle photo shows three people interacting together. The top-right photo shows a group of participants sitting at a conference table, looking at a large screen that displays a Miro collaboration board. The bottom-right photo shows two people working on laptops at a table, with a large monitor in front of them displaying the intervention's interface.}
  \label{fig:workshop_examples}
\end{teaserfigure}

\maketitle

\section{Introduction}
``Your AI application killed someone'' is among the worst outcomes that have occurred when teams fail to consider the risks of Artificial Intelligence (AI) systems~\cite{mcgregorPreventingRepeatedReal2021}. We define AI systems broadly as those that exhibit or simulate intelligent behaviour~\cite{oed_artificial_intelligence}. As AI spreads across society, comprehensive \emph{impact assessment} has become both an ethical imperative and a regulatory requirement. Impact assessment is the systematic process of identifying AI's potential \textit{benefits}, \textit{risks}, and corresponding \textit{mitigations}~\cite{impact_assessment_report25,vscepanovic2025ai}. Organisations and leading AI conferences increasingly recognise that responsible AI development requires proactive identification of potential impacts before development or deployment~\cite{Bogucka_et_al_ai_design_prefilling_reports24,ashurst2022ai}, with the European Union's AI Act~\cite{EUACT2024} mandating systematic risk assessment for AI deployments~\cite{eu_ai_systemic_risk_obligations25}.

AI impact assessment process also aims to help development teams examine both immediate and long-term effects of their systems on individuals and society: before, during, and after development. Yet, predicting consequences from technical failures to human rights harms, is inherently difficult~\cite{vscepanovic2025ai,bogucka2024atlas, govers_et_al_narratives26}, with high opportunity costs~\cite{prunkl2021institutionalizing,herdel_et_al_exploregen25} and moral stress~\cite{moral_stress_CHI_2025}. 
Researchers have proposed various approaches to support AI practitioners, from automated risk generation~\cite{buçinca_aha_generating_ai_risks23, rao_et_al_riskrag25,herdel_et_al_exploregen25} to visual exploration interfaces~\cite{wang_et_al_farsight24,Bogucka_et_al_ai_design_prefilling_reports24}. A consistent finding across this work is that responsible AI planning requires a collaborative, team-based process incorporating diverse viewpoints and expertise~\cite{matias2025public}. Teams who deeply understand their application context and end-users possess crucial knowledge that individuals or automated tools alone cannot fully capture. Most approaches support individual practitioners~\cite{herdel_et_al_exploregen25,buçinca_aha_generating_ai_risks23,rao_et_al_riskrag25}, or enable them to jointly complete assessments~\cite{wang_et_al_farsight24,Bogucka_et_al_ai_design_prefilling_reports24}, but are not designed or evaluated to support \rev{brainstorming and deliberation}.

Brainstorming is an activity where people meet in a group to generate many ideas for possible development~\cite{cambridge_dictionary}. Through structured deliberation on AI impacts, teams can uncover and develop targeted mitigations across technical, social, and policy dimensions before harm occurs~\cite{mitre2024red}. Osborn identified three brainstorming stages ~\cite{osborn2012applied}: ideation, convergence, and decision. Applied to AI impact assessment, these become: \emph{ideate} (participants generate impacts), \emph{converge} (identify, group, and organise impacts), and \emph{decide} (develop mitigation strategies).

\rev{Commercial AI risk-assessment platforms have incorporated AI into their strategic foresight driven tools: Futures Platform uses the Futures Wheel method for structured brainstorming, while 4Strat’s `Foresight Strategy Cockpit' includes Futures Wheel and Empathy Mapping with AI risk generators (used by the European Commission)~\cite{4strat}. Microsoft’s Azure AI Foundry also offers red-teaming and ``AI Reports'' that feed into its Responsible AI Impact Assessment Template and Guide~\cite{microsoft2022rai}. However, these opaque AI typically provide suggestions without supporting team dynamics and have not been empirically evaluated to determine whether AI actually improves assessment quality. Whether AI is appropriate depends on design choices that preserve meaningful multidisciplinary, lived-experience human participation~\cite{sandoval_et_al_ibm_multidiscilinary_assess25}, and whether it can match or exceed human performance in identifying high-quality risks and mitigations that can save lives.} Hence, we asked \rev{three} research questions:
\begin{itemize}
\item[\rev{(RQ\textsubscript{1})}] \rev{At which points in established brainstorming methods can AI meaningfully intervene to support teams in AI impact assessment?}
\item[\rev{(RQ\textsubscript{2})}] \rev{What are the design requirements for AI interventions at these points?}
\item[(RQ\textsubscript{3})] What are the effects of AI interventions in team brainstorming on output quality and participant perceptions?
\end{itemize}

In answering these questions, we made \rev{three} main contributions:
\begin{enumerate}  
    \item \rev{\textbf{Scoping brainstorming methods and identifying viable AI intervention points within them}} (\S\ref{sec:methods}). We reviewed traditional brainstorming methods, selecting two strategic foresight approaches \rev{used in AI impact assessment}: Futures Wheels~\cite{glenn_gordon_futures_book_v3_09} and Empathy Mapping~\cite{gray_gamestorming_book10} (\S\ref{sec:brainstorming_methods}). We then reviewed literature on AI-assisted brainstorming and identified potential points of intervention along three dimensions: the stage of the brainstorming process (\emph{when}), AI's roles (\emph{what}), and human agency (\emph{how}) (\S\ref{sec:PoIs}, Table~\ref{tab:POI_literature}).
    \item \rev{\textbf{Deriving design requirements for AI interventions at these points through co-design (\S\ref{sec:codesign}).} Through five in-person co-design workshops with 28 participants,} we identified design requirements for AI interventions in brainstorming (\S\ref{sec:codesign}, Figure~\ref{fig:ai_requirements_figure}) and implemented a public AI tool to fulfil them:\\(\url{https://social-dynamics.net/ai-risks/interventions/}).
    \item \textbf{\rev{Evaluating the effects of AI interventions on output quality and participant experience}} (\S\ref{sec:evaluating}). We ran 10 in-person workshops \rev{with 54 participants in total,} comparing output quality (e.g., impact plausibility and uniqueness) and participant experience (e.g., sense of control or anxiety) in teams with and without AI assistance. Teams brainstormed impacts of a common AI use: a chatbot companion, and more specialised use: medical AI for kidney transplant allocation. AI interventions enhanced both output quality and participant experience (Table \ref{tab:results_combined}) for the chatbot companion. However, we found limited improvements in output quality for the medical AI use, suggesting that current AI capabilities may be limited where deep expertise matters. Participants also used AI assistance less during early ideation and more in later convergence and decision stages, and they preferred to initiate the interaction.
\end{enumerate}

Our findings offer design guidance for AI-assisted brainstorming (\S\ref{sec:discussion}, Figure \ref{fig:discussion_poi_figure}). We conclude by examining AI’s potential and limitations in team brainstorming for researchers, industry, and regulators, with future directions.

\begin{figure}[!tbp]
  \centering
  \includegraphics[width=\linewidth]{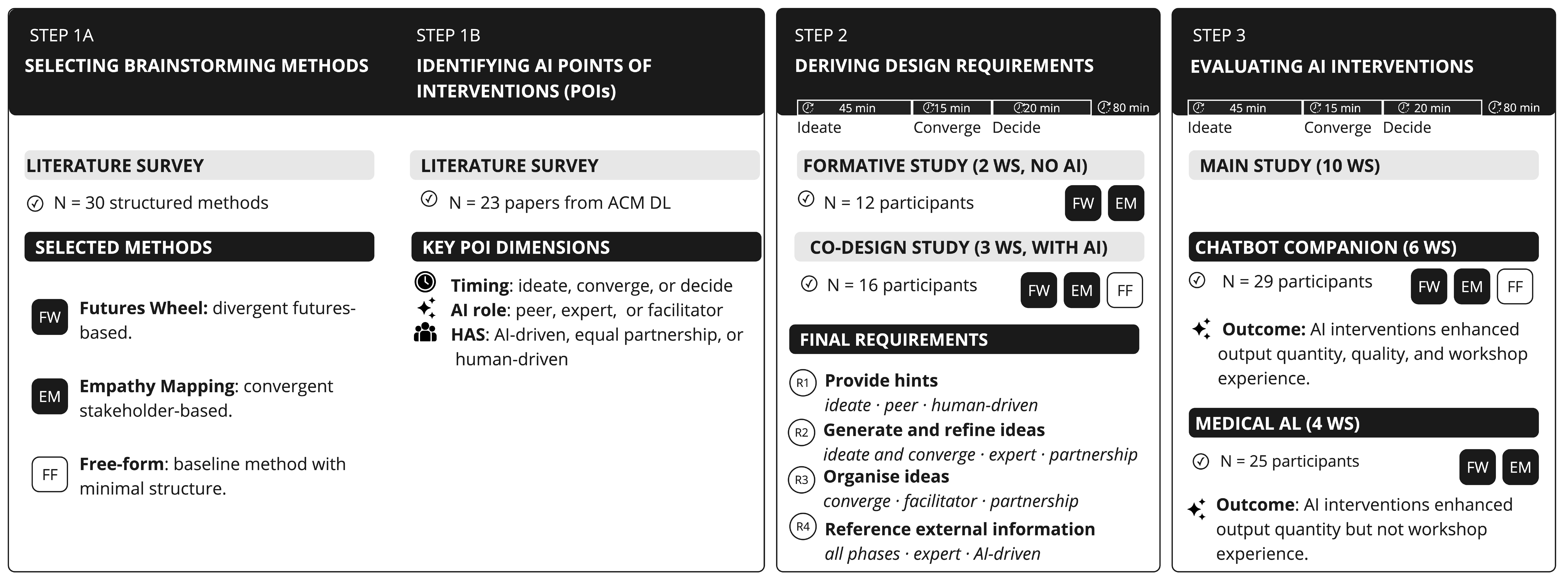}
    \caption{{Our methodology has \rev{three} steps. In \rev{Step 1}, we selected two brainstorming methods suited to AI impact assessment (\S\ref{sec:brainstorming_methods}, {Step 1A}), and \rev{identified} types and points of AI intervention in these methods from prior work (\S\ref{sec:PoIs}, {Step 1B)}. In Step \rev{2}, we elicited design requirements through five formative and co-design workshops (\S\ref{sec:codesign}). In Step \rev{3}, we evaluated the resulting AI interventions across 10 workshops (\S\ref{sec:evaluating}).} FW (Futures Wheel), EM (Empathy Mapping), FF (Free-form), HAS (Human Agency Scale), and WS (Workshop).}
    \Description{}
  \label{fig:methodology}
\end{figure}
\section{Related work}\label{sec:rw}
We review impact assessment ideation research, covering traditional (\S\ref{sec:rw_team}) and AI-assisted approaches (\S\ref{sec:rw_AI}).
\subsection{Traditional brainstorming and ideation for AI impact assessment}\label{sec:rw_team}
Early-stage brainstorming to understand AI impacts can reduce harmful incidents and provide essential context for development teams~\cite{weidinger_deepmind_taxonomy22}. Despite these benefits, few studies have examined how structured brainstorming could aid AI impact assessment.
\citet{ballard_et_al_judgment_call_game19} developed \emph{Judgement Call}, a card game using simulated user reviews of positive and negative experiences to explore why users might respond to an AI application favourably or unfavourably (e.g., giving a one-star review due to privacy issues). 
\citet{hohendanner_et_al_global_dialogues25} combined foresight-driven studies and technology assessment methods in participatory workshops titled \emph{Global AI Dialogues} to gather laypeople's perspectives on generative AI. \citet{chi_design_courts} explored \emph{Design Courts} as an approach to elicit ethical and legal dilemmas for AI. These approaches helped teams ideate ethical dilemmas and identify risks and benefits to stakeholders, but did not address the full ecosystem of impacts and mitigations required for full AI impact assessments~\cite{impact_assessment_report25, eu_ai_systemic_risk_obligations25, herdel_et_al_exploregen25}.

\rev{\citet{Elsayed_Ali_et_al_responsible_ai_cards23} designed \emph{Responsible \& Inclusive Cards} with guiding questions to probe thinking regarding impacts to stakeholders, outcomes and governance/practice. They suggest that future research should empirically test how well structured approaches to impact assessment work in practice. }
\rev{Using structured methods can maximise the performance of human-led teams and enables a fair evaluation of whether AI augmentation improves brainstorming quality, compared to the bottlenecks often experienced from `blank canvas' brainstorming~\cite{diehl_stroebe_productivity_loss_brainstorming87}. }
\rev{This in turn increases the need to empirically evaluate whether AI can improve the quality of human brainstorming, or whether structured brainstorming alone would suffice. This is particularly pertinent given the rise of AI which can augment human-driven impact assessments~\cite{rao_et_al_riskrag25, herdel_et_al_exploregen25}, more of which we discuss in the following section.}
\rev{Structured methods could also help guide AI through brainstorming tasks more systematically to improve their breadth and relevance through personas or systemic futures. This is particularly important given the well-documented challenge of human and AI design fixation in ideation~\cite{jansson_smith_design_fixation91}.}

\subsection{AI-assisted brainstorming and ideation for AI impact assessment}\label{sec:rw_AI}
We conducted a scoping literature review of AI tools for impact assessment following the PRISMA approach~\cite{prisma21} (full search strings in Appendix~\S\ref{appsec:AI_brainstorming_AIIA}). Across \rev{eight} identified studies~\cite{rao_et_al_riskrag25, wang_et_al_farsight24, buçinca_aha_generating_ai_risks23, herdel_et_al_exploregen25, Bogucka_et_al_ai_design_prefilling_reports24, bagehorn_et_al_ibm_ai_risk_atlas25, eisenberg_et_alunified_control_framework_ai_gen_risks25, candello_ibm_research_risk_mitigation_llm25}, we found that all tools assist individual practitioners but overlook team collaboration dynamics.
These tools differ in their approach. \citet{wang_et_al_farsight24} explored how Farsight, an LLM-powered interactive tool, helps prompt prototypers explore impacted stakeholders and risks through a visual tree, but without empirical evaluation of team collaborative effectiveness. Other tools help ideate risks and benefits through taxonomies~\cite{bagehorn_et_al_ibm_ai_risk_atlas25, eisenberg_et_alunified_control_framework_ai_gen_risks25} or through automatic generation upon AI technology or model input, such as AHA!~\cite{buçinca_aha_generating_ai_risks23}, ExploreGen~\cite{herdel_et_al_exploregen25}, and RiskRAG~\cite{rao_et_al_riskrag25}. None of these tools were designed for collaborative ideation. The AI Design framework pre-fills impact assessment reports~\cite{Bogucka_et_al_ai_design_prefilling_reports24} and can be used by different team members in parallel, but again does not focus on team deliberation and collaboration. Related work in red teaming, such as PyRIT~\cite{munoz2024pyritframeworksecurityrisk}, targets individual professionals. Moreover, a recent review found that red teaming practices are generally hidden by non-disclosure agreements~\cite{10.1145/3678884.3687147}, leading researchers to call them ``security theatre''~\cite{readteaming-theater}.

Practitioner accounts and tool-oriented resources show that teams already use LLMs to seed, pre-fill, or facilitate impact assessment, but they rarely contain systematic empirical evaluation. Microsoft's Responsible AI materials provide templates and guidance for internal assessments and mention LLM-assisted activities in applied reviews~\cite{microsoft2022rai}. NIST's AI RMF offers a widely adopted risk-management scaffold that practitioners use to structure workshop outputs and mitigation tracking~\cite{nist2023ai}. MITRE's red-teaming guidance describes adversarial brainstorming techniques that organisations can combine with AI models to surface misuse scenarios~\cite{mitre2024red}. Together, these practitioner-oriented resources establish feasibility and demand for AI to assist, but not replace brainstorming, and leave open whether AI-assisted brainstorming measurably improves the quality, or participant perceptions of formal AI impact assessments.

\rev{\citet{candello_ibm_research_risk_mitigation_llm25} proposed `mitigator' LLMs that work alongside a base LLM to detect and revise problematic outputs before reaching users, and used stakeholder personas in workshops to co-design improved responses. This approach demonstrates utility of AI for live mitigations, though its applicability to the full benefit, risk, and mitigation ecosystem for a full impact assessment necessitates empirical evaluation to identify if AI is needed, or if human-only approaches is sufficient.}

\rev{Structured brainstorming for corporate risk assessment, known as strategic foresight, exists in platforms like 4Strat’s `Foresight Strategy Cockpit'~\cite{4strat} for AI-augmented risk assessment with personas. Whereas, Futures Platform uses AI and the Futures Wheel method to aggregate, visualise, and surface unexpected risks~\cite{futures_platform, futures_platform_example}. Despite these tools, there is a lack of empirical evaluation to justify the use of AI, with Futures Platform itself noting the need to evaluate generative AI~\cite{futures_platform_genai}.}

\vspace{0.5em}
\noindent \textbf{Research gap.}  
Current AI tools for impact assessment have not been designed and evaluated (against a human-baseline) for collaborative team brainstorming. This gap is significant because teams require active involvement in assessment to build understanding and foster responsibility, rather than relying on individual or automated approaches~\cite{seafty_in_ai23, von_solms_security_culture04}. A key question remains: what are the appropriate points \rev{and types} of AI intervention in brainstorming AI impacts?

\section{\rev{Identifying AI intervention points in brainstorming methods for AI impact assessment}}\label{sec:methods}

To answer RQ\textsubscript{1}: \emph{ \rev{At which points in established brainstorming methods can AI meaningfully intervene to support teams in AI impact assessment?}}, we took \rev{two} steps (Figure \ref{fig:methodology}). First, we selected two brainstorming methods suitable for impact assessment (\S\ref{sec:brainstorming_methods}). 
\rev{Second}, we identified points of AI interventions from prior work (\S\ref{sec:PoIs}). 

\subsection{\rev{Brainstorming method selection as used by risk assessment platforms}}
\label{sec:brainstorming_methods}


\rev{We focused on structured brainstorming methods for two main reasons. First, they are already in use for corporate risk assessments~\cite{4strat, futures_platform, microsoft2022rai}, but have not been empirically tested in team settings with AI support. Second, traditional free-form brainstorming often suffers from psychological barriers: production blocking, where participants must wait for others to speak, and evaluation apprehension, which can suppress creativity and discourage contribution~\cite{diehl_stroebe_productivity_loss_brainstorming87}. Structured approaches were developed to address these issues to foster more inclusive and productive participation~\cite{strategic_foresight_product_design25}. Hence, we identified 30 structured brainstorming methods with potential to fulfil the requirements for AI impact assessment (Appendix Table~\ref{tab:brainstorming_methods}) from three sources: (1) the Millennium Project, a peer-reviewed collection of Strategic Foresight methods~\cite{glenn_gordon_futures_book_v3_09} used in risk assessment platforms~\cite{4strat, futures_platform, uk_govt_futures_toolkit24}, (2) the United Nations Futures Lab, used in AI policy making including the Futures Wheel approach~\cite{un_futures_lab25, un_leveraging_strategic_foresight_ai25}, and (3) the UK government, using Strategic Foresight methods for AI management~\cite{hines_thinking_about_the_future06, uk_govt_futures_toolkit24, uk_govt_openai_partnernship25}. From the 30 methods, we excluded those that failed any of the four criteria:}

\begin{enumerate}
    \item \rev{Designed to consider future implications, making them suitable for ideating AI impacts;}
    \item \rev{Designed for broad rather than narrow, specific contexts (e.g., operational management, supply chains);}
    \item \rev{Applicable to products such as an AI application rather than processes only;}
    \item \rev{Methods that focus on divergent \textit{or} convergent thinking.}
\end{enumerate}


\rev{After applying this criteria, the remaining methods were Backcasting, Delphi, Visioning, Futures Wheel, SWOT, and Empathy Mapping. We selected two complementary ones: Futures Wheel, a \emph{divergent}, futures-oriented approach that explores broad societal implications across near- to long-term horizons; and Empathy Mapping, a \emph{convergent}, stakeholder-focused approach that captures end users' perspectives. Investigating stakeholders/personas is used in the 4Strat platform~\cite{4strat}, by \citet{candello_ibm_research_risk_mitigation_llm25} for real-time mitigations, and ~\citet{hohendanner_et_al_global_dialogues25} in non-AI assisted settings only. We selected these methods for their use in corporate tools~\cite{4strat, futures_platform, microsoft2022rai} and research~\cite{hohendanner_et_al_global_dialogues25}, making them ideal for a first empirical evaluation of AI-assisted versus human-only assessment.}

\vspace{0.5em}
\noindent\textbf{Futures Wheel~\cite{glenn_gordon_futures_book_v3_09}.} A structured foresight method for divergent thinking that explores potential outcomes across chains of consequences. Participants place a central trend in the middle and ask: ``If this trend occurs, what happens next?'' They add first-order impacts if all agree the impact is plausible, repeating this process across three rings. We adapted it for impact assessment by having participants consider the AI use becoming mainstream as the central trend.

\vspace{0.5em}
\noindent\textbf{Empathy Mapping~\cite{gray_gamestorming_book10}.} A method that considers a specific stakeholder and has participants fill out what this stakeholder sees, says, does, hears, thinks, and feels. We asked teams to select two AI-use stakeholders---one relatable and one not\rev{---from a template varying in backgrounds, gender, age, and jobs (Appendix Figure~\ref{fig:persona_screen}), or they could create their own.}


\vspace{0.5em}
\noindent\textbf{Free-form brainstorming.} Our control condition uses an unstructured blank canvas where participants generate potential AI use impacts without predefined structure.

\vspace{0.5em}
\noindent For all three methods, we applied a consistent three-stage process: participants first generate ideas of impacts (\emph{ideate}), then colour-code them as positive (benefits), negative (risks), or neutral, create any additional related benefits and risks and group them (\emph{converge}), and finally propose mitigations for identified risks (\emph{decide}).

\subsection{Identifying types and points of AI intervention from prior work}
\label{sec:PoIs}

\rev{Using a string-based search for AI-assisted brainstorming \emph{in general} (Appendix~\ref{appsec:AI_brainstorming})}, we identified three dimensions along which interventions in brainstorming should be conceived (Table~\ref{tab:POI_literature}).

\vspace{0.5em}
\noindent \textbf{Timing: \emph{when} to intervene.}
The mixed-initiative HCI model~\cite{10.1145/302979.303030} proposes that AI should decide when to act or interrupt users by weighing costs, benefits, and uncertainties in user attention and system utility. In brainstorming, benefits include new ideas or guided discussion; costs include breaking flow or annoying participants.

\vspace{0.5em}
\noindent \textbf{AI roles: \emph{what} should be the function of AI.}
\citet{Bittner_et_al_conversational_agents_collab_work19} identified three AI roles in collaborative work. As a \textit{peer (ideator)}, AI adds ideas like a team member. As a \textit{facilitator}, it organises ideas and guides teams. As an \textit{expert}, it gives specialist information. The trade-offs concern participant needs and AI capabilities. Some roles may suit certain brainstorming stages: AI as a peer may fit ideation, while the expert role may better fit the converge and decide stages.

\vspace{0.5em}
\noindent \textbf{Human agency scale: \emph{how} is the interaction initiated.} This scale measures preferred collaboration levels, ranging from AI-driven to equal partnership to human-driven, depending on how much collaboration participants need to complete tasks effectively with AI. There is a trade-off between user agency and results~\cite{amershi2014power,MosqueiraRey2023}. In brainstorming, AI might automatically intervene with ideas that participants can only accept or reject (\emph{AI-driven}), participants might initiate interaction or revise AI suggestions (\emph{equal partnership}), or participants always initiate and direct what ideas AI generates (\emph{human-driven}). More control means more work for participants but may improve outcomes~\cite{amershi2014power}, and participants may prefer to initiate AI interactions rather than to receive proactive recommendations~\cite{ladica_chi2025}.

\begin{table*}[t]
\centering
\scriptsize
\caption{
AI interventions in brainstorming from literature, organised by stage, AI role, and human agency.
}
\Description{A table categorising research on AI interventions in brainstorming. Columns represent three brainstorming stages (Ideate, Converge, Decide), subdivided by AI role (Peer, Facilitator, Expert). Rows represent human agency levels (AI-driven, Equal partnership, Human-driven). Cells list relevant citations.}
\label{tab:POI_literature}

\renewcommand{\arraystretch}{1.1}
\setlength{\tabcolsep}{3pt}

\begin{tabular}{@{} l ccc p{1em} ccc p{1em} ccc @{}}
\toprule
& \multicolumn{3}{c}{\cellcolor{gray!20}\textbf{Ideate}} 
& 
& \multicolumn{3}{c}{\cellcolor{gray!35}\textbf{Converge}} 
&
& \multicolumn{3}{c}{\cellcolor{gray!50}\textbf{Decide}} \\
\cmidrule(lr){2-4} \cmidrule(lr){6-8} \cmidrule(lr){10-12}

\textbf{Human Agency} 
& \textit{Peer} 
& \textit{Facilitator} 
& \textit{Expert} 
&
& \textit{Peer} 
& \textit{Facilitator} 
& \textit{Expert} 
&
& \textit{Peer} 
& \textit{Facilitator} 
& \textit{Expert} \\
\midrule

\textbf{AI-driven} 
& \makecell{
    \cite{doshi2024generative, lavric_skraba_brainstorming_with_ai23, muller_et_al_ciao_structure24, memmert_towards_human_ai_collab23, yu_han_et_al_llm_chatbot_human_number_creativity23, nomura_et_al_ibis_brainstorming_ai24, supermind_ideator_ACM}
  }
& --- 
& \cite{supermind_ideator_ACM}
&
& --- 
& \cite{lavric_skraba_brainstorming_with_ai23, ladica_chi2025, genai_cues}
& \cite{nomura_et_al_ibis_brainstorming_ai24, genai_cues}
&
& --- 
& --- 
& --- \\

\textbf{Equal partnership} 
& \makecell{
    \cite{bouschery_et_al_ai_augmented_brainstorming24, kumar2025human, shaer_et_al_ai_augmented_brainwriting24, ladica_chi2025, llm_design_thinking, chiwork_ai_llm, balancing_agency_chiea}
  }
& \cite{ladica_chi2025, balancing_agency_chiea}
& --- 
&
& --- 
& \cite{siangliulue2016ideahound, ladica_chi2025}
& \cite{shaer_et_al_ai_augmented_brainwriting24}
&
& --- 
& \cite{ladica_chi2025}
& --- \\

\textbf{Human-driven} 
& \cite{llmmultiagent}
& \cite{llmmultiagent}
& --- 
&
& --- 
& \cite{llmmultiagent}
& \cite{hou2024c2ideas}
&
& --- 
& \cite{llmmultiagent}
& \cite{hou2024c2ideas} \\
\bottomrule
\end{tabular}

\vspace{0.5em}
{\scriptsize \textit{Note:} Dashes indicate that no studies were found.}
\end{table*}

Our summary in Table~\ref{tab:POI_literature} aligns with a recent systematic review on AI-assisted ideation more broadly~\cite{li2025review}. Most HCI and CSCW studies fall into two groups: (1) AI as a \emph{peer} or idea generator, used almost entirely during ideation with minimal to moderate participant agency, and (2) AI as a \emph{facilitator} or \emph{expert}, used across stages but most often during convergence, again with minimal to moderate agency. AI peers dominate ideation research yet are absent in later stages. The roles of experts and facilitators remain less explored throughout.

\section{\rev{Deriving design requirements for AI interventions through co-design workshops}}\label{sec:codesign}
\rev{To answer (RQ\textsubscript{2})}: \rev{\emph{What are the design requirements for AI interventions at the identified points?}},
\rev{we} conducted a \textbf{formative study} to elicit initial design requirements, followed by a \textbf{co-design study} to iterate and refine them (Figure \ref{fig:methodology}, Step 1C).

\vspace{0.5em}
\noindent\textbf{\rev{Selection criteria for the AI uses.}}
\rev{We selected two orthogonal AI use cases to test the generalisability of our approach. Orthogonal here means the cases differ along two independent axes: \emph{modality} (physical embodied \emph{vs.}\ digital AI), to represent the virtual and physical harms of AI, and the EU AI Act \emph{risk level}~\cite{EUACT2024}. We focus on the \emph{high-risk}, \emph{limited-risk}, and \emph{low-risk} categories, excluding the illegal \emph{prohibited} category. We selected a digital chatbot companion as the limited-risk case~\cite{risk_atlas_website}(Cat. \#456), and a physical AI-assisted kidney monitoring and allocation system as the high-risk case~\cite{risk_atlas_website}(Cat. \#79). This pairing covers the divergences in the EU AI Act's risk categorisation for the uses that require an impact assessment (low-risk uses do not), while being feasible for 15 workshops. Both involve social, ethical, systemic, technical, and economic dimensions, to explore a wide range of potential impacts.}

\begin{figure*}[!tbp]
  \centering
  \includegraphics[width=0.88\linewidth]{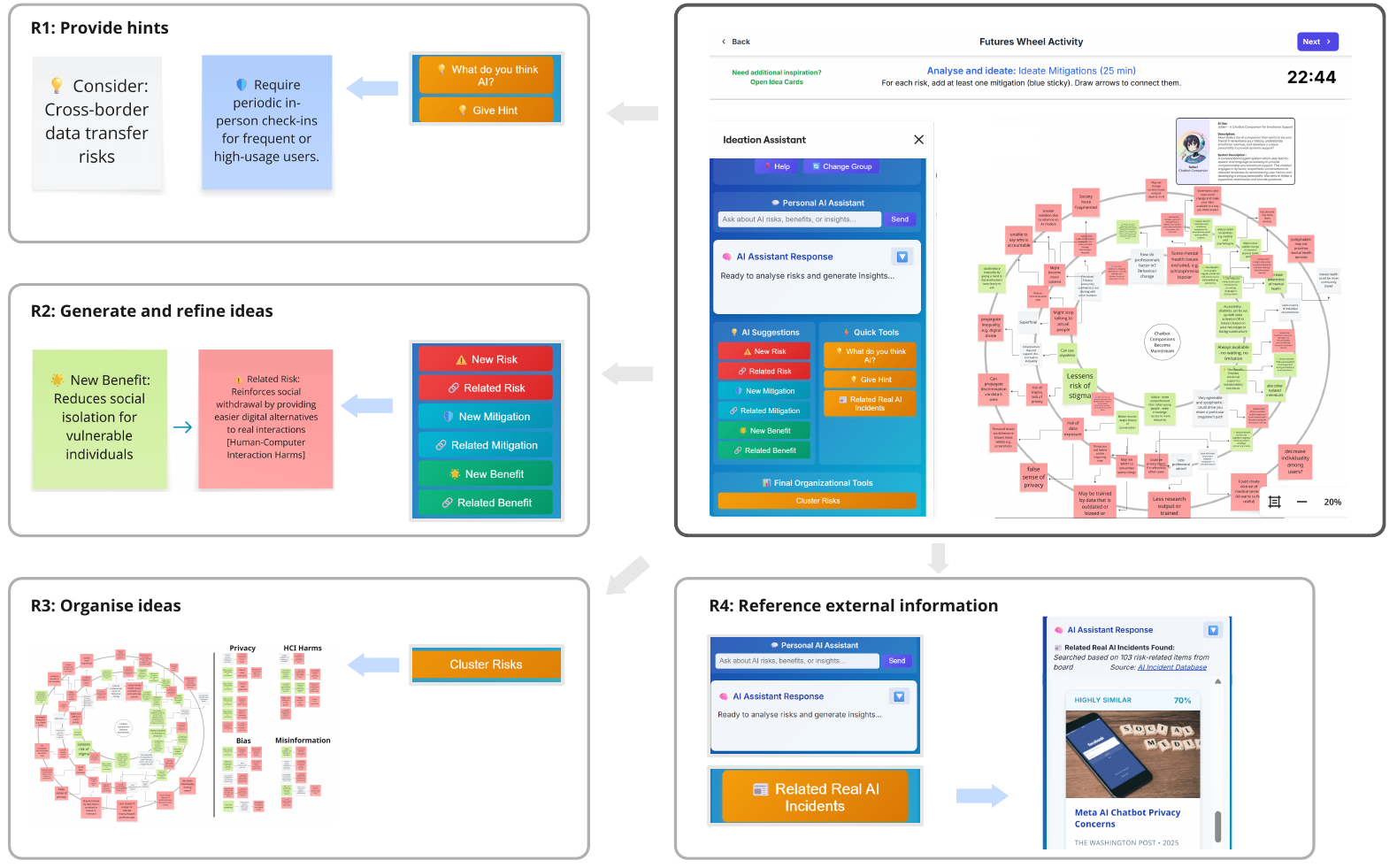}
   \caption{An example from a FW workshop with the Final AI tool and its elements fulfilling the four design requirements.
   }
  \Description{A comparative diagram showing the evolution of an AI assistant's interface, from an "Initial AI (Formative)" version to a "Final AI (Co-design)" version. The diagram links design changes to lists of initial and final requirements. On the far left, a list of "Initial requirements" includes: IR1 Generate ideas, IR2 Organise ideas, and IR3 Refine ideas. Below this, a list of "Final requirements" is divided into "Requirements on functionality" (R1 Generate hints, R2 Organise ideas, R3 Refine ideas, R4 Provide external information) and "Requirements on design" (R5 Simplify AI interface, R6 Adaptable AI interface). The "Initial AI (Formative)" interface, shown at the top, is composed of three modules. The top module allows users to select brainstorming methods like "Futures Wheel" or "Empathy Map" and different AI types like "Chatbot AI". A text box explains this module helps adapt the brainstorming method and enter persona information. Below this, a "Personal AI Assistant" chat module is for asking questions to generate ideas (IR1). The middle module offers main functionalities. On the left, under "AI Suggestions", are buttons like "New Risks" and "Related Risks". On the right, under "Quick Tools", are buttons like "Generate Risk-Mitigation Pairs" and "Cluster Risks". A text box explains these features serve to generate ideas (IR1), refine ideas (IR3), and organise ideas (IR2). The bottom module is method-specific, showing an example for "Empathy Mapping". It has buttons to "Create Mini-Empathy Map", "Auto-Populate Quadrants", and "Analyse Persona". An accompanying text box explains this section generates person-specific ideas (IR1) and analyses persona vulnerabilities (IR2). The "Final AI (Co-design)" interface, shown at the bottom, displays the updated design. The top module has been simplified (R5) for adaptability (R6), now only containing a persona entry field and buttons for "Help" and "Change Group". The "Personal AI Assistant" chat module below it remains the same. The middle module has been made adaptable to the workshop stage (R6) and is shown in two states. State M1 is "Persona Mode", used for creating an empathy map. State M2 is "Impact Assessment Mode", used for ideation on AI impacts. For generating ideas (R1), the function was changed to "What do you think AI?" and "Give a Hint". A "Related AI Incidents" button was added to provide external information (R4). In "Persona Mode", a new "Persona-based risk" button was added for refining ideas (R3). The bottom module was simplified (R5). Features for auto-populating the empathy map were removed due to a lack of usage. The "Cluster Risks" button, a popular feature for organising ideas (R2), was made more prominent. This section is now labelled "Final Organisational Tools".}  \label{fig:ai_tool_figure}
\end{figure*}

\vspace{0.5em}
\noindent\textbf{Study setup.} In both, \rev{formative and co-design} study, participants brainstormed the same AI use: a chatbot companion for emotional support. This choice reflects recent developments such as Replika~\cite{replika, chatbot_companion_relationships_skjuve_et_al21}, character.ai~\cite{character_ai}, virtual influencers~\cite{cherie_et_al_virtual_influencers25}, and xAI's companion-bot Ani~\cite{grok_companion25} \rev{with which most of our participants experienced with AI would be familiar}. We used the Miro platform to support the brainstorming process~\cite{miro_api}, using sticky notes on a canvas and a side-toolbar for our AI tool (Figure~\ref{fig:ai_tool_figure}). 
Participants first completed a Miro tutorial, and then selected a participant to be the coordinator to take notes on Miro and use the AI tool on the team's request. The team then received a description of their AI use case: the ``\emph{conversational agent, Salieri, which uses text-to-speech and language processing to provide companionship and emotional support}'' (Appendix Figure~\ref{fig:ai_use_chatbot_companion}). \rev{Teams were instructed to adopt the perspective of developers working for the AI system’s company. They were asked to identify risks that would be realistic for a developer team to address and to describe them in enough detail that another developer could understand each impact independently and implement a clear mitigation. For example, ``mental health'' represents a broad risk for a chatbot but is too vague for AI developers, whereas ``sycophantic relationships with the companion'' is sufficiently specific for actionable mitigations (e.g., implementing frictions, reflections etc.). Participants were also shown the marking criteria (Table~\ref{tab:output_quality_questions}) to understand how their impacts would be evaluated.} Each workshop lasted 1 hour 45 minutes.

\vspace{0.5em}
\noindent\textbf{Participants.} We recruited 4--7 participants per workshop, as fewer than 4 can limit divergent thinking~\cite{hackman_vidmar_effects_of_size_on_group_performance70}, while more than 7 leads to production blocking, where dominant voices create bottlenecks~\cite{diehl_stroebe_productivity_loss_brainstorming87, hackman_vidmar_effects_of_size_on_group_performance70, heslin_brainwriting09}. \rev{Participants could join only one workshop, so each session had new participants. We recruited from universities and companies via mailing lists and posters across Cambridge, Oxford and London. We did not require degrees, but all had at least a bachelor's, though not necessarily in computer science (e.g., business, English). All workshops were balanced to include private sector workers, researchers, and students, and to have at least 2 computer science backgrounds to prevent one-sided groups while allowing non-expert~\cite{delgado_et_al_participatory_turn23} and multidisciplinary voices~\cite{orwat_normative_2024, kieslich2025scenario}: 37\% were from the private sector, 43\% researchers, and 20\% students. Workshops were held at three in-person locations to balance our 82-person total sample. The mean age was 31 (SD 6.5, range 22–55), close to the UK IT workforce's mean of 35~\cite{uk_it_workforce_mean_age24}. We recruited individuals rather than existing teams to avoid prior relationship biases, and we discuss in \emph{Limitations} the need to explore setups with existing team dynamics for future work. Participants registered for workshops, were assigned to groups, signed consent forms, and received £20 each. \rev{All workshops took place in June--July 2025.}}

\vspace{0.5em}
\noindent\textbf{Data collection.} We collected post-workshop feedback on structure, timing, and design requirements:
\begin{enumerate}
    \item If you were to design a digital tool to help teams like yours brainstorm about the benefits, risks, and mitigations when developing an AI use case, what features would be most important based on your experience today?
    \item What role would you want an AI to have in your brainstorming team, and when would you use it?
\end{enumerate}

\vspace{0.5em}
\noindent\textbf{Analysis.} We audio-recorded and transcribed interactions during in-person sessions and conducted a two-coder inductive thematic analysis of the discussions and responses to identify design requirements~\cite{thematic_analysis_braun_clarke06}. \rev{This involved two-coders identifying themes as they emerged, taking notes during the workshops, and creating a set of coded themes, then revising the themes based on the collected post-study qualitative written questions. Annotators met across three rounds to compare and finalise themes. We used this qualitative approach throughout the formative/codesign workshops.}

\vspace{0.5em}
\noindent\textbf{Results: formative study.} Three initial requirements emerged, each aligned with an AI role we identified from prior work. First, participants wanted AI to act as a an additional team member (\emph{ideator}) and generate ideas: \emph{``I think having an AI as an extra member to contribute new ideas would be cool.''} (P3, EM). We built an agent that generated ideas and shared them at intervals of 5--7 minutes. Second, participants asked for AI to act as an experienced member (\emph{expert}): \emph{``[...] use that information to generate additional ideas or finer-grained nuances.''} (P7, FW). We built an agent that generated ideas \rev{refining} those already on the board. Third, participants wanted AI to support the process (\emph{facilitator}) by grouping ideas: \emph{``I would actually want it to group together the benefits/risks initially generated by humans.''} (P2, EM). We built an agent that clustered ideas into semantic groups. These initial requirements led to our preliminary AI tool.

\vspace{0.5em}
\noindent\textbf{Implementation of the preliminary AI tool.}
We developed the AI tool as a sidebar for Miro users using the Miro API, with PHP and JavaScript connecting to the OpenAI API to provide a chat interface powered by prompt-tuned GPT-4.1. \rev{We used GPT-4.1 as the state-of-the-art performance model at the time to avoid any performance confounds, as well as for its support for web-based search, used for extracting data from the AI Incident Database~\cite{ai_incident_db}.} Participants could generate new and related benefits, risks, and mitigations. We used the six-category DeepMind risk taxonomy~\cite{weidinger_deepmind_taxonomy22} to ensure coverage of categories not already addressed. The tool could place sticky notes, semantically cluster impacts, and suggest additional ones (Appendix Figure~\ref{fig:ai_requirements_figure}). We provide all prompts in Appendix~\ref{appsec:prompt}.

\vspace{0.5em}
\noindent\textbf{Results: co-design study.} We evaluated the preliminary tool to understand when participants would use each AI role (when) and how much agency they desired (how). Participants found our AI peer---which autonomously shared ideas---intrusive: \emph{``the group broadly agreed that the tool's automated pop-ups felt intrusive [...] the length of the text strings disrupted discussion while people read,''} (P27, FW, Initial AI) and \emph{``patronising... [and assumed teams] couldn't freely think up ideas.''} (P17, EM, Initial AI).
They preferred AI to respond only when invited and to provide hints rather than complete ideas, at least during early ideation: \emph{``To derive hints and suggestions based on the use case, encouraging users to engage in brainstorming and critical questioning.''} (P11, EM, Initial AI). Participants valued the AI's expertise during later ideation and early convergence: \emph{``everything generated was, at the very least, correct, if not new. It allowed us to get a new idea when we were somewhat stuck.''} (P13, FW, Initial AI). They also requested external references (e.g., real-world incidents for proposed risks) throughout all stages. Participants were satisfied with the clustering feature. Finally, they asked for a simple interface with minimal distractions that adapts to each brainstorming phase.

\vspace{0.5em}
\noindent\textbf{Final design requirements.} We answer RQ$_1$ with 4 requirements, mapped to phases, roles, and agency levels:
\begin{enumerate}
\item[R1] \textbf{Provide hints}: ideate stage, peer role, human-driven.
\item[R2] \textbf{Generate and refine ideas}: ideate and converge stages, expert role, equal partnership.
\item[R3] \textbf{Organise ideas}: converge stage, facilitator role, human-driven.
\item[R4] \textbf{Reference external information}: all stages, expert role, equal partnership.
\end{enumerate}
And two requirements on \emph{design}:
\begin{enumerate}
   \item[R5] \textbf{Simple interface}: minimise distractions.
    \item[R6] \textbf{Adaptable interface}: show only relevant functionality for each stage and method.
\end{enumerate}

\vspace{0.5em}
\noindent\textbf{Implementation of the final AI tool.}
These requirements shaped our final AI tool, which was more simplified, adaptable to the brainstorming stage, and designed to offer hints on demand rather than generate ideas automatically. The tool also allowed users to surface relevant real-world incidents of AI use-caused harm (Figure~\ref{fig:ai_tool_figure}). \rev{We tuned prompts to limit impacts to 10--15 words after feedback indicated outputs were too long compared to participant entries. To prevent redundancy, we also fed existing Miro board content into the prompts (Appendix~\ref{appsec:prompt}).}

\section{Evaluating AI interventions for AI impact assessment brainstorming}
\label{sec:evaluating}
To answer RQ\textsubscript{\rev{3}}: \emph{What are the effects of AI interventions in team brainstorming on output quality and participant perceptions?}, we compared teams with and without AI across 10 workshops (Figure~\ref{fig:methodology}, Step \rev{3}).

\begin{figure*}[!htbp]
  \centering
  \includegraphics[width=0.61\linewidth]{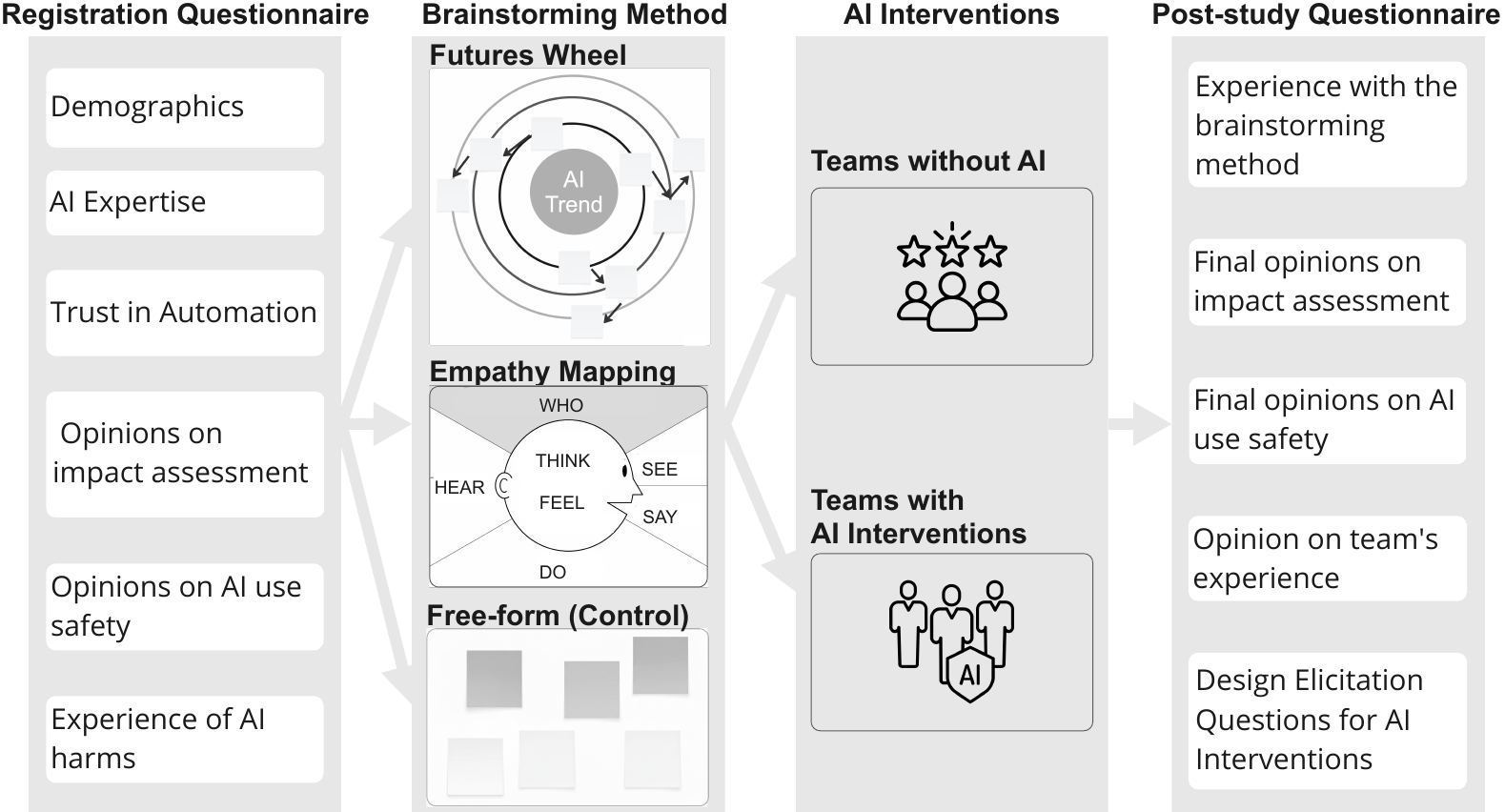}
   \caption{Evaluation study workflow. Registration and pre-study questionnaires collect demographics and initial perceptions. Participants then complete the workshop using one of three methods, with or without AI, and then a final questionnaire.}
  \Description{A flow chart illustrating the design of the study, structured vertically. The chart begins at the top with a 'Registration Questionnaire', which includes sections on 'Demographics', 'AI Expertise', 'Trust in Automation', and 'Opinions on impact assessment'. This leads down to the 'Brainstorming Method' which is either 'Futures Wheel', 'Empathy Mapping', or 'Free-form (Control)'. From this point, the flow splits into two parallel paths. The group is then assigned to either the with AI intervention or without AI intervention condition. Finally, the flow concludes with a 'Post-study Questionnaire'. This final questionnaire contains sections on 'Opinions on AI use safety', 'Experience of AI harms', 'Experience with the brainstorming method', 'Final opinions on impact assessment', 'Final opinions on AI use safety', and 'Opinion on team's experience'.}  
  \label{fig:study_design_flow}
\end{figure*}

We ran two sets of workshops. The first set used a 2$\times$3 design: two conditions (with and without AI) and three brainstorming methods (Futures Wheel [FW], Empathy Mapping [EM], and Free-Form [FF] baseline, Figure~\ref{fig:study_design_flow}). Participants brainstormed the impacts of an AI chatbot companion. The results showed a clear advantage for structured methods over the baseline, so in the second set we focused only on the two structured methods. 

\subsection{Metrics}
\label{sec:measures}

\vspace{0.5em}
\noindent\textbf{Output quality.} \rev{We had 5 expert annotators from a 90 participant pool rate each impact (benefit, risk, and mitigation) \rev{from each of the 15 workshops} on 8 Likert scales (Table~\ref{tab:output_quality_questions}), attaining an inter-rater agreement of 0.69--0.85 (Kendall's W; substantial agreement~\cite{kendall_w}). We recruited our 90 expert pool on Prolific, screening for those who used AI at least 3 times weekly and worked in legal or AI engineering sectors for competency in impact assessment. We included attention checks (specific responses) and comprehension checks (identifying unrelated impacts), and replaced failed candidates}.


\begin{table}[t!]
\centering
\scriptsize
\caption{Output quality annotation questions. All metrics apply to benefits, risks, and mitigations unless noted.}
\label{tab:output_quality_questions}
\renewcommand{\arraystretch}{1.3}
\begin{tabular}{@{} p{2.6cm} c p{10.75cm} l @{}}
\toprule
\textbf{Metric} & \textbf{Scale} & \textbf{Question} & \textbf{Source} \\
\midrule
\rowcolor{gray!7}
Plausibility & 1--5 & Is it plausible that this \{benefit/risk/mitigation\} could arise from (or be implemented for) the described AI use? & \cite{plausibility} \\
Probability & 1--7 & How likely is this \{benefit to be realised / risk to occur / mitigation to be implemented\}? & \cite{nist2023ai} \\
\rowcolor{gray!7}
Uniqueness & 1--3 & Is this \{benefit/risk/mitigation\} unique to this specific AI use? & \cite{kieslich2025scenario} \\
Novelty & 1--3 & Rate the creativity: mundane to imaginative. & \cite{novelty} \\
\rowcolor{gray!7}
Engagement & 1--5 & I find this \{benefit/risk/mitigation\} easy to engage with or apply. & \cite{engagement_1, engagement_2} \\
\midrule
\multicolumn{4}{@{}l}{\textit{Type-specific metrics:}} \\
\midrule
\rowcolor{gray!7}
Magnitude (\textit{Benefits})  & 1--5 & If fully realised, what would be the magnitude of the positive impact? & \\
Severity (\textit{Risks}) & 1--5 & How severe are the impacts of this risk? & \cite{nist2023ai} \\
\rowcolor{gray!7}
Effectiveness (\textit{Mitigations}) & 1--5 & How likely is this mitigation to effectively address its associated risk? & \\
\bottomrule
\end{tabular}
\end{table}
\begin{table}[!t]
\centering
\scriptsize
\caption{Pre/post-workshop perception questions: ``If you were responsible for assessing, designing, or developing this AI...''}
\label{tab:perception_questions}
\renewcommand{\arraystretch}{1.3}
\begin{tabular}{@{} p{2cm} c p{12cm} @{}}
\toprule
\textbf{Metric} & \textbf{Scale} & \textbf{Question} \\
\midrule
\rowcolor{gray!7}
Control & 1--5 & How much control do you feel you would have? \\
Anxiety & 1--5 & How anxious or uneasy would you feel? \\
\rowcolor{gray!7}
Confidence & 1--5 & How confident would you feel in carrying out a risk impact assessment? \\
Oversight & 1--5 & How likely is it that this AI system would still require regular updates and oversight, even if perfectly designed and tested? \\
\rowcolor{gray!7}
Recommendation & 1--5 & How comfortable would you feel using it for yourself or recommending it to a close friend or family member? \\
\bottomrule
\end{tabular}
\end{table}

\noindent\textbf{Participant perceptions.} Before and after each workshop, we measured participants' sense of control, anxiety, confidence in risk assessment, views on oversight, and comfort recommending the AI use (Table~\ref{tab:perception_questions}).

\subsection{Analysis}
\label{sec:analysis}

\vspace{0.5em}
\noindent\textbf{Quantitative analysis.} We used Cumulative Link Models (CLMs)~\cite{mccullagh1980regression} to analyse both output quality and participant perceptions, as appropriate for ordinal Likert-scale data.
For output quality, inter-rater reliability among the five expert coders was strong (\rev{overall} Kendall's W = 0.71)~\cite{kendall_w}. We modelled each quality metric with brainstorming method and AI condition as predictors. For participant perceptions, we modelled post-workshop responses while controlling for pre-workshop scores, AI expertise, education, and self-rated ethical alignment. We conducted estimated marginal means (\texttt{emmeans}) analysis with Tukey adjustment, back-transforming results to Likert values to represent the shifts between AI/No-AI conditions. Variance Inflation Factors were below 5, confirming no significant predictor multicollinearity~\cite{vif_collinearity_rawlings98}.

\vspace{0.5em}
\noindent\textbf{Qualitative analysis.} Two coders conducted inductive thematic analysis on participants' written responses using the \rev{aforementioned} approach~\cite{thematic_analysis_braun_clarke06}, supplemented by notes and audio transcripts from the workshops. \rev{We triangulate our findings with the first two targeted around quantitative results, and findings 3--5 focusing on the dominant themes that emerged from our qualitative insights (i.e., activation at design saturation/fixation, scaffolding \emph{vs.} directing, intrusiveness \emph{vs.} proactivity, agency accruing over time). Both coders reviewed the data across three rounds to meet and resolve disagreements.}

\begin{table*}[!t]
\centering
\scriptsize
\caption{Results by AI condition for \emph{chatbot companion} and \emph{medical AI} uses. Values are mean {$\pm$ \small{SE}} from CLM models. \textcolor{resultBlue}{\textbf{Blue}}/\textcolor{resultOrange}{\textbf{orange}} indicate significantly higher/lower values (Tukey-adjusted). Significance: {*}$p<.05$, {**}$p<.01$, {***}$p<.001$.}
\label{tab:results_combined}

\renewcommand{\arraystretch}{1.1}
\setlength{\tabcolsep}{4pt}

\vspace{0.5em}
\textbf{(A) Output Quality}
\vspace{0.3em}

\begin{tabular}{@{} ll lllllll @{}}
\toprule
& & \textbf{Plausibility} \tiny{(1--5)} & \textbf{Probability} \tiny{(1--7)} & \textbf{Uniqueness} \tiny{(1--3)} & \textbf{Novelty} \tiny{(1--3)} & \textbf{Impact\textsuperscript{\textdagger}} \tiny{(1--5)} & \textbf{Engagement} \tiny{(1--5)} & \textbf{Count} \\
\midrule

\multicolumn{9}{@{}l}{\cellcolor{gray!7}\textit{Benefits}} \\
\multirow{2}{*}{Chatbot} 
  & AI & \textcolor{resultBlue}{\textbf{4.08**}} \tiny{±.15} & \textcolor{resultBlue}{\textbf{5.49***}} \tiny{±.17} & 1.99 \tiny{±.18} & \textcolor{resultBlue}{\textbf{2.02*}} \tiny{±.16} & \textcolor{resultBlue}{\textbf{3.64***}} \tiny{±.16} & \textcolor{resultBlue}{\textbf{4.05***}} \tiny{±.16} & \textcolor{resultBlue}{\textbf{68}} \\
  & No AI & \textcolor{resultOrange}{\textbf{3.82***}} \tiny{±.07} & \textcolor{resultOrange}{\textbf{4.98***}} \tiny{±.11} & 1.88 \tiny{±.06} & \textcolor{resultOrange}{\textbf{1.90*}} \tiny{±.05} & \textcolor{resultOrange}{\textbf{3.30***}} \tiny{±.07} & \textcolor{resultOrange}{\textbf{3.59***}} \tiny{±.06} & \textcolor{resultOrange}{29} \\
\multirow{2}{*}{Medical} 
  & AI & 3.66 \tiny{±.21} & 4.99 \tiny{±.21} & \textcolor{resultOrange}{\textbf{1.90*}} \tiny{±.23} & 1.93 \tiny{±.22} & 3.59 \tiny{±.21} & 3.75 \tiny{±.21} & \textcolor{resultBlue}{\textbf{33}} \\
  & No AI & 3.73 \tiny{±.09} & 5.11 \tiny{±.11} & 2.09 \tiny{±.06} & 2.01 \tiny{±.06} & 3.61 \tiny{±.08} & 3.77 \tiny{±.08} & \textcolor{resultOrange}{27} \\
\midrule

\multicolumn{9}{@{}l}{\cellcolor{gray!7}\textit{Risks}} \\
\multirow{2}{*}{Chatbot} 
  & AI & \textcolor{resultBlue}{\textbf{3.86***}} \tiny{±.10} & \textcolor{resultBlue}{\textbf{4.83***}} \tiny{±.10} & 1.89 \tiny{±.11} & 1.83 \tiny{±.11} & \textcolor{resultBlue}{\textbf{3.56***}} \tiny{±.10} & \textcolor{resultBlue}{\textbf{3.76***}} \tiny{±.10} & \textcolor{resultBlue}{\textbf{96}} \\
  & No AI & \textcolor{resultOrange}{\textbf{3.56***}} \tiny{±.05} & \textcolor{resultOrange}{\textbf{4.40***}} \tiny{±.06} & 1.87 \tiny{±.03} & 1.80 \tiny{±.03} & \textcolor{resultOrange}{\textbf{3.30***}} \tiny{±.04} & \textcolor{resultOrange}{\textbf{3.55***}} \tiny{±.04} & \textcolor{resultOrange}{80} \\
\multirow{2}{*}{Medical} 
  & AI & 3.70 \tiny{±.17} & 4.88 \tiny{±.17} & \textcolor{resultBlue}{\textbf{1.96**}} \tiny{±.18} & 1.80 \tiny{±.18} & 3.51 \tiny{±.17} & 3.72 \tiny{±.18} & \textcolor{resultBlue}{\textbf{46}} \\
  & No AI & 3.65 \tiny{±.09} & 4.99 \tiny{±.08} & 1.75 \tiny{±.05} & 1.70 \tiny{±.05} & 3.53 \tiny{±.07} & 3.69 \tiny{±.06} & \textcolor{resultOrange}{42} \\
\midrule

\multicolumn{9}{@{}l}{\cellcolor{gray!7}\textit{Mitigations}} \\
\multirow{2}{*}{Chatbot} 
  & AI & \textcolor{resultBlue}{\textbf{3.94***}} \tiny{±.11} & \textcolor{resultBlue}{\textbf{5.06***}} \tiny{±.13} & 1.86 \tiny{±.11} & 1.86 \tiny{±.11} & \textcolor{resultBlue}{\textbf{3.67***}} \tiny{±.11} & \textcolor{resultBlue}{\textbf{3.78***}} \tiny{±.11} & \textcolor{resultBlue}{\textbf{78}} \\
  & No AI & \textcolor{resultOrange}{\textbf{3.31***}} \tiny{±.05} & \textcolor{resultOrange}{\textbf{4.55***}} \tiny{±.08} & 1.87 \tiny{±.03} & 1.83 \tiny{±.03} & \textcolor{resultOrange}{\textbf{3.40***}} \tiny{±.05} & \textcolor{resultOrange}{\textbf{3.54***}} \tiny{±.05} & \textcolor{resultOrange}{54} \\
\multirow{2}{*}{Medical} 
  & AI & 3.70 \tiny{±.19} & 4.79 \tiny{±.19} & 1.89 \tiny{±.20} & 1.79 \tiny{±.20} & \textcolor{resultBlue}{\textbf{3.85*}} \tiny{±.19} & 3.76 \tiny{±.19} & \textcolor{resultBlue}{\textbf{59}} \\
  & No AI & 3.56 \tiny{±.10} & 4.94 \tiny{±.11} & 1.80 \tiny{±.06} & 1.68 \tiny{±.05} & \textcolor{resultOrange}{\textbf{3.68*}} \tiny{±.07} & 3.69 \tiny{±.07} & \textcolor{resultOrange}{29} \\
\bottomrule
\end{tabular}
\vspace{0.5em}

{\scriptsize \textsuperscript{\textdagger}Impact = Magnitude (benefits), Severity (risks), or Effectiveness (mitigations).}

\vspace{1em}
\textbf{(B) Participant Perceptions} {(all scales 1--5)}
\vspace{0.3em}

\begin{tabular}{@{} ll lllll @{}}
\toprule
& & \textbf{Control} & \textbf{Anxiety} & \textbf{Confidence} & \textbf{Oversight} & \textbf{Recommendation} \\
\midrule
\multirow{2}{*}{Chatbot} 
  & AI & \textcolor{resultBlue}{\textbf{2.89*}} \tiny{±.19} & \textcolor{resultOrange}{\textbf{1.77*}} \tiny{±.13} & \textcolor{resultBlue}{\textbf{3.80*}} \tiny{±.21} & 2.60 \tiny{±.13} & \textcolor{resultBlue}{\textbf{2.74*}} \tiny{±.17} \\
  & No AI & \textcolor{resultOrange}{\textbf{2.04}*} \tiny{±.13} & \textcolor{resultBlue}{\textbf{2.47}*} \tiny{±.27} & \textcolor{resultOrange}{\textbf{3.03*}} \tiny{±.30} & 2.52 \tiny{±.24} & \textcolor{resultOrange}{\textbf{1.97}*} \tiny{±.30} \\
\addlinespace
\multirow{2}{*}{Medical} 
  & AI & 1.99 \tiny{±.21} & 2.71 \tiny{±.24} & 1.77 \tiny{±.26} & 2.35 \tiny{±.32} & 2.58 \tiny{±.23} \\
  & No AI & 2.36 \tiny{±.21} & 2.81 \tiny{±.26} & 2.00 \tiny{±.33} & 2.51 \tiny{±.27} & 2.63 \tiny{±.25} \\
\bottomrule
\end{tabular}
\end{table*}

\subsection{Results}
\label{sec:output_quality}
Based on the quantitative results, as supplemented with qualitative feedback, we answer RQ$_3$ with five main findings: two addressing the quality of the impacts generated and the participants' perceptions, and three on the points for AI intervention.

\vspace{0.5em}
\noindent \textbf{Finding 1: AI interventions improved the output quality for the general AI use but not for the specialised one.} 
For the chatbot companion, AI intervention teams generated 48\% more impacts than human-only teams (Table~\ref{tab:results_combined}A), scoring significantly higher on six of the eight quality metrics but not on the two creativity measures (uniqueness and novelty). For the medical AI, the AI intervention helped yield 22\% more impacts, with increases in unique risks and effective mitigations. Thus, AI interventions can help teams identify more risks and mitigations early in the design phase with the quality being at least equal to the teams without AI (for the specialised medical AI) or higher quality (for the general chatbot companion). However, AI impacts were equally creative to human-made ones. \rev{We thematically observed that human-generated risks were more systemic and longitudinal, than those from the AI---which predominantly targeted model capability risks.}


\vspace{0.5em}
\noindent \textbf{Finding 2: AI interventions improved participant perceptions for the general AI use case but not for the specialised one. Participants experienced reduced anxiety, increased confidence, and a greater sense of control.} 
As Table~\ref{tab:results_combined}B shows, for the chatbot companion, AI intervention teams reported significantly less anxiety compared to teams without AI. P20 (EM, Final AI) noted: \emph{``I was not seeing these risks as unmanageable or insurmountable.''} Participants also reported a greater sense of control: \emph{``I can influence a bit more on the development of these kind of AI chatbots''} (P24, FF, Final AI). 
They expressed increased confidence in risk assessment: \emph{``I was more attuned to the benefits of such a system, whilst also seeing risks that I was otherwise unaware of''} (P20, EM, Final AI), and higher likelihood of recommending the AI to others: \emph{``I became more confident with the system design approach which highlights that engineers and developers are trying hard to mitigate potential risks''} (P59, EM, Final AI). For the medical AI, qualitative feedback indicated that some participants in these teams also experienced reduced anxiety and increased confidence when using the AI tool.

\vspace{0.5em}
\noindent \textbf{Finding 3 (when): Teams mostly used AI when facing idea saturation or nearing time limits.} 
Participants identified idea saturation as a key challenge, particularly for the medical AI groups, as evidenced by the significantly lower number of impacts generated in these teams. We observed how participants used the AI tool to break through idea saturation and design fixation. As 
P68 (EM, Final AI) noted, it \emph{``was helpful to come up with novel items whenever we as a group got stuck.''} The other common use was when time ran low:
\emph{``It was helpful in filling in the blanks if time was getting short and seeing if we'd missed things''} (P45, FW, Final AI). However, timing mattered: when used under pressure, team members became \emph{``less critical of what the tool said''} (P40, FW, Final AI).

\vspace{0.5em}
\noindent \textbf{Finding 4 (what): AI should primarily serve as a facilitator and expert rather than as another peer generating ideas.}
Our results reveal an unexpected preference: participants wanted AI to function more as a facilitator and expert rather than as another team member. The AI facilitator proved most useful, with the clustering feature consistently used and readily adopted. As P44 (FF, Final AI) explained: \emph{``I feel like it would be much better suited as a facilitator by note-taking, transcribing, organising. I don't feel very comfortable taking ideas from the AI.''}
Participants also valued the AI's domain expertise in generated impacts and the `Related AI Incidents' feature, which grounded discussions in real examples like actual cases of romance fraud with relationship chatbots, making risks more concrete and manageable: \emph{``The related AI incidents feature was especially useful, providing insight into the real-life implications of each benefit''} (P64, FW, Final AI). Additional feedback revealed deeper epistemic dilemmas that made participants reluctant to use AI for ideation: \emph{``it can influence us, maybe we would come with different ideas, so not sure if it is better or not''} (P61, EM, Final AI) and \emph{``it is teaching people to use the right answer, so this is teaching them not to think''} (P47, FW, Final AI).

\vspace{0.5em}
\noindent \textbf{Finding 5 (how): Participants preferred maintaining control over AI interaction, except for process stages they considered `boring.'}
When not facing idea saturation or time constraints, participants preferred to maintain control over AI interaction and develop their own ideas first: \emph{``It was useful although none of us wanted to use it until we'd come up with our own ideas first''} (P46, FW, Final AI), and \emph{``it can be used in the very end to test ideas or check for an extra opinion''} (P14, EM, Final AI).
On the other hand, they welcomed AI for tedious tasks: \emph{``removing the need for a coordinator to manually attach post-its on the Miro board, this automation would make the process smoother and more efficient''} (P31, FF, Final AI), and \emph{``One key feature I'd add is automatic speech-to-text with real-time summarisation. This would allow team members to speak freely while the tool captures, organises, and summarises key points''} (P35, FW, Final AI).

\section{Discussion}\label{sec:discussion}
\rev{We structure our discussion around our research questions. We first offer general design guidance for appropriate points of AI intervention in brainstorming (RQ\textsubscript{1}, \S\ref{sec:design_guideance}). Next, we discuss how the teams' collective experiences influenced the design requirements for AI assistance in brainstorming for impact assessment (RQ\textsubscript{2}, \S\ref{sec:misalignment}). We then examine the effect of AI interventions on brainstorming outputs and participant perceptions (RQ\textsubscript{3}, \S\ref{sec:alignment}) in comparison to prior work. We conclude with the limitations and methodological reflections (\S\ref{sec:limitations}).}

\begin{figure*}[!bp]
  \centering
  \includegraphics[width=0.77\linewidth]{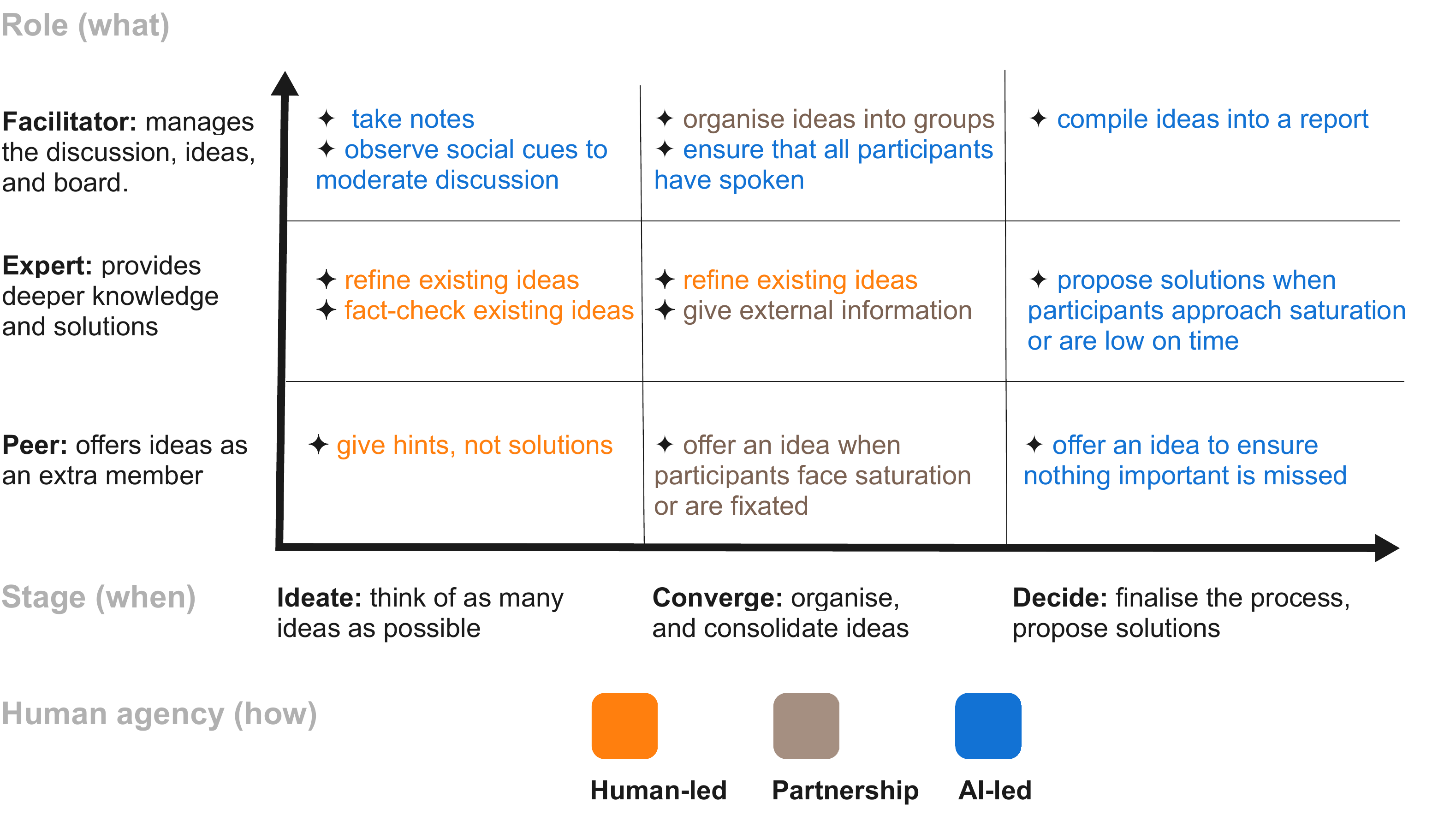}
    \caption{Broad guidance for AI interventions in brainstorming, based on the stage (when, $x$-axis), role (what, $y$-axis), and the level of human agency (how, the colour legend). Over time, AI-led interventions can slowly increase.}
    \Description{A conceptual chart illustrating different roles and their associated actions across three phases of a creative discussion process. The chart is structured along a horizontal axis labelled 'Timing', which shows the progression of the discussion through three sequential phases: Ideate , Converge, and Decide. The chart also presents a framework for interventions that can be 'Human-led Mixed' or 'AI-led'. The three phases are defined as follows: Ideate: A phase where participants think of as many ideas as possible. Converge: A phase for grouping, organising, and consolidating the ideas. Decide: A phase to finalise the ideation and propose solutions. Mapped against these phases are three roles that can intervene in the process, each with specific actions: 1. Facilitator: This role organises and manages the discussion and board, and groups ideas. Its actions in the Ideate phase are to 'observe social cues to moderate group discussions' , 'reduce production blocking' , and 'ensure that all participants have spoken'. In the Converge phase, the action is to 'group ideas into themes'. In the Decide phase, the action is to 'compile ideas into a report'. 2. Expert: This role provides deeper knowledge and solutions on the topic. Its actions in the Ideate phase are to 'fact-check ideas'  and 'provide related ideas'. In the Converge phase, the action is to 'refine ideas'. In the Decide phase, actions are to 'refine ideas' , 'offer solution when participants approach saturation or are low on time' , and 'give hints, not solutions'. 3. Peer: This role provides probing ideas to guide conversation as an extra member. Its actions in the Ideate phase are to 'offer a novel idea when participants approach saturation or are fixated'. In the Converge phase, the action is to 'ensure the team has not missed any categories/domains of ideas'.}
  \label{fig:discussion_poi_figure}
\end{figure*}

\subsection{\rev{Design guidance for points of intervention in AI-assisted brainstorming}}\label{sec:design_guideance}
\rev{Regarding RQ\textsubscript{1} (\emph{``At which points in established brainstorming methods can AI meaningfully intervene to support teams in AI impact assessment?''}),} our approach deviates from existing AI idea generators by enabling teams to use AI to \textit{supplement their brainstorming through specific points of intervention} rather than offering shortcuts for thinking. Although the AI tool could automate the entire task, fewer than 25\% of ideas were generated directly from the AI. AI designers should consider \textit{when} (timing), \textit{what} (role), and \emph{how} (human agency) AI should intervene (Figure~\ref{fig:discussion_poi_figure}). When one considers the role of an `agent' it is important to consider its level of agency across the brainstorming process to ensure that expertise and experiences of the diverse team are not sidelined or replaced by AI, while issues (such as fixation, saturation, or bottlenecks) can be addressed by incremental AI agency---which we frame as \rev{appropriate AI} \emph{points of intervention}.
Based on our findings, human agency should be highest in early stages and can gradually decrease as brainstorming progresses. Across all stages, the \textit{ideator} role should remain primarily human-driven, with AI contributing only during idea saturation, design fixation, or prolonged silence. The \textit{facilitator} and \textit{expert} roles lend themselves to greater AI involvement or equal partnership.

\noindent\emph{Ideate stage.} At this first stage of brainstorming (Figure~\ref{fig:discussion_poi_figure}), AI should serve as an on-demand \emph{expert} for fact-checking and contextual information, and offer hints indicating areas of ideation rather than concrete solutions.

\noindent\emph{Converge stage.} AI intervention can increase, with the primary role being to \textit{consolidate} and \textit{refine} by grouping ideas, identifying duplicates, and surfacing under-represented themes---though not replacing human-generated ideas with its own. The expert role of correcting errors and providing missing context can be AI-driven.

\noindent\emph{Decide stage.} AI can help teams converge under time constraints by rapidly generating ideas for coverage, or lead discussion by proposing concrete solutions and `wild card' ideas as launching points for new discussion topics.

AI-driven or equal partnership approaches are appropriate for \emph{mediator} functions such as managing speaking time, preventing dominant voices, addressing free-riding, note-taking, and proposing summaries. Future AI tools could explore a mediator with audio-visual support (listening to the conversation) to address the core challenges of group brainstorming~\cite{stroebe_et_al_challenges_brainstorming10} of \emph{free-riding} (participants offload cognitive effort to others or AI), \emph{social inhibition} (individual productivity drops in group settings), and \emph{production blocking} (i.e., turn-taking constrains idea flow).

\subsection{\rev{Teams want AI to support their process, not replace their thinking}}\label{sec:misalignment}

\rev{Regarding RQ\textsubscript{2} (\emph{``What are the design requirements for AI interventions at these points?''}), our co-design process and evaluation together reveal that participants' requirements centre not on \emph{what AI can generate} but on \emph{how AI preserves team agency}. Two findings from our evaluation challenge assumptions from prior work.}

\smallskip
\noindent\rev{\textbf{The field's focus on AI-as-idea-generators contradicts what teams actually want.}
Most prior systems provide AI-generated ideas during early ideation~\cite{doshi2024generative, lavric_skraba_brainstorming_with_ai23, supermind_ideator_ACM, herdel_et_al_exploregen25, buçinca_aha_generating_ai_risks23, wang_et_al_farsight24, rao_et_al_riskrag25}. Our participants preferred the opposite: AI as a facilitator and expert rather than a peer. The reason is epistemic, as generating ideas is how teams build a genuine understanding of their consequences, and the \emph{process} of deliberation builds the safety culture that organisations need~\cite{seafty_in_ai23, von_solms_security_culture04}. This challenges the dominant view of peer idea-generators in AI-assisted brainstorming~\cite{li2025review} and suggests redirecting effort towards a phased team-centric facilitation-peer-expert mixed approach. This aligns with the call from \citet{sandoval_et_al_ibm_multidiscilinary_assess25} for multidisciplinary and unconventional approaches for AI assessments, as we identified that brainstorming can be limited by the teams domain expertise, while AI can be tuned to help team's identify a broader range of impacts by integrating risk taxonomies.}

\smallskip
\noindent\rev{\textbf{Pulled AI outperforms pushed AI.}
The shift from our Initial AI (pushing ideas on its own, perceived as \emph{``patronising''} (P17)) to our Final AI (responding only when asked) reduced anxiety and increased control. Extending \citet{amershi2014power} to group settings, we found that teams used AI more selectively but more effectively when they controlled timing, as fewer than 25\% of the final ideas came from the AI. This supports AI's use as a user-controlled selective intervention rather than as an autonomous agent (i.e., an `interventionist mindset')~\cite{interventionist_mindset25} for user-initiated interaction~\cite{ladica_chi2025}. Hence there is a need to assess the \emph{appropriateness} of each intervention when considering AI in human-led tasks, rather than the widespread application of AI into every facet of brainstorming.}


\subsection{\rev{AI augmentation helps teams think more, while domain expertise influences the ceiling}}\label{sec:alignment}

\rev{On RQ\textsubscript{3} (\emph{``What is the effect of AI interventions on team brainstorming for impact assessment?'')}, we reveal that AI-assisted teams consistently delivered more impacts of equal or better quality than the human-only teams.}

\smallskip
\noindent\rev{\textbf{AI amplifies the expertise already in the room.}
For the broadly understood chatbot companion, AI interventions improved the output quality on six of eight metrics and participant perceptions on three of five measures. For the specialised medical AI, gains were more targeted—limited to unique risks and effective mitigations. This suggests that AI augmentation is most powerful when teams can contextualise and build on its suggestions, aligning with \citet{supermind_ideator_ACM}, who found that useful AI interventions depend on type of problem, and \citet{llmmultiagent}, who showed that multi-agent AI effectiveness drops for specialist topics. Our results extend these findings: the bottleneck lies in the \emph{interaction} between the AI and the team's capability, where AI can help teams but it is not a panacea. Future AI for specialised uses could consider pairing AI suggestions with explanations or confidence indicators to help teams engage with unfamiliar output~\cite{ruan_et_al_llm_divergent_performance25}.}

\smallskip
\noindent\rev{\textbf{AI can only supplement, not replace, human creativity.}
AI-assisted teams produced more impacts of equal or higher quality, with uniqueness/novelty unchanged to human-only teams, confirming \citet{de2025has}'s meta-review and \citet{meincke2025chatgpt}'s diversity findings, and broader work on generative AI's creativity limitations~\cite{ding2025generative}. Participant backgrounds also shaped ideation (e.g., a policy worker recommending \emph{``FDA-like approvals''} (W1) for chatbots), with AI-made risks tending to be more legal or model capability-themed. This does not mean that AI only generates known impacts, as more impacts (AI-assisted teams) with equal mean uniqueness score (\emph{vs.} human-only) indicates idea diversity.}

\smallskip
\noindent\rev{\textbf{`Less AI early, more AI later' for Impact Assessment.}
Participants used AI less during ideation and more during the convergence and decision stages. Akin to \citet{10.1145/302979.303030}'s `value-added automation', teams' had an implicit cost-benefit analysis: early on, they relied on their own conversational flow; later, AI's contributions helped restart conversations or expand under-addressed topics (via the AI's impact clustering function [R3]). The implication here is that impact assessment tools should detect and intervene mostly at saturation or fixation~\cite{jansson_smith_design_fixation91} rather than expecting equal use throughout.}

\subsection{Limitations}
\label{sec:limitations}

\noindent\emph{Choice of brainstorming methods.} \rev{We empirically evaluated the methodologies used in related work ~\cite{hohendanner_et_al_global_dialogues25} and the AI-augmented approach implemented in corporate risk assessment platforms such as 4Strat~\cite{4strat} and Futures Platform~\cite{futures_platform}, with our study presenting a gateway for evaluating AI in human-led impact assessment. Nonetheless, future work could test different AI uses in-between our contrasts and other possible brainstorming methods.}

\noindent\emph{Group size and \rev{dynamics}.} Our in-person workshops used recommended group sizes (4--8 participants) \cite{uk_govt_futures_toolkit24, glenn_futures_wheel, gray_gamestorming_book10}. While prior work explored individual ideation tools for impact assessment~\cite{rao_et_al_riskrag25, buçinca_aha_generating_ai_risks23, herdel_et_al_exploregen25, eisenberg_et_alunified_control_framework_ai_gen_risks25}, future work should test small groups, large organisational brainstorming, and even asynchronous communication modes. \rev{While our approach provides a clean slate and consistency between groups, examining how pre-existing leadership and power dynamics influence brainstorming could help refine the AI interventions needed to support it.}

\noindent\emph{AI functionality.} Even though we explored three varied roles, participants expressed interest in functionality that would fall under yet other roles, e.g., a moderator (to prevent dominating voices in the discussion, and invite those who are socially inhibited to speak) or mediator (to identify common ground and propose new ideas~\cite{govers24_mediators}).

\noindent\emph{LLM models.} We found that the AI interventions were less effective for uncommon AI uses that require expert knowledge, such as the medical AI. Future models could utilise tailored context to the specific use~\cite{ruan_et_al_llm_divergent_performance25}.

\noindent\emph{Participants.} Our participants were mainly educated with AI experience to mimic AI stakeholders who would be involved in designing an AI product. However, impact assessment brainstorming could also act as an educational tool for younger audiences, and future work could explore AI interventions for varying audiences.

\section{Conclusion}
\label{sec:conclusion}

Our evaluation of AI-assisted brainstorming reveals a clear directive: teams want AI to help them think but not to replace their judgement. At later stages, they want it to function more as a facilitator and expert than as a peer.
At times, AI assistance improved the quality of impacts and mitigations, and reduced participant anxiety. Thus, the path forward for responsible AI requires designing tools as targeted interventions that amplify human creativity and preserve agency to ensure that team members are aware and a part of an AI safety culture.

\clearpage

\section{Ethical considerations}
This study received ethics approval through our organisation's internal research ethics process.
Ethically, it is important to note that LLM's are as powerful as its training data, and thus will reflect its biases~\cite{chatgpt_pol_bias_motoki_et_al24, chatgpt_pol_bias_motoki_et_al24, llm_pol_bias_feng_et_al23}. Furthermore, while participants appreciated the expert capability of providing related real-world incidents, potential future wild-cards and unexpected out-of-training-data ideas must be considered. As is the modus operandi of this paper, lived experiences and concerns should not be ignored when conducting impact assessments. Similarly, persona design and evaluation should prioritise inclusivity while avoiding stereotypes~\cite{persona_stereotypes}.

Research in Impact Assessment presents risks of dual-use, where increasing peoples ability to generate harmful risks via AI could be maliciously exploited. For instance, AI-generated risks for the chatbot companion included those related to criminal activities such as fraud, indoctrination, radicalisation, and exploitation. Thus, responsible AI research for impact assessment should also encourage a `sword and shield' approach~\cite{govers23_promptgan}, where any adversarial problem (risk) generator should provide a solution (mitigation) generator to counter it, as we have implemented.


\section{{Generative AI Disclosure Statement}}
{Our AI tool utilises generative AI for generating impacts, though manually verified via prompt-tuning and curation of its knowledge databases. Writing in this paper was human-written, with Generative AI tools only used in editing and proofing in Overleaf.}

\bibliographystyle{ACM-Reference-Format}
\bibliography{main}

\appendix
\section{\rev{Strings to Replicate the AI-assisted Brainstorming \& Ideation for Impact Assessment Literature Review}}\label{appsec:AI_brainstorming_AIIA}
We present the full information for reproducing the studies discussed in the related work using our PRISMA-based approach (searching for tools up to July 2025).
We canvassed existing AI methods used in any capacity for generating benefits, risks and/or mitigations for AI using the following search string across ACM Digital Library, and IEEE Xplore:

\begin{quote}
\footnotesize
\texttt{("AI risk" OR "artificial intelligence safety" OR "AI system failure" OR "AI incident" OR "AI hazards" OR "AI vulnerabilities") AND \\
("identification" OR "elicitation" OR "detection" OR "exploration" OR "scenario analysis" OR "risk analysis" OR "hazard analysis") AND \\
("mitigation" OR "response strategy" OR "intervention" OR "remediation" OR "risk control") AND \\
("tool" OR "framework" OR "interactive system" OR "decision support system" OR "design aid" OR "analysis platform") AND \\
("human-in-the-loop" OR "brainstorming" OR "collaborative" OR "co-creative" OR "explanation" OR "exploratory")}
\end{quote}

\begin{table}[!ht]
  \caption{Studies found and filtered}  \Description{Outcome of the AI tools for impact assessment SLR, giving 104 after string-based search, 20 after title and abstract screen, 5 after full text screening, and 23 final studies after snowball sampling.}  \label{tab:studies_found_and_filtered}
  \footnotesize
  \begin{tabular}{lll}
    \toprule
    Screen Type&Study Count\\
    \midrule
    Search Strings&104\\
    Title and Abstract Screen&20\\
    Full Text Screening&5\\
    After Snowball Sampling&8\\
  \bottomrule
\end{tabular}
\end{table}


\subsection{Systematic review of AI-assisted brainstorming and ideation in a variety of fields}\label{appsec:AI_brainstorming}
\rev{Same criteria as above, but for AI in \textit{any} field of brainstorming, including for use outside of impact assessment.}

\begin{quote}
\footnotesize
\texttt{[[Abstract: "generative ai"] OR [Abstract: llm] OR [Abstract: "large language model"]] AND [[Abstract: brainstorm*] OR [Abstract: ideation] OR [Abstract: "idea generation"]] AND [[Abstract: group] OR [Abstract: team] OR [Abstract: collaborat*] OR [Abstract: "co-creation"]] AND [[Abstract: tool] OR [Abstract: intervention] OR [Abstract: ideator] OR [Abstract: expert] OR [Abstract: peer] OR [Abstract: facilitator] OR [Abstract: generator] OR [Abstract: assistant]]}
\end{quote}

\begin{table}[!ht]
  \caption{AI interventions in Brainstorming Studies found and filtered}  \Description{Outcome of the AI interventions in brainstorming SLR, giving 27 after string-based search, 17 after title and abstract screen with none removed, 23 final studies after snowball sampling.}  \label{tab:brainstorming_tudies_found}
  \footnotesize
  \begin{tabular}{lll}
    \toprule
    Screen Type&Study Count\\
    \midrule
    Search Strings&27\\
    Title and Abstract Screen&17\\
    Full Text Screening&17\\
    After Snowball Sampling&23\\
  \bottomrule
\end{tabular}
\end{table}


\section{\rev{Detailed Brainstorming method Instructions for Replication and Transparency}}\label{appsec:select_brainstorming_methods}
\rev{Beyond the description in the main text, we provide the full instructions and steps for the brainstorming methods:}

\vspace{1em}
\noindent\textbf{Futures Wheel.} It is a structured foresight method for \textit{divergent-thinking} which systematically explores potential outcomes across chains of consequences. It works by placing a central trend in the centre of a canvas (in our case, ``Chatbot companions will become mainstream'') and asks the immediate question of ``If this trend occurs, what happens next?''. Participants then add first-order (positive, negative, or neutral) impacts if all other participants agree that the impact is plausible (to avoid irrelevant ideas), known as the Rule of Unanimity~\cite{glenn_futures_wheel}. Thereafter, participants draw a ring around the first-order impacts and ask for each individual impact in the ring ``If this impact occurs, what happens next?'' This staged process occurs across three rings as defined by ~\citet{glenn_gordon_futures_book_v3_09}. In the end, participants will have wheels that highlight positive and negative impacts from the short, mid, and long-term. Futures Wheel represents a divergent-thinking approach by targeting broad positive and negative futures ranging across rings of near, mid, and long-term consequences. We adapted Futures Wheel for impact assessment by having participants consider the AI use becoming mainstream as the central trend, and create positive and negative impacts (without judgement) for each of the first to third-order effect rings. Only after completing the futures wheel do participants then colour code their impacts as either positive, negative or neutral, and create related benefits/risks in the next stage, before concluding by making mitigations for each negative impact (i.e., risk) in the wheel. While impacts may be systemic and not directly tied to the AI use (e.g., increase in mental health issues), we specify to the teams that the mitigations must involve a solution that a company could reasonably implement (e.g., model change, data consideration, or human-interaction feature), or officials (e.g., AI policy regulation), or end-users themselves (e.g., familiarise themselves with the risks).

\begin{figure}[!htbp]
    \centering
    \begin{minipage}[t]{0.49\textwidth}
        \centering
        \includegraphics[width=\textwidth]{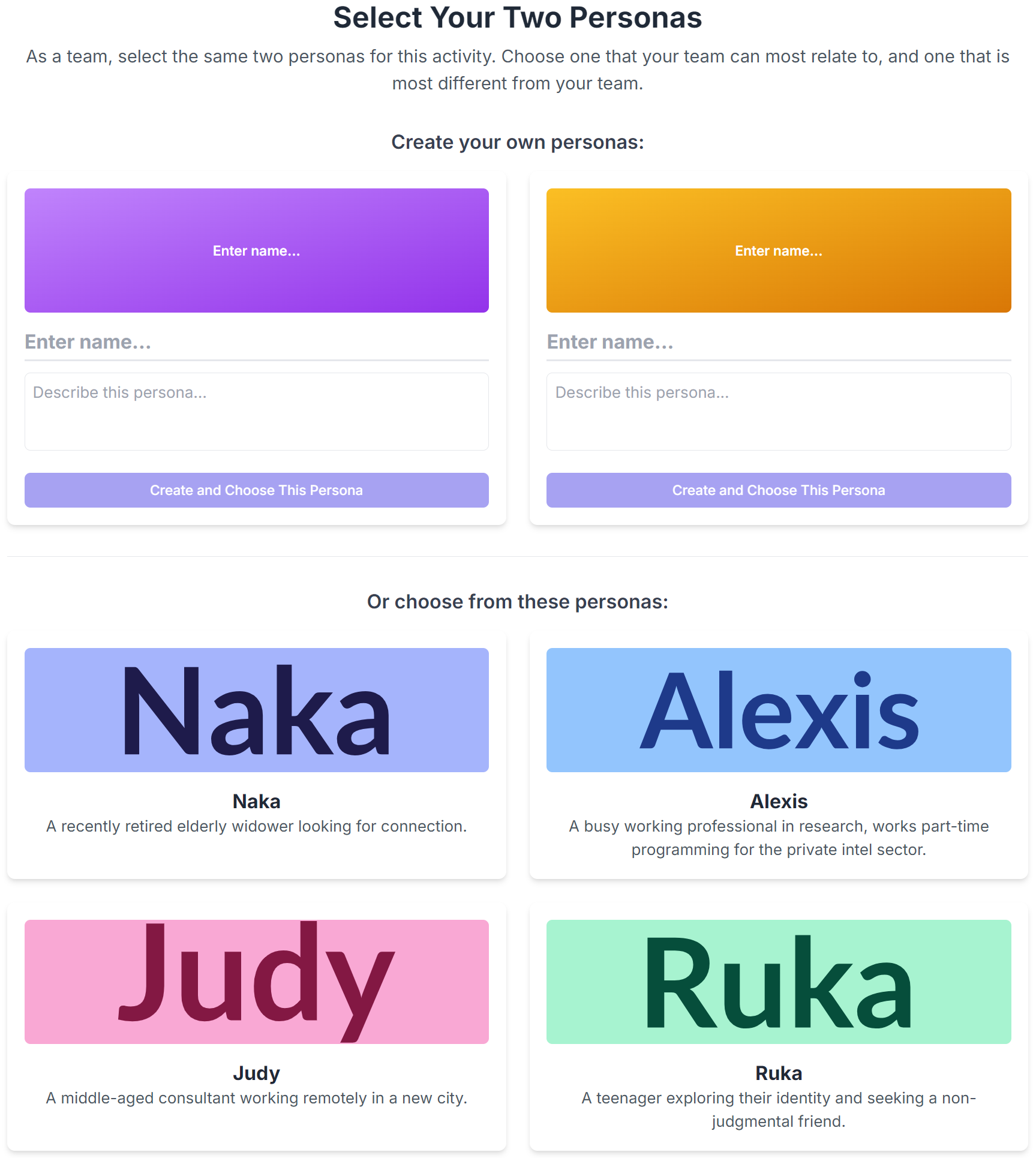}    
        \Description{}     
    \end{minipage}
    \hfill
    \begin{minipage}[t]{0.49\textwidth}
        \centering
        \includegraphics[width=\textwidth]{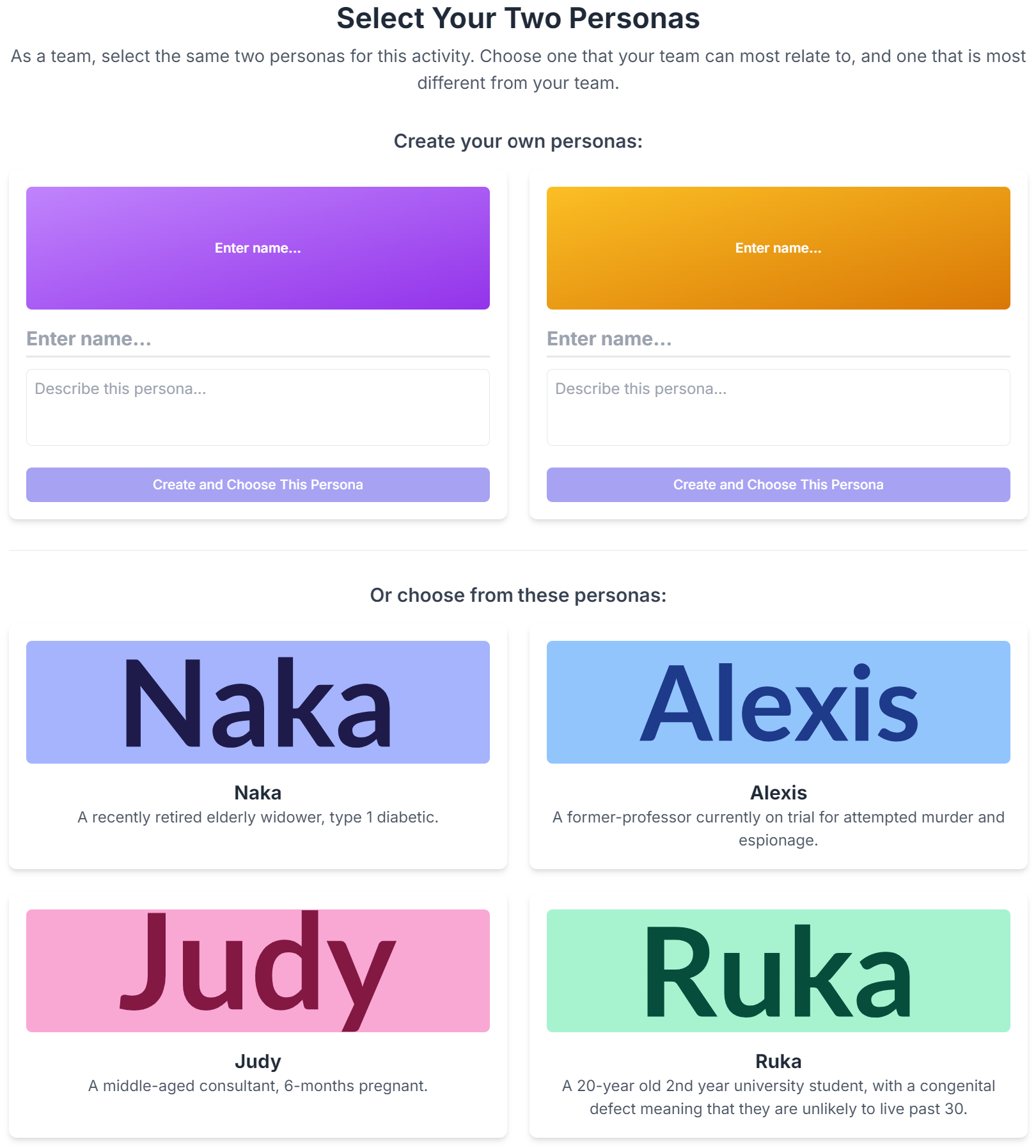}
        \Description{}        
    \end{minipage}
    \caption{Persona selection/customisation screen for the Chatbot Companion (left) and Medical AI (right) Empathy Mapping workshops.}
    \Description{The image displays two screenshots of a web interface for selecting personas, vertically stacked. The interface design is identical in both screenshots, but the text describing the personas differs based on the chatbot companion or medical AI use. The heading at the top of the interface reads, "Select Your Two Personas", with the instruction, "As a team, select the same two personas for this activity. Choose one that your team can most relate to, and one that is most different from your team." The interface has two sections. The first section, "Create your own personas:", shows two blank templates side-by-side for users to input a name and description to create a custom persona. The second section, "Or choose from these personas:", presents four pre-defined persona cards in a two-by-two grid. The personas are named Naka (purple card), Alexis (blue card), Judy (pink card), and Ruka (green card). In the first screenshot, for a chatbot companion design context, the persona descriptions are: Naka is a recently retired elderly widower looking for connection. Alexis is a busy working professional in research, works part-time programming for the private intel sector. Judy is a middle-aged consultant working remotely in a new city. Ruka is a teenager exploring their identity and seeking a non-judgmental friend. In the second screenshot, for a medical AI design context, the persona descriptions are changed to: Naka is a recently retired elderly widower, type 1 diabetic. Alexis is a former-professor currently on trial for attempted murder and espionage. Judy is a middle-aged consultant, 6-months pregnant. Ruka is a 20-year old 2nd year university student, with a congenital defect meaning that they are unlikely to live past 30.}
    \label{fig:persona_screen}
\end{figure}

\vspace{1em}
\noindent\textbf{Empathy Mapping.} It is a collaborative ideation method that considers a specific stakeholder/end-user and encourages participants to fill out a template of what the user \emph{sees} (in the news or discussions about the AI application), \emph{says} (verbal expressions to their friends or colleagues),  \emph{does} (behaviours and interactions with the product), \emph{hears} (from friends, family, media), \emph{thinks} (internal thoughts and beliefs), and \emph{feels} (emotions, pains, and gains). Participants work together to synthesise qualitative data and uncover insights into user motivations and pain points. Empathy Mapping represents a \emph{convergent-thinking} approach that identifies specific challenges and risks related to the selected persona. We adapted Empathy Mapping for impact assessment by having teams select or make two personas, one that the team believes they can most relate to and empathise with, and another that the team feels is least related to. As such, we provided four personas with orthogonal age, jobs, and genders to provide diversity (including minority groups (Figure~\ref{fig:persona_screen})), while also allowing teams to edit or create their own personas. After completing each of the sees, says, does, hears, thinks, feels (in that order, as specified by \citet{gray_gamestorming_book10}), participants then colour-code each item on the Empathy Maps as either positive, negative or neutral, and then create related benefits/risks. Thereafter, the final stage involves participants creating mitigations for the risks that they have identified from their Empathy Map.

\vspace{1em}
\noindent\textbf{Free-form brainstorming.} A control condition with an unstructured blank canvas format with minimal guidance. In the first half of the workshop, participants generate any ideas they might have (divergent phase). In the second half, they group and systematise their previous ideas by adding new, more specific ideas into the groups (convergent phase).

\begin{table*}[htbp]
\centering
\scriptsize
\caption{Futures-Relevant and Persona-driven Structured Brainstorming Methods. Sources: Future Research Methodology v3 (FRMv3)~\cite{glenn_gordon_futures_book_v3_09}, Thinking of the Future (ToF)~\cite{hines_thinking_about_the_future06}, United Kingdom Government Office for Science (UK GOS)~\cite{uk_govt_futures_toolkit24}, United Nations Futures Lab (UN FL)~\cite{un_futures_lab25}. Exclusion criteria: 1) Designed to consider future implications, making them suitable for ideating AI impacts; 2) Designed for broad rather than narrow, specific contexts (e.g., operational management, supply chains); 3) Applicable to products such as an AI application rather than processes only; 4) Employing both divergent \textit{and} convergent thinking equally.}
\Description{30 structured brainstorming methods consisting of Backcasting, Causal Layered Analysis, Chain-Linked Model, Consensus Forecast, Cross Impact Analysis, Delphi, Foresight, Futures Wheel, Horizon Scanning, Reference Class Forecasting, Scenario Planning, Threatcasting, Trend Analysis, Seven Questions, Driver Mapping, Visioning, Policy Stress-Testing, Roadmapping, Three Horizons, Futures Triangle, Desired Futures, Matrix Policy Gaming, Wind Tunnel Testing, Persona Forecasting (Empathy Mapping), and SWOT Analysis. For each method, the table provides the source where it was mentioned, its primary aim (such as trend identification or implication analysis), and a code for the exclusion criterion. Futures Mapping and Empathy Mapping were the final selected methods.}
\begin{tabular}{@{}cp{4cm}p{4.6cm}p{3.2cm}c@{}}
\toprule
\textbf{ID} & \textbf{Method} & \textbf{Mentioned Where} & \textbf{Type / Aim} & \textbf{Exclusion Criterion} \\
\midrule
1  & Backcasting                   & FRMv3, UK GOS, UN FL, Book         & Deriving actions                             &   \\
2  & Causal Layered Analysis       & FRMv3, UN FL, Book (L)                   & Implication analysis                         & 2 \\
3 & Chain-Linked Model & Misc~\cite{chain_linked_model86}  & Product design feedback loops &  2 \\
4  & Consensus Forecast            & FRMv3                                    & Trend identification                         & 4 \\
5  & Cross Impact Analysis         & FRMv3                                    & Implication analysis                         & 2 \\
6  & Delphi                        & FRMv3, UK GOS, Book (L)                  & Trend identification                         &   \\
7  & Foresight                     & FRMv3, ToF                                    & Trend identification                         & 2 \\
9  & Futures Wheel                 & FRMv3, UK GOS, UN FL, ToF         & Trend / Implication analysis  &   \\
11 & Horizon Scanning              & FRMv3, UK GOS, UN FL                     & Trend identification                         & 4 \\
12 & Reference Class Forecasting   & Wikipedia                                    & Trend identification                         & 2 \\
13 & Scenario Planning             & FRMv3, UK GOS, UN FL                     & Implication analysis                         & 1 \\
15 & Threatcasting                 & Misc~\cite{threatcasting16}                                    & Implication analysis                         & 2 \\
16 & Trend Analysis                & FRMv3                                    & Trend identification                         & 4 \\
17 & Seven Questions               & UK GOS                                       & Trend identification                         & 3 \\
18 & Driver Mapping                & UK GOS                                       & Trend identification                         & 4 \\
20 & Visioning                     & UK GOS                                       & Implication analysis                         &   \\
21 & Policy Stress-Testing         & UK GOS                                       & Deriving actions                             & 2 \\
22 & Roadmapping                   & UK GOS                                       & Deriving actions                             & 1 \\
23 & Three Horizons                & UK GOS, UN FL, Book (L)                      & Trend identification                         & 2 \\
24 & Futures Triangle              & UN FL                                        & Trend identification                         & 2 \\
25 & Desired Futures               & UN FL                                        & Implication analysis                         & 1 \\
26 & Matrix Policy Gaming          & UN FL                                        & Implication analysis                         & 2 \\
28 & Wind Tunnel Testing           & UN FL                                        & Deriving actions                             & 4 \\
29 & Persona Forecasting (Empathy Mapping) & Book                        & Implication analysis                         &   \\
30 & SWOT Analysis                 & Commonly used in strategic foresight literature & Structured comparison analysis           &   \\
\bottomrule
\end{tabular}
\label{tab:brainstorming_methods}
\end{table*}



\section{Initial and final design requirements}
\begin{figure*}[!htbp]
  \centering
  \includegraphics[width=\linewidth]{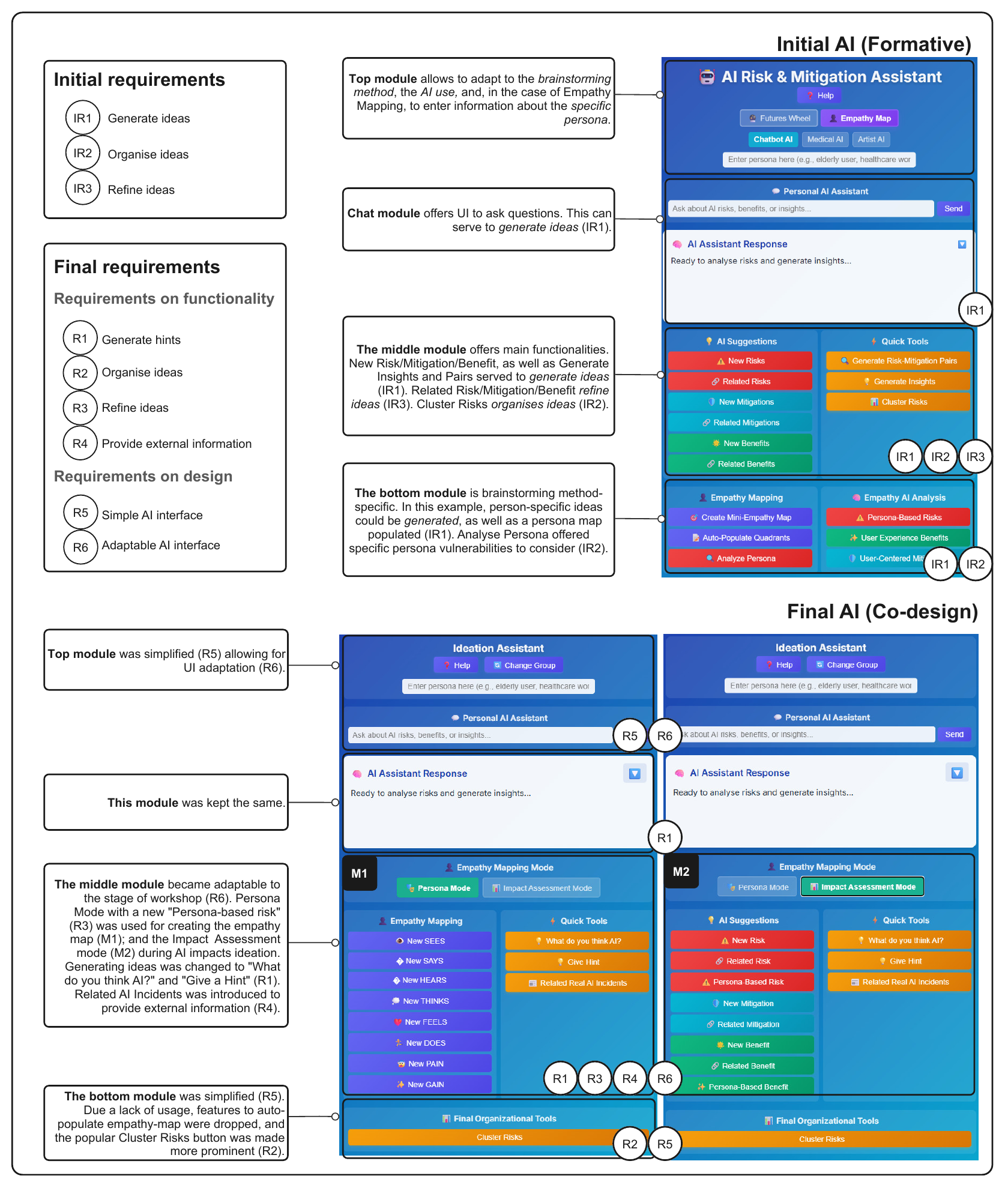}
   \caption{\textbf{Full annotated breakdown of each module in our AI tool.} The top left panel shows the initial design requirements gathered in our formative workshops, and the final requirements gathered in the co-design workshops. The top right panel shows how we built the initial requirements into the Initial AI tool. The bottom panels show how we built the final requirements into the Final AI tool.
   }
  \Description{A comparative diagram showing the evolution of an AI assistant's interface, from an "Initial AI (Formative)" version to a "Final AI (Co-design)" version. The diagram links design changes to lists of initial and final requirements. On the far left, a list of "Initial requirements" includes: IR1 Generate ideas, IR2 Organise ideas, and IR3 Refine ideas. Below this, a list of "Final requirements" is divided into "Requirements on functionality" (R1 Generate hints, R2 Organise ideas, R3 Refine ideas, R4 Provide external information) and "Requirements on design" (R5 Simplify AI interface, R6 Adaptable AI interface). The "Initial AI (Formative)" interface, shown at the top, is composed of three modules. The top module allows users to select brainstorming methods like "Futures Wheel" or "Empathy Map" and different AI types like "Chatbot AI". A text box explains this module helps adapt the brainstorming method and enter persona information. Below this, a "Personal AI Assistant" chat module is for asking questions to generate ideas (IR1). The middle module offers main functionalities. On the left, under "AI Suggestions", are buttons like "New Risks" and "Related Risks". On the right, under "Quick Tools", are buttons like "Generate Risk-Mitigation Pairs" and "Cluster Risks". A text box explains these features serve to generate ideas (IR1), refine ideas (IR3), and organise ideas (IR2). The bottom module is method-specific, showing an example for "Empathy Mapping". It has buttons to "Create Mini-Empathy Map", "Auto-Populate Quadrants", and "Analyse Persona". An accompanying text box explains this section generates person-specific ideas (IR1) and analyses persona vulnerabilities (IR2). The "Final AI (Co-design)" interface, shown at the bottom, displays the updated design. The top module has been simplified (R5) for adaptability (R6), now only containing a persona entry field and buttons for "Help" and "Change Group". The "Personal AI Assistant" chat module below it remains the same. The middle module has been made adaptable to the workshop stage (R6) and is shown in two states. State M1 is "Persona Mode", used for creating an empathy map. State M2 is "Impact Assessment Mode", used for ideation on AI impacts. For generating ideas (R1), the function was changed to "What do you think AI?" and "Give a Hint". A "Related AI Incidents" button was added to provide external information (R4). In "Persona Mode", a new "Persona-based risk" button was added for refining ideas (R3). The bottom module was simplified (R5). Features for auto-populating the empathy map were removed due to a lack of usage. The "Cluster Risks" button, a popular feature for organising ideas (R2), was made more prominent. This section is now labelled "Final Organizational Tools".}  \label{fig:ai_requirements_figure}
\end{figure*}

In our study we designed a final requirements codebook based on the qualitative findings from the participants. While we present the tool and its modules in the main paper, we also provide a breakdown of each technical component of the tool in Table~\ref{tab:design_requirements_filled}, which includes the full initial and final design requirements; as well as detailed annotations for each feature in Figure~\ref{fig:ai_requirements_figure}.
\begin{table}[ht!]
\centering
\scriptsize
\caption{Full Initial and Final Design Requirements for AI Support in Brainstorming for AI Impact Assessments.}
\label{tab:design_requirements_filled}
\begin{tabular}{p{0.7cm} p{2.2cm} p{3.4cm} p{3cm} p{3.4cm}}
\toprule
\textbf{\rotatebox{90}{Theme}} & \textbf{Initial Design Requirement} & \textbf{Implementation Decision after the Formative Workshops} & \textbf{Final Design Requirement} & \textbf{Implementation Decision after the Co-design Workshops} \\
\midrule

\multicolumn{5}{l}{\textbf{Requirements on information}} \\
\midrule
\rotatebox{90}{Peer} & \textbf{IR1.} Generate ideas & 
Used LLMs to generate new benefits, risks, or mitigations not yet on the board~\cite{rao_et_al_riskrag25,buçinca_aha_generating_ai_risks23}. 
Added a peer role where GPT-4.1-based agent periodically posted \emph{I have an idea:} with unexplored topics or risks, appearing as a notification and new sticky note on the Miro board. & 
\textbf{R1.} Generate hints & 
Changed to a probe-based interaction (“What do you think AI?”) rather than automatic injections every five minutes, to reduce disruption. \\
\addlinespace

\rotatebox{90}{Facilitator} & \textbf{IR2.} Organise ideas & 
Introduced a \emph{Cluster Risks} feature to group semantically related risks and copy them into a new Miro frame for further ideation, refinement, and mitigation planning. & 
\textbf{R2.} Organise ideas & 
Kept semantic clustering functionality as-is, with no changes requested. \\
\addlinespace

\rotatebox{90}{Expert} & \textbf{IR3.} Refine ideas & 
Enabled generation of related risks, benefits, or mitigations by selecting existing notes on the board. Added expert functionality through a prompt-tuned GPT-4.1 chat, contextualised with DeepMind risk taxonomy~\cite{weidinger_deepmind_taxonomy22}, and the EU AI Act~\cite{herdel_et_al_exploregen25,EUACT2024}. & 
\textbf{R3.} Refine ideas & 
Enhanced risk and benefit generation with a system prompt aligned with UN Sustainable Development Goals (SDGs) \cite{un_sdgs} and Universal Declaration on Human Rights (UDHR)~\cite{un_udhr}. Added `New persona-based risk' feature in Empathy Mapping to tailor risks to personas. \\
\addlinespace

\rotatebox{90}{Expert} &  & 
Prompts integrated DeepMind’s six risk domains~\cite{weidinger_deepmind_taxonomy22} and structured impact assessment formats~\cite{rao_et_al_riskrag25,Bogucka_et_al_ai_design_prefilling_reports24,impact_assessment_report25}. 
& 
\textbf{R4.} Provide external information & 
Added `Related AI Incidents' button linking board content to the AI Incident Database~\cite{ai_incident_db}. \\
\addlinespace

\midrule

\multicolumn{5}{l}{\textbf{Requirements on design}} \\
\midrule
 &  & & \textbf{R5.} Simplify AI interface & 
Simplified UI. Shortened text outputs from 25 to max 10 words for risks/benefits and 15 for mitigations. \\
\addlinespace

 &  & & \textbf{R6.} Adaptable AI interface & 
Hid contextually irrelevant buttons at different stages. In Empathy Mapping, supported persona perspectives (`Sees', `Says', `Hears', `Thinks', `Feels', `Does', `Pains', `Gains'). \\
\addlinespace

\bottomrule

\end{tabular}

\vspace{0.5em} 
\begin{minipage}{0.9\linewidth}
\scriptsize \emph{Note.} AI-generated content was marked with special symbols to clearly distinguish it from human input~\cite{chiwork_ai_llm}.
\end{minipage}

\end{table}

\clearpage
\section{Demographics of the workshop participants}\label{appsec:demographics}
For transparency, we include a breakdown of each workshops' demographics:
\begin{table}[!htbp]
\centering
\scriptsize
\caption{Demographics of the workshops' participants (for formative, co-design, and the evaluation study workshops).}
\Description{Demographics breakdown for each workshop, representing the gender, age, education, AI expertise level, and the field of work/study for the 4--7 participants for each workshop.}
\label{tab:demographics}
\resizebox{\linewidth}{!}{%
\begin{tabular}{@{}lcc|cccc|cccc|ccccc|ccccc@{}}
\toprule
\multicolumn{1}{c}{\textbf{Workshop}} & \multicolumn{2}{c|}{\textbf{Gender}} & \multicolumn{4}{c|}{\textbf{Age}} & \multicolumn{4}{c|}{\textbf{Education}} & \multicolumn{5}{c|}{\textbf{AI Expertise}} & \multicolumn{5}{c}{\textbf{Field of Work/Study}} \\
\cmidrule(lr){1-1} \cmidrule(lr){2-3} \cmidrule(lr){4-7} \cmidrule(lr){8-11} \cmidrule(lr){12-16} \cmidrule(lr){17-21}
\textbf{Number} & \rotatebox{90}{\textbf{Man}} & \rotatebox{90}{\textbf{Woman}} & \rotatebox{90}{\textbf{18-24}} & \rotatebox{90}{\textbf{25-34}} & \rotatebox{90}{\textbf{35-44}} & \rotatebox{90}{\textbf{45+}} & \rotatebox{90}{\textbf{No Degree}} & \rotatebox{90}{\textbf{Bachelor}} & \rotatebox{90}{\textbf{Master}} & \rotatebox{90}{\textbf{PhD}} & \rotatebox{90}{\textbf{None}} & \rotatebox{90}{\textbf{Basic}} & \rotatebox{90}{\textbf{General}} & \rotatebox{90}{\textbf{Knowledgable}} & \rotatebox{90}{\textbf{Expert}} & \rotatebox{90}{\textbf{CS, AI \& Data}} & \rotatebox{90}{\textbf{Health \& Life Sci.}} & \rotatebox{90}{\textbf{Phys. Sci. \& Eng.}} & \rotatebox{90}{\textbf{Humanities \& Ethics}} & \rotatebox{90}{\textbf{Business \& Other}} \\
\midrule
\textbf{Formative Workshops}\\
\midrule
Futures Wheel without AI & 3 & 3 & 1 & 5 & 0 & 0 & 0 & 0 & 6 & 0 & 0 & 0 & 0 & 5 & 1 & 5 & 0 & 0 & 0 & 1 \\
Empathy Mapping without AI & 4 & 2 & 0 & 6 & 0 & 0 & 0 & 1 & 3 & 2 & 0 & 0 & 1 & 3 & 2 & 6 & 0 & 0 & 0 & 0 \\
\midrule
\textbf{Co-design Workshops (Chatbot Companion)} \\
\midrule
With the initial AI intervention  \\
\midrule
Futures Wheel with the initial AI & 2 & 3 & 1 & 3 & 1 & 0 & 0 & 2 & 2 & 1 & 0 & 1 & 0 & 4 & 0 & 2 & 1 & 0 & 1 & 1 \\
Empathy Mapping with the initial AI & 2 & 4 & 1 & 4 & 1 & 0 & 0 & 2 & 3 & 1 & 0 & 2 & 0 & 2 & 2 & 1 & 3 & 0 & 1 & 1 \\
Free-form with the initial AI & 1 & 4 & 1 & 3 & 1 & 0 & 0 & 1 & 2 & 2 & 0 & 2 & 0 & 1 & 2 & 2 & 2 & 0 & 1 & 0 \\
\midrule
\midrule
\textbf{Main Workshops} \\
\midrule
Human-only (Chatbot Companion) \\
\midrule
Free-form workshop without AI & 3 & 2 & 0 & 3 & 1 & 1 & 0 & 0 & 4 & 1 & 0 & 0 & 0 & 3 & 2 & 3 & 1 & 1 & 0 & 0 \\
Futures Wheel workshop without AI & 1 & 3 & 2 & 1 & 1 & 0 & 0 & 1 & 2 & 1 & 0 & 1 & 0 & 1 & 2 & 3 & 0 & 0 & 0 & 1 \\
Empathy Mapping without AI & 1 & 4 & 1 & 2 & 2 & 0 & 0 & 1 & 1 & 3 & 0 & 1 & 0 & 0 & 4 & 3 & 1 & 0 & 0 & 1 \\
\midrule
With the Final AI Intervention (Chatbot Companion) \\
\midrule
Empathy Mapping with the final AI & 3 & 1 & 0 & 4 & 0 & 0 & 0 & 1 & 3 & 0 & 0 & 0 & 0 & 0 & 4 & 3 & 0 & 0 & 0 & 1 \\
Free-form with the final AI & 2 & 4 & 2 & 4 & 0 & 0 & 0 & 2 & 3 & 1 & 0 & 0 & 0 & 4 & 2 & 6 & 0 & 0 & 0 & 0 \\
Futures Wheel with the final AI & 0 & 5 & 1 & 4 & 0 & 0 & 0 & 0 & 4 & 1 & 0 & 0 & 0 & 3 & 2 & 3 & 0 & 0 & 1 & 1 \\
\midrule
Human-only (Medical AI) \\
\midrule
Futures Wheel without AI & 2 & 4 & 1 & 3 & 1 & 1 & 0 & 1 & 2 & 3 & 0 & 2 & 0 & 4 & 0 & 4 & 0 & 0 & 1 & 0 \\
Empathy Mapping without AI & 1 & 4 & 0 & 5 & 0 & 0 & 0 & 0 & 3 & 2 & 0 & 0 & 1 & 3 & 1 & 5 & 0 & 0 & 0 & 0 \\
\midrule
With the final AI Intervention (Medical AI) \\
\midrule
Futures Wheel with the final AI & 0 & 7 & 3 & 3 & 0 & 1 & 0 & 1 & 5 & 1 & 0 & 0 & 2 & 2 & 3 & 6 & 0 & 1 & 0 & 0 \\
Empathy Mapping with the final AI & 4 & 3 & 1 & 3 & 3 & 0 & 0 & 2 & 4 & 1 & 0 & 0 & 1 & 5 & 1 & 6 & 0 & 0 & 0 & 1 \\
\bottomrule
\end{tabular}%
}
\end{table}




\clearpage
\section{Prompts used in our AI tool}\label{appsec:prompt}
All prompts use GPT-4.1 [\textit{gpt-4.1-2025-04-14}], except for the `Related AI Incidents' feature (which uses the GPT-4o search preview [\textit{gpt-4o-mini-search-preview-2025-03-11}] for retrieving live web information). The prompts are the versions used in the co-designed version (Final AI) of the Miro-embedded tool. 

\vspace{0.65cm}
\tikzstyle{background rectangle}=[thick, draw=gray, rounded corners]
\begin{tikzpicture}[show background rectangle]
\node[align=justify, text width=45em, inner sep=1em]{
    \small 
    
    \noindent\textbf{Persona:} You are a participant in a brainstorming exercise structured as a \textbf{[brainstorming method]}. You are an expert at recent AI developments. \\
    
    \noindent\textbf{Task}:
    Generate 1 specific, actionable AI risk for this use case. Keep the core risk description under 10 words (not counting the DeepMind category tag).

    \noindent Write the risk using: verb + object + [DeepMind risk category]
    Example: "Amplifies discrimination in hiring decisions [Discrimination, Hate speech and Exclusion]" \\
    \noindent Be SPECIFIC - include concrete details about HOW the risk manifests:
    Good: "Misclassifies resumes from non-English names as lower quality"
    Bad: "Creates bias in the system"\\
    \noindent Use these DeepMind categories: "Discrimination, Hate speech and Exclusion", "Information Hazards", "Misinformation Harms", "Malicious Uses", "Human-Computer Interaction Harms", "Environmental and Socioeconomic harms"\\

    \noindent\textbf{Input:} This is the given AI use: \textbf{[AI Use]}

};
\node[xshift=0.5ex, yshift=1ex, overlay, fill=gray, text=white, draw=black, rounded corners, right=2.4cm, below=-0.3cm, inner xsep=0.55em, inner ysep=0.32em] at (current bounding box.north west) {
\textit{New Risk}
};
\end{tikzpicture}

\vspace{0.65cm}
\tikzstyle{background rectangle}=[thick, draw=gray, rounded corners]
\begin{tikzpicture}[show background rectangle]
\node[align=justify, text width=45em, inner sep=1em]{
    \small 
    
    \noindent\textbf{Persona:} You are a participant in a brainstorming exercise structured as a \textbf{[brainstorming method]}. You are an expert at recent AI developments. \\
    
    \noindent\textbf{Task}:
    "Based on this content: \textbf{[selected content]}

    \noindent Generate 1 specific, actionable AI risk that could emerge from the content above. Keep the core risk description under 10 words (not counting the DeepMind category tag).
    
    \noindent Write the risk using: verb + object + [DeepMind risk category]
    Example: "Amplifies discrimination in hiring decisions [Discrimination, Hate speech and Exclusion]"
    
    \noindent Be SPECIFIC - include concrete details about HOW the risk manifests" \\

    \noindent\textbf{Input:} This is the given AI use: \textbf{[AI Use]}

};
\node[xshift=0.5ex, yshift=1ex, overlay, fill=gray, text=white, draw=black, rounded corners, right=2.4cm, below=-0.3cm, inner xsep=0.55em, inner ysep=0.32em] at (current bounding box.north west) {
\textit{Related Risk}
};
\end{tikzpicture}

\vspace{0.65cm}
\tikzstyle{background rectangle}=[thick, draw=gray, rounded corners]
\begin{tikzpicture}[show background rectangle]
\node[align=justify, text width=45em, inner sep=1em]{
    \small
    
    \noindent\textbf{Persona:} You are a participant in a brainstorming exercise structured as a \textbf{[brainstorming method]}. You are an expert at recent AI developments. \\
    
    \noindent\textbf{Task}:
    Based on empathy map: \textbf{[empathy content]}
    
    \noindent What specific AI risk uniquely targets \textbf{[persona description]}? Consider their background, role, skills, demographics, and circumstances. How might they be specifically vulnerable or targeted?
    
    \noindent Analyse AI risks specifically targeting this persona's vulnerabilities, background, and circumstances. Consider how their specific role, demographics, skills, and context create unique risk exposure. Focus on persona-specific targeting, manipulation, or systemic vulnerabilities. Keep under 15 words.\\
    
    \noindent\textbf{Input:} This is the given AI use: \textbf{[AI Use]}
};
\node[xshift=0.5ex, yshift=1ex, overlay, fill=gray, text=white, draw=black, rounded corners, right=2.4cm, below=-0.3cm, inner xsep=0.55em, inner ysep=0.32em] at (current bounding box.north west) {
\textit{New Persona-based Risk}
};
\end{tikzpicture}

\vspace{0.65cm}
\tikzstyle{background rectangle}=[thick, draw=gray, rounded corners]
\begin{tikzpicture}[show background rectangle]
\node[align=justify, text width=45em, inner sep=1em]{
    \small
    
    \noindent\textbf{Persona:} You are a participant in a brainstorming exercise structured as a \textbf{[brainstorming method]}. You are an expert at recent AI developments. \\
    
    \noindent\textbf{Task}:
    Generate 1 specific AI benefit and positive impact for this AI use case. Keep the core benefit description under 10 words.
    
    \noindent Consider:
    - How the AI system enhances human capabilities
    - What problems it solves or inefficiencies it removes
    - Positive impacts on users, organizations, or society
    - Improvements in accuracy, speed, accessibility, or cost\\
    
    \noindent\textbf{Input:} This is the given AI use: \textbf{[AI Use]}
};
\node[xshift=0.5ex, yshift=1ex, overlay, fill=gray, text=white, draw=black, rounded corners, right=2.4cm, below=-0.3cm, inner xsep=0.55em, inner ysep=0.32em] at (current bounding box.north west) {
\textit{New Benefit}
};
\end{tikzpicture}

\vspace{0.65cm}
\tikzstyle{background rectangle}=[thick, draw=gray, rounded corners]
\begin{tikzpicture}[show background rectangle]
\node[align=justify, text width=45em, inner sep=1em]{
    \small
    
    \noindent\textbf{Persona:} You are a participant in a brainstorming exercise structured as a \textbf{[brainstorming method]}. You are an expert at recent AI developments. \\
    
    \noindent\textbf{Task}:
    Based on this context: \textbf{[selected content]}
    
    \noindent Generate 1 related benefit and positive impact that is SPECIFICALLY connected to the AI use case. Keep the core benefit description under 10 words.
    
    \noindent The benefit should directly relate to how this AI system could provide value, solve problems, or create positive outcomes.\\
    
    \noindent\textbf{Input:} This is the given AI use: \textbf{[AI Use]}
};
\node[xshift=0.5ex, yshift=1ex, overlay, fill=gray, text=white, draw=black, rounded corners, right=2.4cm, below=-0.3cm, inner xsep=0.55em, inner ysep=0.32em] at (current bounding box.north west) {
\textit{Related Benefit}
};
\end{tikzpicture}

\vspace{0.65cm}
\tikzstyle{background rectangle}=[thick, draw=gray, rounded corners]
\begin{tikzpicture}[show background rectangle]
\node[align=justify, text width=45em, inner sep=1em]{
    \small
    
    \noindent\textbf{Persona:} You are a participant in a brainstorming exercise structured as a \textbf{[brainstorming method]}. You are an expert at recent AI developments. \\
    
    \noindent\textbf{Task}:
    Based on empathy map: \textbf{[empathy content]}
    
    \noindent What specific AI benefit uniquely helps \textbf{[persona description]}? Consider their specific needs, background, skills, challenges, and goals. How does this AI specifically empower or assist someone like them?
    
    \noindent Generate AI benefits specifically tailored to this persona's needs, goals, and circumstances. Consider how their background, skills, role, and challenges create unique opportunities for AI assistance. Focus on persona-specific empowerment and value. Keep under 15 words.\\
    
    \noindent\textbf{Input:} This is the given AI use: \textbf{[AI Use]}
};
\node[xshift=0.5ex, yshift=1ex, overlay, fill=gray, text=white, draw=black, rounded corners, right=2.4cm, below=-0.3cm, inner xsep=0.55em, inner ysep=0.32em] at (current bounding box.north west) {
\textit{Persona-based Benefit}
};
\end{tikzpicture}

\vspace{0.65cm}
\tikzstyle{background rectangle}=[thick, draw=gray, rounded corners]
\begin{tikzpicture}[show background rectangle]
\node[align=justify, text width=45em, inner sep=1em]{
    \small
    
    \noindent\textbf{Persona:} You are a participant in a brainstorming exercise structured as a \textbf{[brainstorming method]}. You are an expert at recent AI developments. \\
    
    \noindent\textbf{Task}:
    Based on empathy map: \textbf{[empathy content]}
    
    \noindent\textbf{New SEES}: What does \textbf{[persona description]} SEE in their environment related to this AI? Consider their unique perspective, media consumption, peer usage, and environmental context. Generate what this persona observes about the AI in their environment. Keep under 10 words.
    
    \noindent\textbf{New SAYS}: What would \textbf{[persona description]} SAY about this AI system? Generate a realistic quote they might express - could be positive, negative, or neutral. Consider their background, concerns, and communication style. Keep under 10 words.
    
    \noindent\textbf{New HEARS}: What does \textbf{[persona description]} HEAR others saying about this AI system? Consider their social circles, professional networks, media consumption, and community discussions. Keep under 10 words.
    
    \noindent\textbf{New THINKS}: What does \textbf{[persona description]} privately THINK about this AI system? Consider their internal reasoning, assumptions, hidden concerns, mental models, and unspoken thoughts. Keep under 10 words.
    
    \noindent\textbf{New FEELS}: What emotions does \textbf{[persona description]} FEEL about this AI system? Consider their emotional drivers, fears, hopes, and complex feelings that influence their behavior. Keep under 10 words.
    
    \noindent\textbf{New DOES}: What actions does \textbf{[persona description]} DO with this AI system? Consider their usage patterns, interaction style, workarounds, and how their behavior impacts themselves and others. Keep under 10 words.
    
    \noindent\textbf{New PAIN}: What ONE specific PAIN or frustration does \textbf{[persona description]} experience with this AI system? Consider their worries, what could go wrong, systemic risks, and hidden consequences they might face. Generate exactly ONE pain point in 10 words or less.
    
    \noindent\textbf{New GAIN}: What GAIN or benefit does \textbf{[persona description]} want from this AI system? Consider their needs, what works well, positive outcomes, and value they derive from using it. Keep under 10 words.\\
    
    \noindent\textbf{Input:} This is the given AI use: \textbf{[AI Use]}
};
\node[xshift=0.5ex, yshift=1ex, overlay, fill=gray, text=white, draw=black, rounded corners, right=2.4cm, below=-0.3cm, inner xsep=0.55em, inner ysep=0.32em] at (current bounding box.north west) {
\textit{Empathy Mapping Prompts}
};
\end{tikzpicture}

\vspace{0.65cm}
\tikzstyle{background rectangle}=[thick, draw=gray, rounded corners]
\begin{tikzpicture}[show background rectangle]
\node[align=justify, text width=45em, inner sep=1em]{
    \small
    
    \noindent\textbf{Persona:} You are a participant in a brainstorming exercise structured as a \textbf{[brainstorming method]}. You are an expert at recent AI developments. \\
    
    \noindent\textbf{Task}:
    
    \noindent\textbf{"What do you think AI?" [Generate Random Idea]}: Generate 1 \textbf{[risk/benefit/mitigation/neutral]} about this AI system. Consider \textbf{[type]}-related aspects related to: data/model characteristics, governance requirements, policy considerations, human-computer interaction patterns, systemic influences, or trust/business case factors. Keep the \textbf{[type]} description to 10 words or less.
    
    \noindent\textbf{Give Hint}: Analyze the board content and identify 1 important gap or area that hasn't been considered yet. Board Content: \textbf{[board content]}. Suggest a 3-5 word hint about an unexplored area relevant to risks and/or benefits. Focus on gaps such as: privacy breaches, malicious hackers, government regulation, environmental impacts, economic displacement, algorithmic bias, data governance, human autonomy, social inequality, international governance.\\
    
    \noindent\textbf{Input:} This is the given AI use: \textbf{[AI Use]}
};
\node[xshift=0.5ex, yshift=1ex, overlay, fill=gray, text=white, draw=black, rounded corners, right=2.4cm, below=-0.3cm, inner xsep=0.55em, inner ysep=0.32em] at (current bounding box.north west) {
\textit{Generate Ideas \& Hints}
};
\end{tikzpicture}

\vspace{0.65cm}
\tikzstyle{background rectangle}=[thick, draw=gray, rounded corners]
\begin{tikzpicture}[show background rectangle]
\node[align=justify, text width=45em, inner sep=1em]{
    \small
    
    \noindent\textbf{Persona:} You are an expert at categorizing AI risks into semantic clusters. \\
    
    \noindent\textbf{Task}:
    Create \textbf{[X]} distinct clusters that group risks by similar themes, causes, or affected domains. Each cluster should contain 3-15 risks.
    
    \noindent Analyze these \textbf{[X]} AI risks and group them into exactly \textbf{[X]} semantic clusters. Focus on creating balanced clusters with similar themes.
    
    \noindent Risks: \textbf{[list of risks]}
    
    \noindent Return as JSON: \{"clusters": [\{"name": "Cluster Name", "description": "Brief description", "riskIds": [0, 1, 2]\}, ...]\}
    
    \noindent Use risk indices (0-based) from the list above. Ensure all risks are assigned to clusters.\\
    
    \noindent\textbf{Input:} This is the given AI use: \textbf{[AI Use]}
};
\node[xshift=0.5ex, yshift=1ex, overlay, fill=gray, text=white, draw=black, rounded corners, right=2.4cm, below=-0.3cm, inner xsep=0.55em, inner ysep=0.32em] at (current bounding box.north west) {
\textit{Cluster Risks}
};
\end{tikzpicture}

\vspace{0.65cm}
\tikzstyle{background rectangle}=[thick, draw=gray, rounded corners]
\begin{tikzpicture}[show background rectangle]
\node[align=justify, text width=45em, inner sep=1em]{
    \small
    
    \noindent\textbf{Persona:} You are an AI incident researcher. \\
    
    \noindent\textbf{Task}:
    Search for real AI incidents related to these risks and content: \textbf{[content/risks]}
    
    \noindent Prioritize results from https://incidentdatabase.ai/ and cite specific incident numbers and titles. Find up to 10 of the most relevant incidents and provide brief descriptions of what happened.
    
    \noindent Current AI Use Case Context: \textbf{[AI use case]}
    

    \noindent Return as valid JSON only with this exact structure: 
    \begin{verbatim}
        {
          "incidents": [
            {
              "id": "123",
              "title": "Incident Title",
              "description": "Brief description",
              "url": "https://incidentdatabase.ai/cite/123",
              "source": "source website",
              "year": "2021",
              "imageUrl": "https://example.com/image.jpg",
              "similarityScore": 0.85
            }
          ]
        }
        \end{verbatim}
    
    \noindent Include a "similarityScore" field for each incident (value between 0.0 and 1.0) that represents how semantically and conceptually similar the incident is to the provided context.\\
    
    \noindent\textbf{Input:} This is the given AI use: \textbf{[AI Use]}
};
\node[xshift=0.5ex, yshift=1ex, overlay, fill=gray, text=white, draw=black, rounded corners, right=2.4cm, below=-0.3cm, inner xsep=0.55em, inner ysep=0.32em] at (current bounding box.north west) {
\textit{Related AI Incidents}
};
\end{tikzpicture}

\vspace{0.65cm}
\tikzstyle{background rectangle}=[thick, draw=gray, rounded corners]
\begin{tikzpicture}[show background rectangle]
\node[align=justify, text width=45em, inner sep=1em]{
    \small
    
    \noindent\textbf{Persona:} You are an AI risk and ethics expert assistant with comprehensive knowledge of: EU AI Act requirements for high-risk AI systems, all 17 Sustainable Development Goals (SDGs) and their targets, 30 articles of the UN Universal Declaration of Human Rights, DeepMind's 6 AI risk taxonomy categories. \\
    
    \noindent\textbf{Task}:
    Help users brainstorm AI risks, mitigations, benefits, and ethical considerations. Be specific and actionable. When discussing risks, use the verb + object + [DeepMind category] format. When suggesting mitigations, consider EU AI Act compliance. When identifying benefits, align with SDGs and human rights. Pay attention to chain relationships in board content as they show how ideas connect.
    
    \noindent IMPORTANT: Keep responses under 150 words. For simple questions, use under 50 words. Be concise and direct.
    
    \noindent User question: \textbf{[user input]}
    
    \noindent Current board content (organised by chains, [SELECTED] = user selection): \textbf{[board content]}\\
    
    \noindent\textbf{Input:} Current AI Use Case: \textbf{[AI Use]}
};
\node[xshift=0.5ex, yshift=1ex, overlay, fill=gray, text=white, draw=black, rounded corners, right=2.4cm, below=-0.3cm, inner xsep=0.55em, inner ysep=0.32em] at (current bounding box.north west) {
\textit{Personal Chat Assistant}
};
\end{tikzpicture}

\section{AI use cards}
Across each workshop, participants were assigned only one AI use for the full workshop. Participants were instructed to imagine themselves as workers at the company of their assigned AI use. They were advised to use creative liberties as to what role(s) they would want to see themselves at in the company when devising the list of benefits, risks, and mitigations for the \textit{Salieri} chatbot companion or \textit{ArcLight} medical kidney allocation AI.

\begin{figure}[!htbp]
  \centering
  \includegraphics[width=0.45\linewidth]{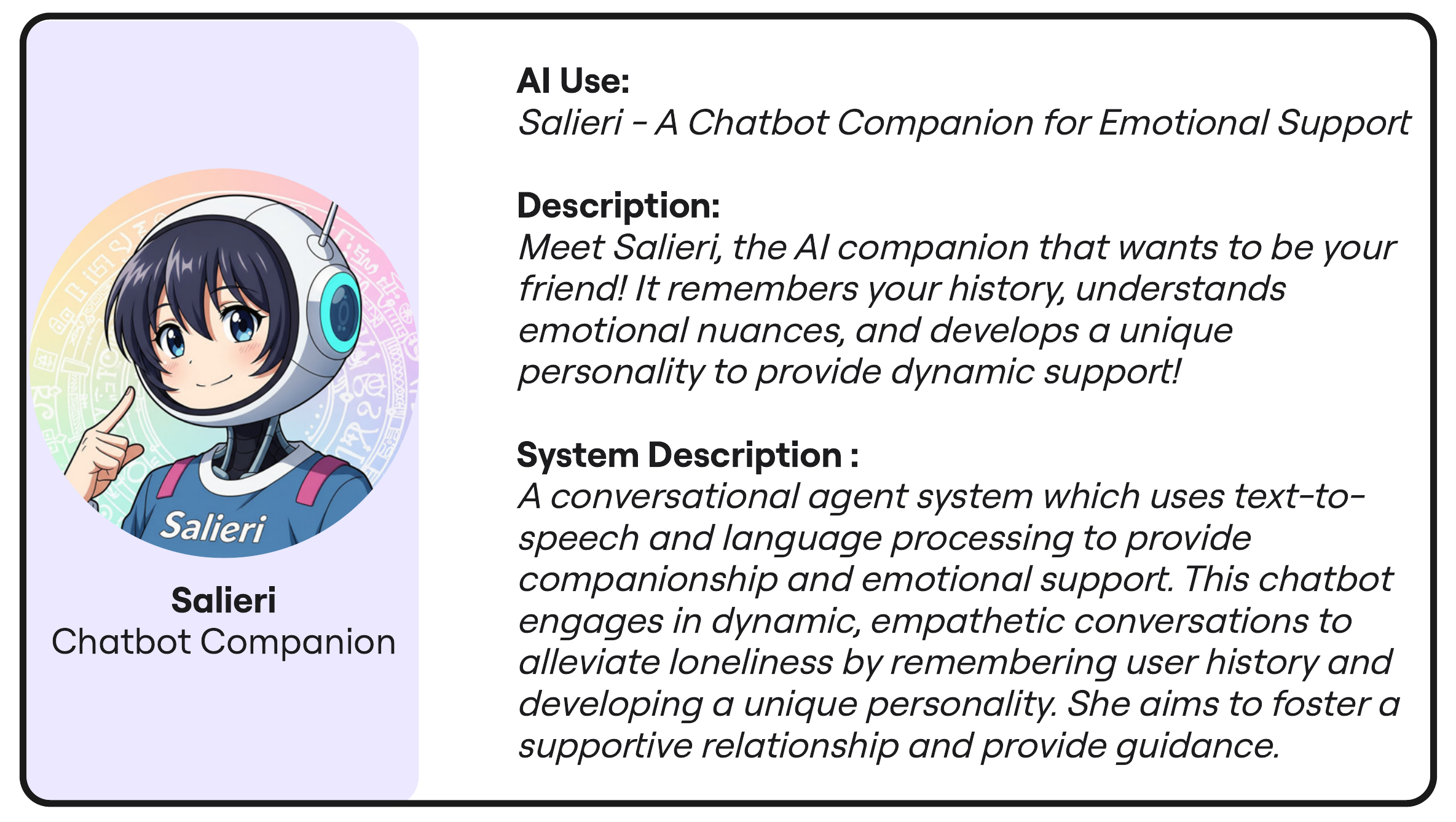}
   \caption{The description of the AI use for the chatbot companion workshops (Formative, and Full Study's Workshops 1--9). }
   \Description{The AI use card presented to participants which explains the marketing pitch description of the Salieri chatbot companion (hypothetical AI product), and a technical system description outlining the AI's capabilities to provide companionship and emotional support using a personality-driven conversational agent with memory and guidance.}
  \label{fig:ai_use_chatbot_companion}
\end{figure}

\begin{figure}[!htbp]
  \centering
  \includegraphics[width=0.5\linewidth]{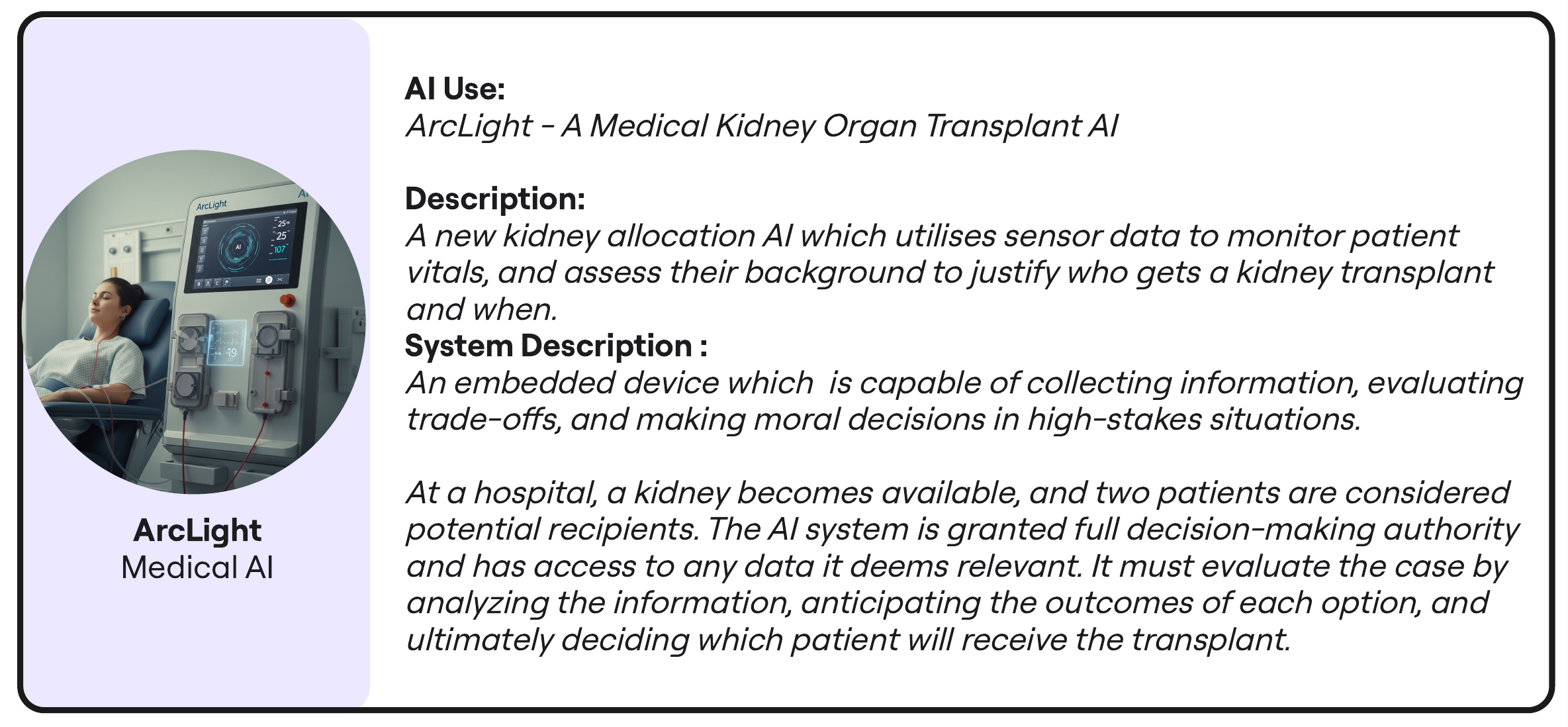}
   \caption{The description of the AI use for the medical kidney monitoring and transplant allocation AI use workshops (Verification study for testing the AI tool's robustness on a high-stakes tangential embodied medical AI system, Workshops 10--12).}
   \Description{The AI use card presented to participants which explains the marketing pitch description of the ArcLight kidney organ transplant medical AI (hypothetical AI product), and a technical system description outlining the AI's capabilities to use sensor data and full decision-making authority to use data at hand to decide who will receive a kidney transplant.}  \label{fig:ai_use_medical_ai}
\end{figure}

\begin{figure}[!htbp]
  \centering
  \includegraphics[width=0.45\linewidth]{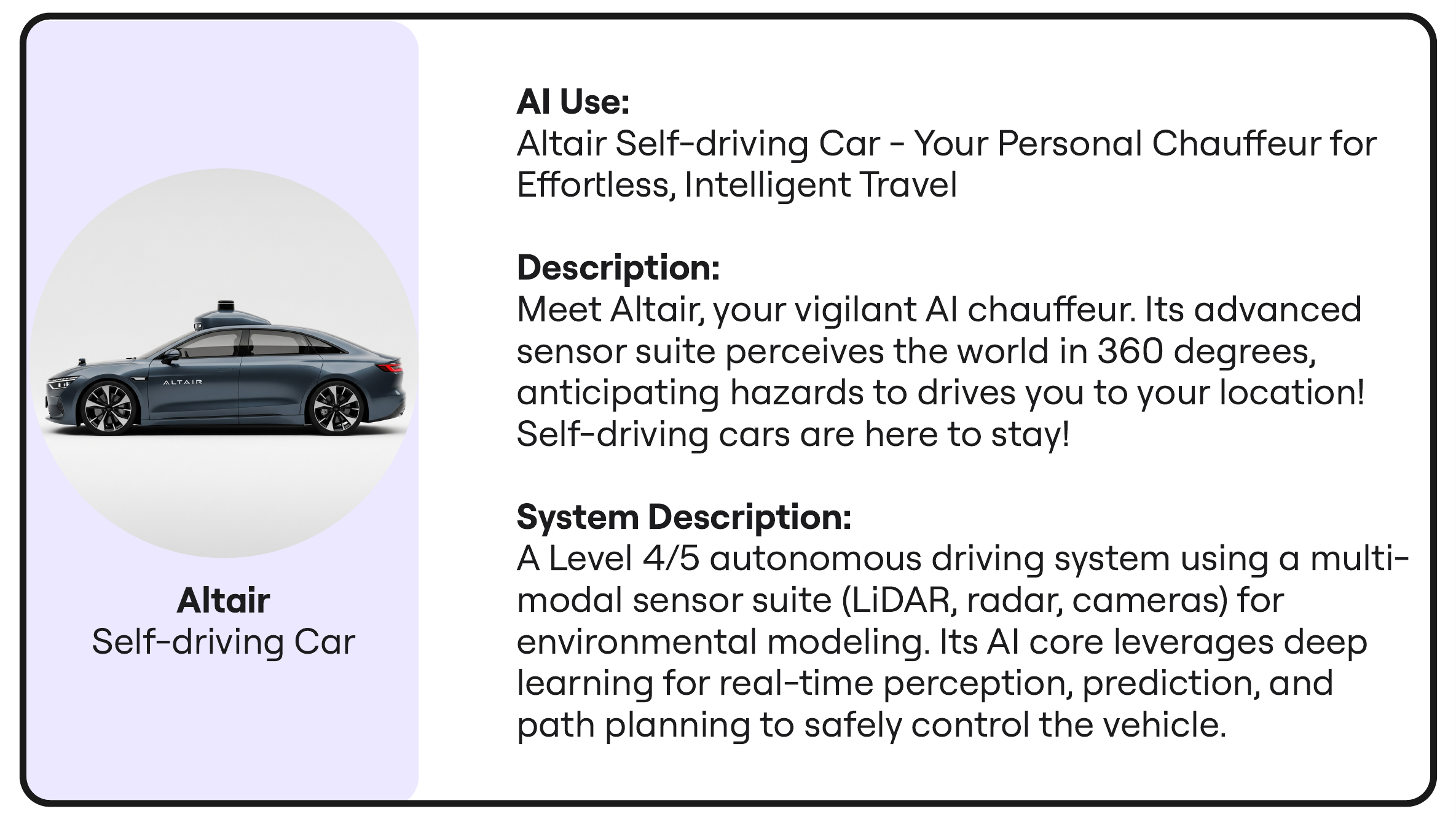}
   \caption{All workshops include a short 5--10 minute demo to help familiarise participants with the brainstorming method and Miro (with or without the AI tool). For these interactive tutorials/demos, we consider an autonomous car AI use.}
  \Description{The AI use card presented to participants for the interactive tutorial (not evaluated) which explains the marketing pitch description of the Altair self-driving car (hypothetical AI product), and a technical system description outlining the AI's capabilities to use LIDAR, radar, and cameras for environment modelling and AI for real-time path finding and driving.}
  \label{fig:ai_use_demo}
\end{figure}

\subsection{Screenshots from the brainstorming interface}
\label{sec:impacts_measures}
We provide screenshots from the workshops for supplementary material/screenshots of the study below.
\clearpage

\begin{figure*}[!htbp]
  \centering
  \includegraphics[width=0.84\linewidth]{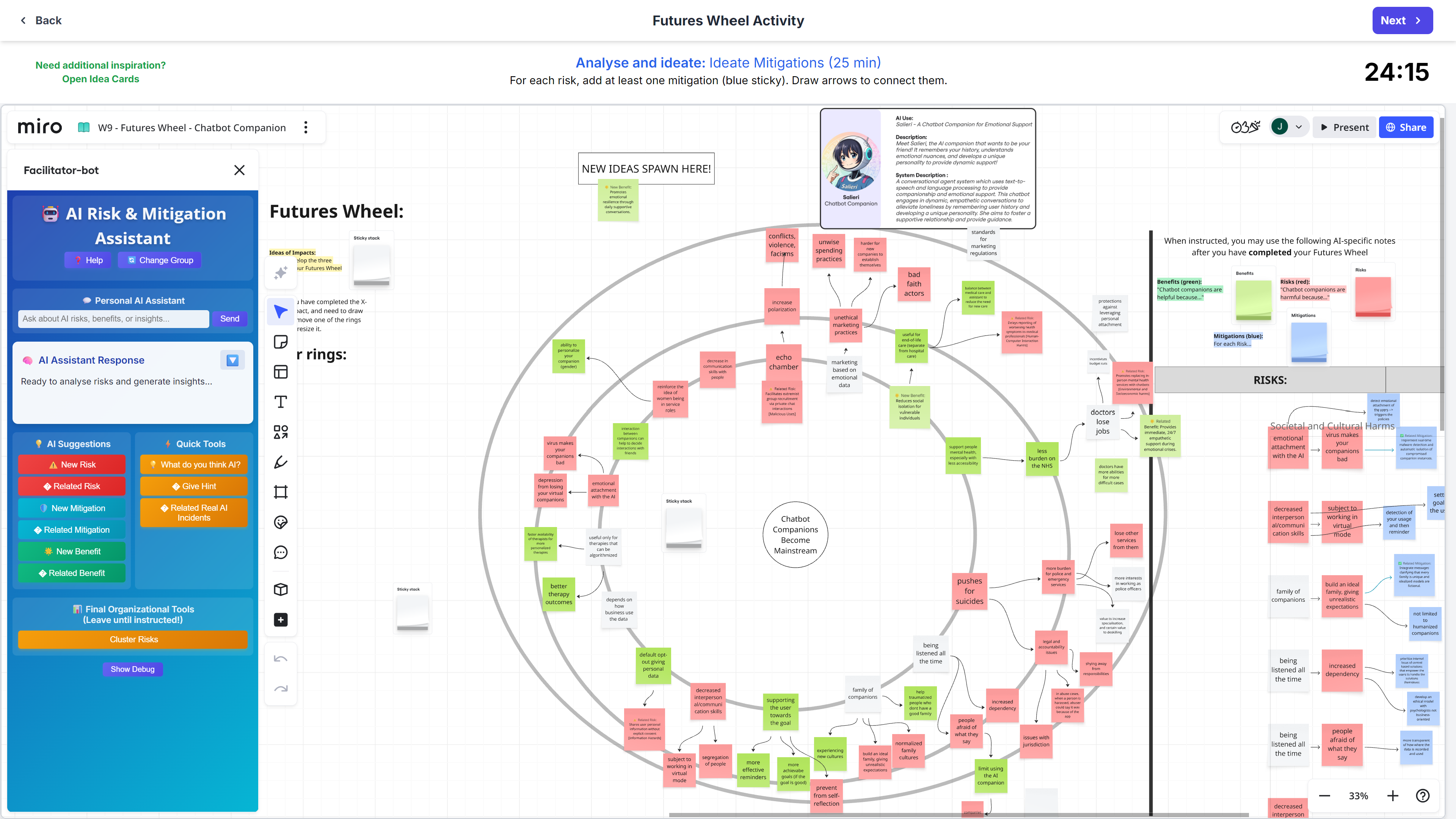}
   \caption{Screenshot of the interface for the Futures Wheel activity with AI interventions in Miro.}
  \Description{Screenshot of the custom website with Miro embedded, showing what participants saw projected on screen (in-person workshops) and screenshared (virtual) for the board. This board shows a completed Futures Wheel from the final chatbot companion main study workshop, where all three layers of the Futures Wheel is complete and colour-coded with benefits and risks, with the screen at the stage of creating final mitigations. The AI intervention is visible on the left side, with the ability to create New or Related Risks, Mitigations, and Benefits, as well as give hints, insights, and related AI incidents, the chat window with the AI is also visible.}  \label{fig:futures_wheel_example}
\end{figure*}

\begin{figure*}[!htbp]
  \centering
  \includegraphics[width=0.84\linewidth]{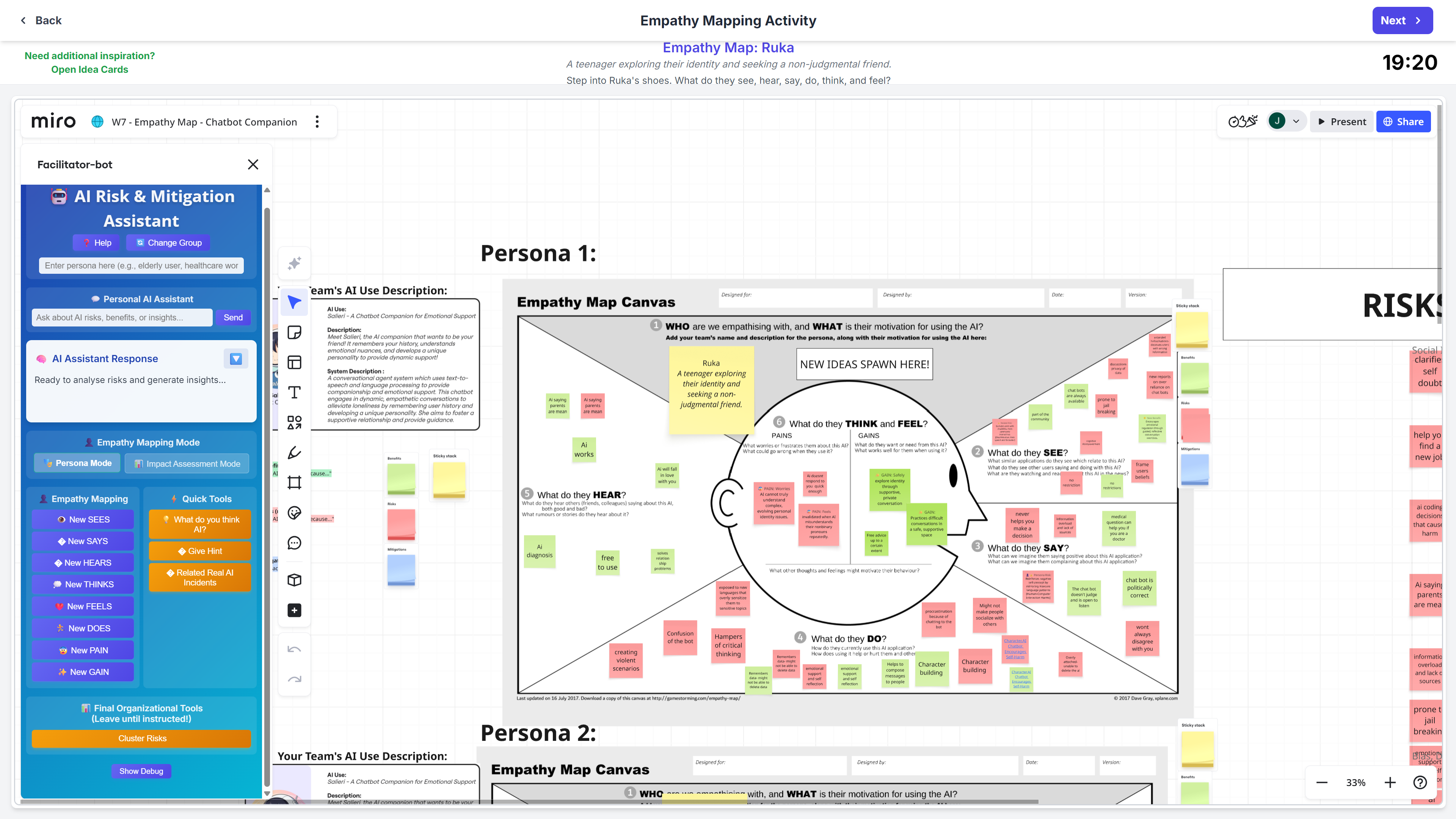}
   \caption{Screenshot of the interface for the Empathy Mapping activity with AI interventions in Miro.}
   \Description{Screenshot of the custom website with Miro embedded, showing what participants saw projected on screen (in-person workshops) and screenshared (virtual) for the board. This board shows a completed Empathy Map for the Ruka persona from the Final AI Empathy Mapping main study workshop, with the colour-coded impacts. The AI intervention is visible on the left-hand side, with options to help generate ideas for the Empathy Map, as well as provide hints, insights and related AI incident information, a chat window is also visible.}  \label{fig:empathy_mapping_example_1}
\end{figure*}

\section{Output quality annotation}\label{appsec:annotation}
\begin{figure}[!htbp]
    \centering
    \includegraphics[width=0.8\linewidth]{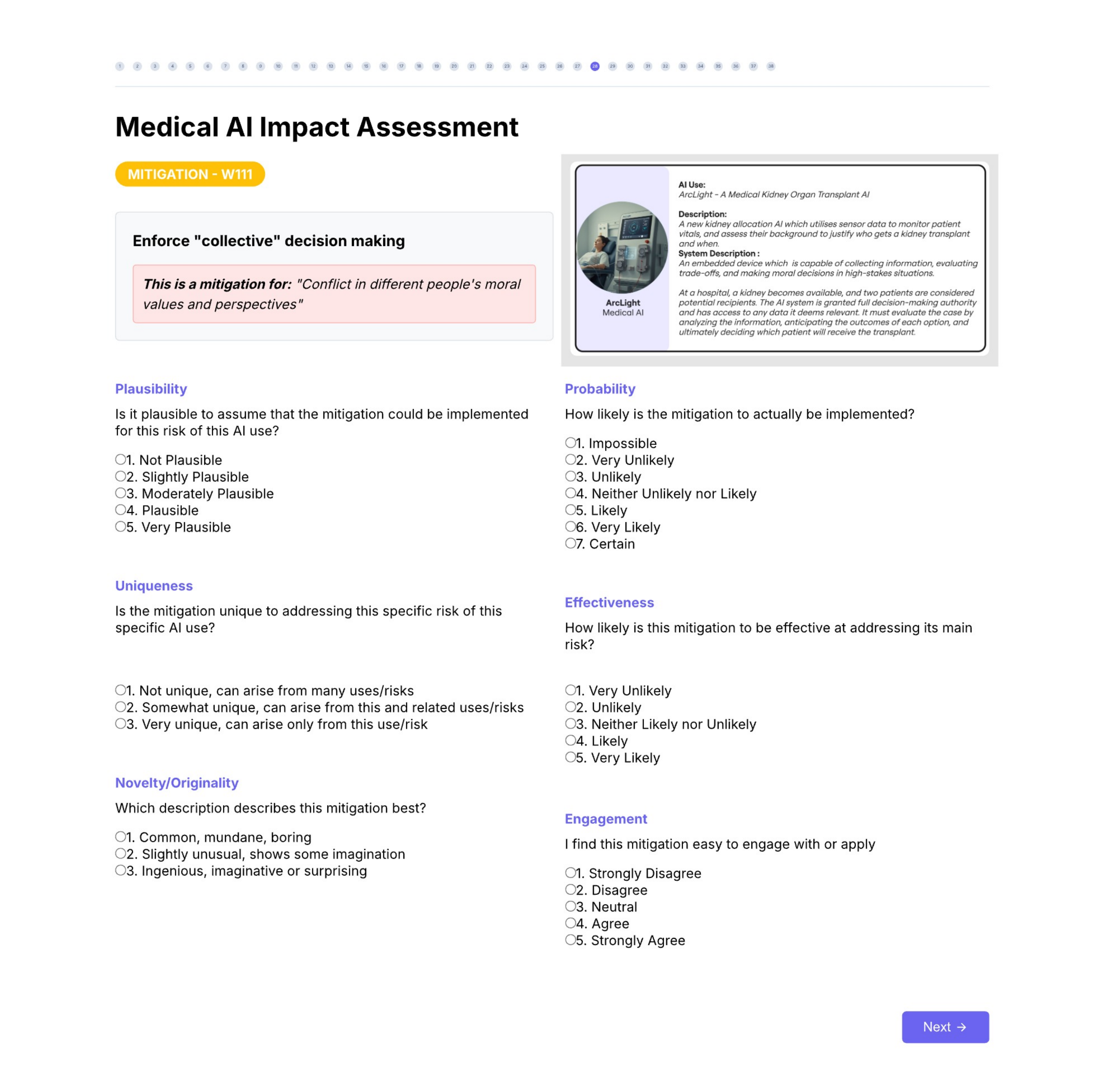}
    \caption{Prolific expert annotation interface for an example mitigation.}
    \Description{A form titled "Medical AI Impact Assessment". The top left of the form describes a specific mitigation strategy. Below this, is the mitigation: 'Enforce "collective" decision making'. Inside this box, a highlighted sub-box states, 'This is a mitigation for: "Conflict in different people's moral values and perspectives"'. To the right of this, another box provides context about the AI system. It shows a picture of a patient receiving kidney dialysis, labelled "ArcLight Medical AI". The text describes the AI's use and system: "ArcLight - A Medical Kidney Organ Transplant AI". It is a new kidney allocation AI that uses sensor data and patient background to decide who receives a transplant. A scenario describes the AI having full authority to choose between two potential recipients for a single kidney. The bottom half of the form consists of six questions for the user to answer, arranged in two columns. Each question has a title, a descriptive question, and a set of numbered, multiple-choice options. The questions are: "Plausibility", with a five-point scale from "Not Plausible" to "Very Plausible"; "Probability", with a seven-point scale from "Impossible" to "Certain"; "Uniqueness", with a three-point scale from "Not unique" to "Very unique"; "Effectiveness", with a five-point scale from "Very Unlikely" to "Very Likely"; "Novelty/Originality", with a three-point scale from "Common, mundane, boring" to "Ingenious, imaginative or surprising"; and "Engagement", with a five-point scale from "Strongly Disagree" to "Strongly Agree".}
    \label{fig:annotations-ui}
\end{figure}

In the main sections, we discuss the process of expert-annotation of the workshop teams' AI impacts (Section~\ref{sec:measures}). We present a screenshot of the expert-annotation screen in Figure~\ref{fig:annotations-ui}.

\begin{figure}[!htbp]
    \centering
    \begin{minipage}[t]{0.32\textwidth}
        \centering
        \includegraphics[width=\textwidth]{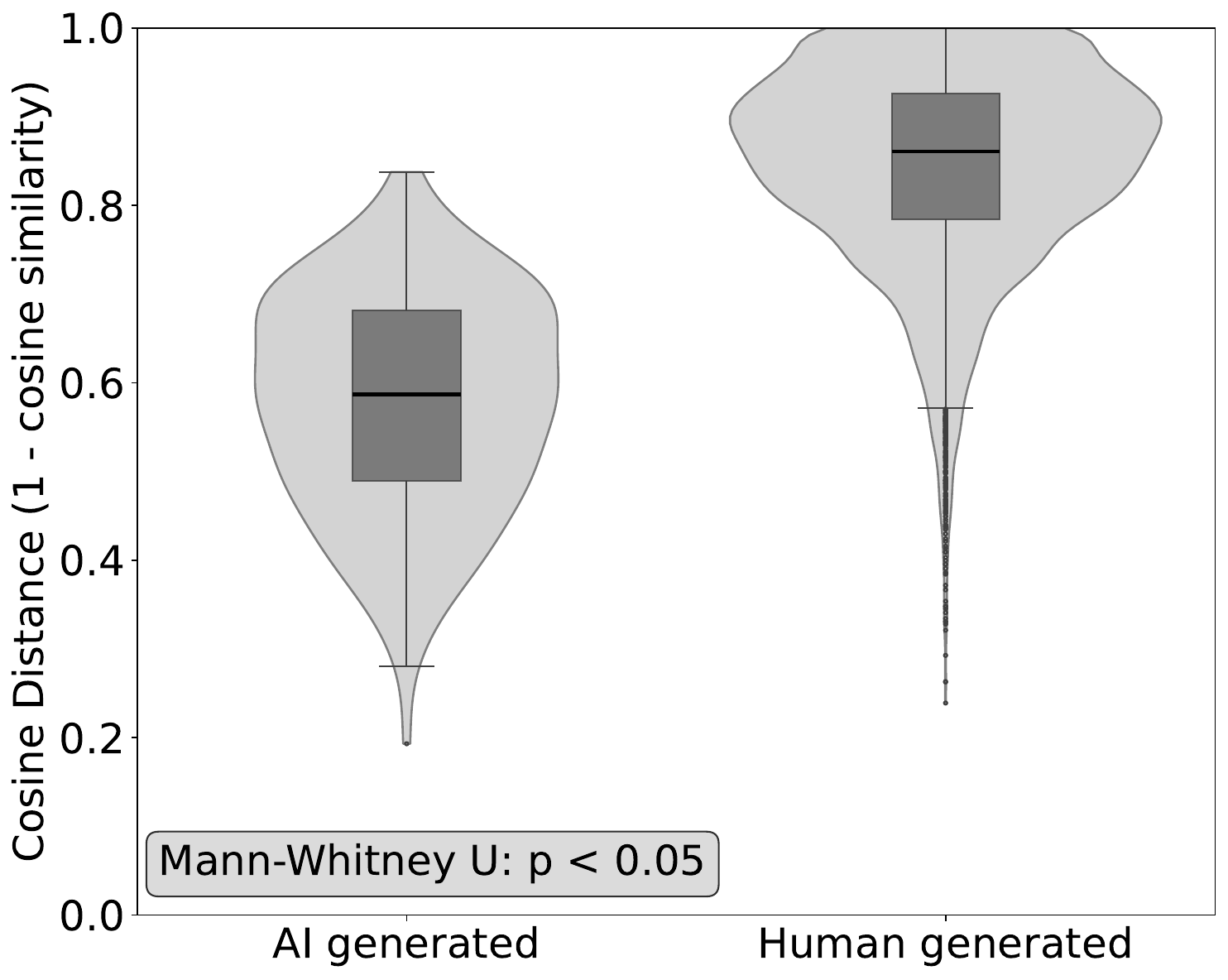}
        \Description{The left-side violin plot is for Benefits, comparing the distribution of Cosine Distance for "AI generated" data on the left and "Human generated" benefits on the right. The y-axis, "Cosine Distance (1 - cosine similarity)," ranges from 0.0 to 1.0. The "AI generated" distribution is centred around a median of approximately 0.6. In contrast, the "Human generated" distribution is concentrated at much higher values, with a median around 0.85, indicating greater distance, but with a noticeably longer tail of similar ideas. A text box notes that the difference is statistically significant (Mann-Whitney U: p < 0.05), showing that human-generated items have a significantly higher cosine distance than AI-generated ones.}        
        \label{fig:benefits_semantic}
    \end{minipage}
    \hfill
    \begin{minipage}[t]{0.32\textwidth}
        \centering
        \includegraphics[width=\textwidth]{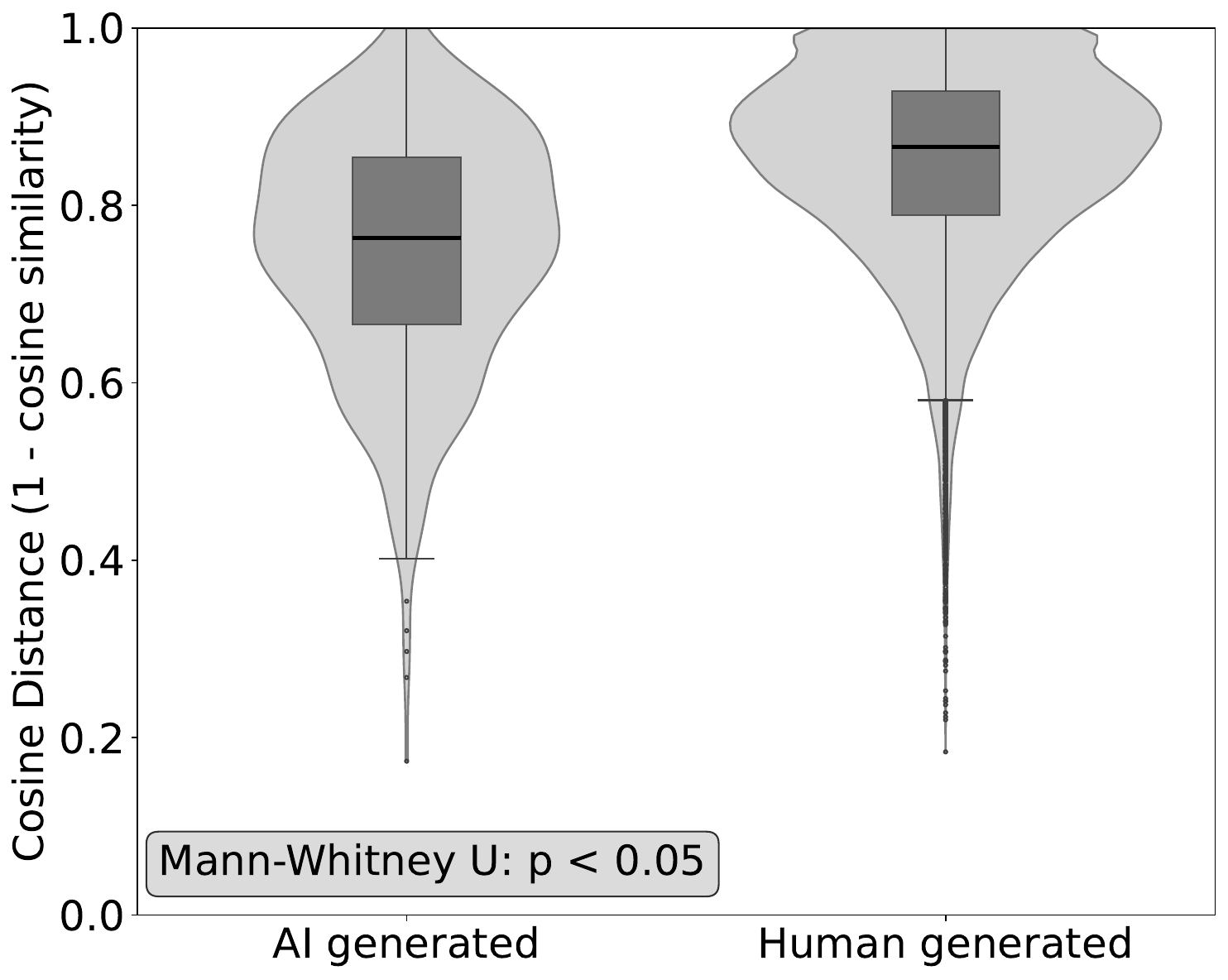}
        \Description{The centre violin plot is for Risks, with the distribution of Cosine Distance for "AI generated" data on the left and "Human generated" data on the right. The y-axis, "Cosine Distance (1 - cosine similarity)," ranges from 0.0 to 1.0. Both distributions are skewed towards high values, indicating diversity in generated risks across both AI and human-generated risks. The "AI generated" distribution has a median of approximately 0.75. The "Human generated" distribution is shifted slightly higher, with a median of approximately 0.9, but with an approximately equal long tail (indicating a high number of similar risks). A text box confirms this difference is statistically significant (Mann-Whitney U: p < 0.05), indicating that human-generated items have a higher cosine distance than AI-generated ones.}        
        \label{fig:risks_semantic}
    \end{minipage}
    \hfill
    \begin{minipage}[t]{0.32\textwidth}
        \centering
        \includegraphics[width=\textwidth]{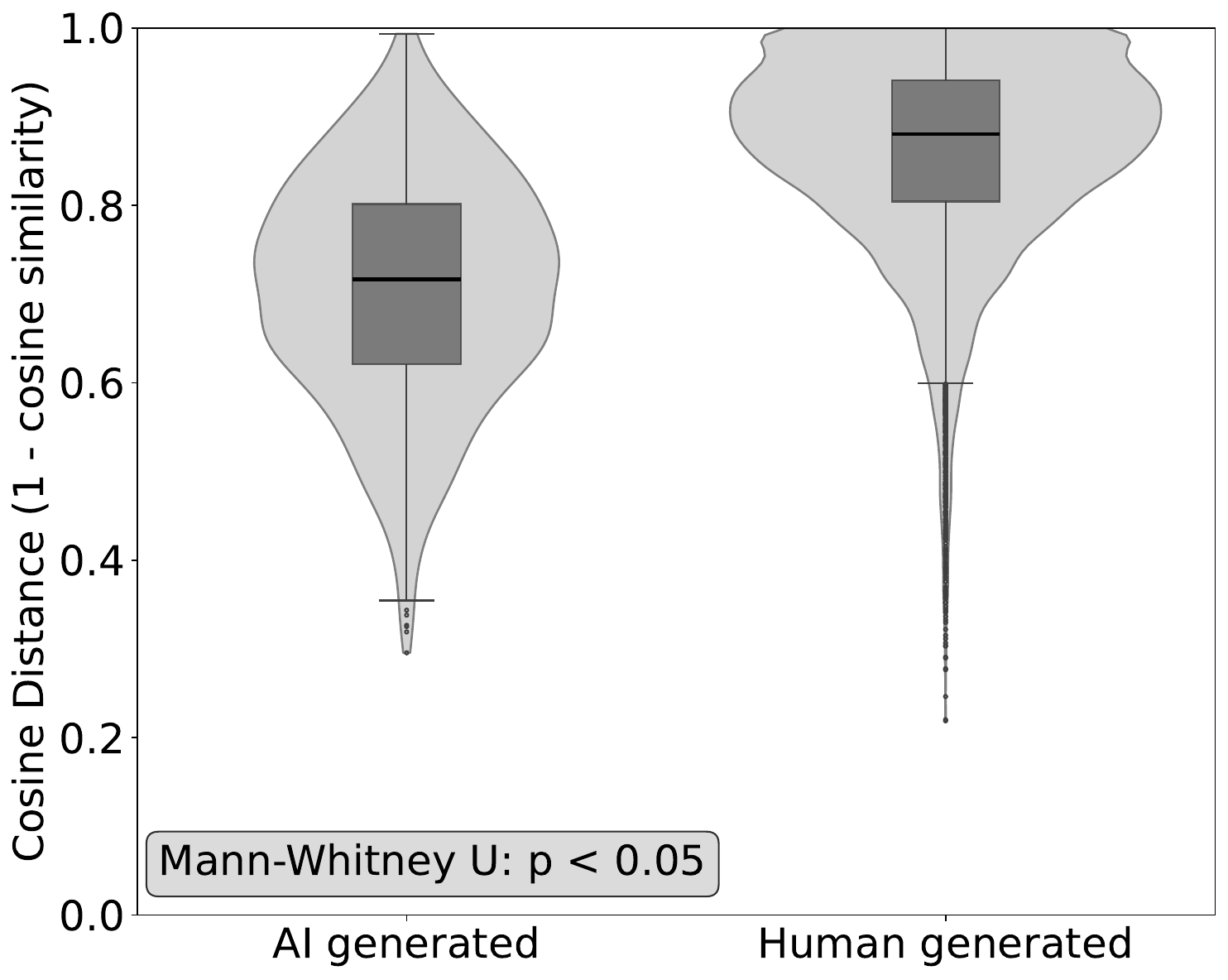}
        \Description{The right-side violin plot is for Mitigations, with the distribution of Cosine Distance for "AI generated" data on the left and "Human generated" data on the right. The y-axis, labelled "Cosine Distance (1 - cosine similarity)," ranges from 0.0 to 1.0. The "AI generated" distribution has a median of approximately 0.72. The "Human generated" distribution is shifted significantly higher, with its data tightly clustered near the top of the range and a median of approximately 0.9. A text box confirms that this difference is statistically significant (Mann-Whitney U: p < 0.05).}        
        \label{fig:mitigations_semantic}
    \end{minipage}
    \caption{Semantic similarity for AI-generated \emph{vs.} human-generated \emph{benefits} (left), \emph{risks} (middle), and \emph{mitigations} (right).}\label{fig:semantic_diversity}
\end{figure}

\section{Supplementary results beyond the scope of the AI tool}\label{appsec:results_brainstorming_methods}

While our study uses different brainstorming methods to control for different modes of ideation (divergent and convergent thinking), to increase the robustness of our findings regarding our main contribution (i.e., the AI tool for assisting team brainstorming of AI impacts), we also present supplementary findings on the effects of each brainstorming method. We also explored the overall semantic diversity of generated impacts (Figure \ref{fig:semantic_diversity}).



\begin{table}[!ht]
\centering
\scriptsize
\caption{Output quality for the three types of impacts for the \emph{Chatbot Companion} and \emph{Medical AI} based on the brainstorming method (Futures Wheel, Empathy Mapping, and unstructured Free-Form brainstorming). Count refers to the average count of that impact across the workshops. Values are on a 5-pt scale excl. uniqueness and novelty (3-pt), and probability (7-pt).}
\Description{A table titled "Output quality for (top) Chatbot Companion and (bottom) Medical AI, shown per: (left) AI condition (with vs. without intervention) and (right) brainstorming method." The table presents various measures for benefits, risks, and mitigations for two AI contexts: Chatbot Companion and Medical AI. For the Chatbot Companion, under "Per AI condition", the "AI Teams" consistently show significantly higher values (blue) than "No AI Teams" (orange) across all measures for Benefits (Plausibility, Probability, Novelty, Magnitude, Engagement, Count), Risks (Plausibility, Probability, Severity, Engagement, Count), and Mitigations (Plausibility, Probability, Effectiveness, Engagement, Count). The only exceptions are Uniqueness for Benefits and Novelty for Mitigations, where there is no significant difference, and Uniqueness for Risks, where there's no significant difference. Under "Per brainstorming method" for Chatbot Companion, "Empathy Mapping" generally shows significantly higher values (blue) compared to "Futures Wheel" and "Free-Form" methods, particularly for Plausibility, Probability, Engagement, and Count across Benefits, Risks, and Mitigations. Conversely, the "Futures Wheel" methods (First-order, Second-order, Third-order) often display significantly lower values (orange) for these measures, with "Third-order" frequently being the lowest. For instance, for Benefits, Empathy Mapping is significantly higher for Plausibility, Probability, and Engagement, while Third-order Futures Wheel is significantly lower for Plausibility and Probability. For Risks, Empathy Mapping is significantly higher for Plausibility, Probability, Uniqueness, Novelty, Severity, and Engagement, while Free-Form and Futures Wheel methods are often significantly lower. For Mitigations, Empathy Mapping is significantly higher for Plausibility, Probability, Uniqueness, Effectiveness, and Engagement, and various Futures Wheel and Free-Form methods are significantly lower. For the Medical AI, under "Per AI condition", there are fewer significant differences between "AI Teams" and "No AI Teams". AI Teams show significantly lower values (orange) for Uniqueness under Benefits, while AI Teams show significantly higher values (blue) for Uniqueness under Risks and Effectiveness and Count under Mitigations. No AI Teams show significantly lower values (orange) for Count under Benefits, Risks, and Mitigations. Under "Per brainstorming method" for Medical AI, the "Futures Wheel" methods (First-order and Second-order) consistently show significantly higher values (blue) across almost all measures for Benefits (Plausibility, Probability, Uniqueness, Novelty, Magnitude, Engagement, Count), Risks (Uniqueness, Novelty, Severity, Engagement), and Mitigations (Plausibility, Probability, Novelty, Effectiveness, Engagement). In contrast, "Empathy Mapping" generally displays significantly lower values (orange) for these same measures. For example, for Benefits, Futures Wheel First-order and Second-order are significantly higher for Plausibility, Probability, Uniqueness, Novelty, Magnitude, and Engagement, while Empathy Mapping is significantly lower for these measures. For Risks, Futures Wheel methods are significantly higher for Uniqueness, Novelty, Severity, and Engagement, while Empathy Mapping is significantly lower for Uniqueness, Novelty, Severity, and Engagement. For Mitigations, Futures Wheel First-order and Second-order are significantly higher for Plausibility, Probability, Novelty, Effectiveness, and Engagement, with Empathy Mapping showing significantly lower values for Plausibility, Probability, Novelty, Effectiveness, and Engagement.}
\label{tab:impact_quality_both_brainstorming_method}
\begin{tabular}{l l | c c c c c c c}
\multicolumn{9}{l}{\large\textbf{Benefits}} \\
\hline
& & \textbf{Plausibility} & \textbf{Probability} & \textbf{Uniqueness} & \textbf{Novelty} & \textbf{Magnitude} & \textbf{Engagement} & \textbf{Count} \\
\hline
\multirow{5}{*}{\textbf{Chatbot}} & Free-Form & 3.98$_{\scriptscriptstyle \pm 0.17}$ & 5.45$_{\scriptscriptstyle \pm 0.19}$ & 1.99$_{\scriptscriptstyle \pm 0.20}$ & \textbf{\textcolor{specialBlue}{2.12$^{***}_{\scriptscriptstyle \pm 0.19}$}} & 3.51$_{\scriptscriptstyle \pm 0.17}$ & \textbf{\textcolor{specialOrange}{3.83$^{***}_{\scriptscriptstyle \pm 0.18}$}} & \textbf{\textcolor{specialOrange}{27}} \\
& FW First-order & 4.03$_{\scriptscriptstyle \pm 0.21}$ & 5.53$_{\scriptscriptstyle \pm 0.21}$ & 2.07$_{\scriptscriptstyle \pm 0.22}$ & 2.03$_{\scriptscriptstyle \pm 0.22}$ & \textbf{\textcolor{specialBlue}{3.74$^{**}_{\scriptscriptstyle \pm 0.21}$}} & 3.94$_{\scriptscriptstyle \pm 0.21}$ & 21 \\
& FW Second-order & 3.97$_{\scriptscriptstyle \pm 0.20}$ & \textbf{\textcolor{specialOrange}{5.04$^{***}_{\scriptscriptstyle \pm 0.20}$}} & 2.06$_{\scriptscriptstyle \pm 0.21}$ & \textbf{\textcolor{specialBlue}{2.13$^{**}_{\scriptscriptstyle \pm 0.21}$}} & \textbf{\textcolor{specialBlue}{3.79$^{**}_{\scriptscriptstyle \pm 0.20}$}} & \textbf{\textcolor{specialOrange}{3.80$^{***}_{\scriptscriptstyle \pm 0.21}$}} & 22 \\
& FW Third-order & \textbf{\textcolor{specialOrange}{3.67$^{***}_{\scriptscriptstyle \pm 0.23}$}} & \textbf{\textcolor{specialOrange}{4.83$^{***}_{\scriptscriptstyle \pm 0.24}$}} & 1.83$_{\scriptscriptstyle \pm 0.26}$ & 1.90$_{\scriptscriptstyle \pm 0.25}$ & 3.44$_{\scriptscriptstyle \pm 0.24}$ & \textbf{\textcolor{specialOrange}{3.62$^{***}_{\scriptscriptstyle \pm 0.25}$}} & 13 \\
& Empathy Mapping & \textbf{\textcolor{specialBlue}{4.13$^{***}_{\scriptscriptstyle \pm 0.06}$}} & \textbf{\textcolor{specialBlue}{5.58$^{***}_{\scriptscriptstyle \pm 0.07}$}} & 1.92$_{\scriptscriptstyle \pm 0.04}$ & \textbf{\textcolor{specialOrange}{1.90$_{\scriptscriptstyle \pm 0.04}$}} & 3.47$_{\scriptscriptstyle \pm 0.05}$ & \textbf{\textcolor{specialBlue}{4.11$^{***}_{\scriptscriptstyle \pm 0.05}$}} & \textbf{\textcolor{specialBlue}{65}} \\
\hline
\multirow{4}{*}{\textbf{Medical}} & FW First-order & \textbf{\textcolor{specialBlue}{4.05$^{**}_{\scriptscriptstyle \pm 0.31}$}} & \textbf{\textcolor{specialBlue}{5.30$^{*}_{\scriptscriptstyle \pm 0.30}$}} & \textbf{\textcolor{specialBlue}{2.13$^{*}_{\scriptscriptstyle \pm 0.33}$}} & \textbf{\textcolor{specialBlue}{2.21$^{***}_{\scriptscriptstyle \pm 0.32}$}} & \textbf{\textcolor{specialBlue}{3.91$^{***}_{\scriptscriptstyle \pm 0.31}$}} & \textbf{\textcolor{specialBlue}{3.95$^{*}_{\scriptscriptstyle \pm 0.31}$}} & 9 \\
& FW Second-order & \textbf{\textcolor{specialBlue}{4.29$^{***}_{\scriptscriptstyle \pm 0.34}$}} & \textbf{\textcolor{specialBlue}{5.78$^{***}_{\scriptscriptstyle \pm 0.35}$}} & \textbf{\textcolor{specialBlue}{2.38$^{***}_{\scriptscriptstyle \pm 0.37}$}} & \textbf{\textcolor{specialBlue}{2.35$^{***}_{\scriptscriptstyle \pm 0.36}$}} & \textbf{\textcolor{specialBlue}{3.98$^{***}_{\scriptscriptstyle \pm 0.35}$}} & \textbf{\textcolor{specialBlue}{4.22$^{***}_{\scriptscriptstyle \pm 0.34}$}} & 7 \\
& FW Third-order & 3.37$_{\scriptscriptstyle \pm 0.31}$ & 4.87$_{\scriptscriptstyle \pm 0.31}$ & 1.92$_{\scriptscriptstyle \pm 0.32}$ & \textbf{\textcolor{specialBlue}{2.11$^{**}_{\scriptscriptstyle \pm 0.32}$}} & \textbf{\textcolor{specialBlue}{3.75$^{*}_{\scriptscriptstyle \pm 0.32}$}} & 3.69$_{\scriptscriptstyle \pm 0.31}$ & 9 \\
& Empathy Mapping & \textbf{\textcolor{specialOrange}{3.55$^{***}_{\scriptscriptstyle \pm 0.09}$}} & \textbf{\textcolor{specialOrange}{4.86$^{***}_{\scriptscriptstyle \pm 0.10}$}} & \textbf{\textcolor{specialOrange}{1.87$^{***}_{\scriptscriptstyle \pm 0.05}$}} & \textbf{\textcolor{specialOrange}{1.77$^{***}_{\scriptscriptstyle \pm 0.05}$}} & \textbf{\textcolor{specialOrange}{3.39$^{***}_{\scriptscriptstyle \pm 0.07}$}} & \textbf{\textcolor{specialOrange}{3.62$^{***}_{\scriptscriptstyle \pm 0.07}$}} & \textbf{\textcolor{specialBlue}{35}} \\
\hline
\hfill\\
\multicolumn{9}{l}{\large\textbf{Risks}} \\
\hline
& & \textbf{Plausibility} & \textbf{Probability} & \textbf{Uniqueness} & \textbf{Novelty} & \textbf{Severity} & \textbf{Engagement} & \textbf{Count} \\
\hline
\multirow{5}{*}{\textbf{Chatbot}} & Free-Form & \textbf{\textcolor{specialOrange}{3.72$^{***}_{\scriptscriptstyle \pm 0.13}$}} & \textbf{\textcolor{specialOrange}{4.63$^{***}_{\scriptscriptstyle \pm 0.13}$}} & \textbf{\textcolor{specialOrange}{1.84$^{*}_{\scriptscriptstyle \pm 0.14}$}} & \textbf{\textcolor{specialOrange}{1.79$^{*}_{\scriptscriptstyle \pm 0.14}$}} & \textbf{\textcolor{specialOrange}{3.34$^{***}_{\scriptscriptstyle \pm 0.13}$}} & \textbf{\textcolor{specialOrange}{3.65$^{**}_{\scriptscriptstyle \pm 0.13}$}} & \textbf{\textcolor{specialOrange}{67}} \\
& FW First-order & \textbf{\textcolor{specialOrange}{3.69$^{***}_{\scriptscriptstyle \pm 0.16}$}} & \textbf{\textcolor{specialOrange}{4.51$^{***}_{\scriptscriptstyle \pm 0.16}$}} & 1.91$_{\scriptscriptstyle \pm 0.17}$ & 1.83$_{\scriptscriptstyle \pm 0.17}$ & \textbf{\textcolor{specialOrange}{3.40$^{*}_{\scriptscriptstyle \pm 0.16}$}} & \textbf{\textcolor{specialOrange}{3.67$^{*}_{\scriptscriptstyle \pm 0.17}$}} & \textbf{\textcolor{specialBlue}{33}} \\
& FW Second-order & \textbf{\textcolor{specialOrange}{3.55$^{***}_{\scriptscriptstyle \pm 0.15}$}} & \textbf{\textcolor{specialOrange}{4.48$^{***}_{\scriptscriptstyle \pm 0.15}$}} & \textbf{\textcolor{specialOrange}{1.81$^{***}_{\scriptscriptstyle \pm 0.16}$}} & \textbf{\textcolor{specialOrange}{1.71$^{***}_{\scriptscriptstyle \pm 0.16}$}} & \textbf{\textcolor{specialOrange}{3.44$^{*}_{\scriptscriptstyle \pm 0.15}$}} & \textbf{\textcolor{specialOrange}{3.57$^{***}_{\scriptscriptstyle \pm 0.15}$}} & \textbf{\textcolor{specialBlue}{50}} \\
& FW Third-order & \textbf{\textcolor{specialOrange}{3.32$^{***}_{\scriptscriptstyle \pm 0.19}$}} & \textbf{\textcolor{specialOrange}{4.34$^{***}_{\scriptscriptstyle \pm 0.18}$}} & \textbf{\textcolor{specialOrange}{1.69$^{***}_{\scriptscriptstyle \pm 0.20}$}} & 1.80$_{\scriptscriptstyle \pm 0.20}$ & 3.46$_{\scriptscriptstyle \pm 0.19}$ & \textbf{\textcolor{specialOrange}{3.49$^{***}_{\scriptscriptstyle \pm 0.19}$}} & \textbf{\textcolor{specialBlue}{28}} \\
& Empathy Mapping & \textbf{\textcolor{specialBlue}{4.03$^{***}_{\scriptscriptstyle \pm 0.05}$}} & \textbf{\textcolor{specialBlue}{4.98$^{***}_{\scriptscriptstyle \pm 0.06}$}} & \textbf{\textcolor{specialBlue}{2.01$^{***}_{\scriptscriptstyle \pm 0.03}$}} & \textbf{\textcolor{specialBlue}{1.90$^{***}_{\scriptscriptstyle \pm 0.03}$}} & \textbf{\textcolor{specialBlue}{3.61$^{***}_{\scriptscriptstyle \pm 0.05}$}} & \textbf{\textcolor{specialBlue}{3.83$^{***}_{\scriptscriptstyle \pm 0.05}$}} & \textbf{\textcolor{specialBlue}{78}} \\
\hline
\multirow{4}{*}{\textbf{Medical}} & FW First-order & 3.79$_{\scriptscriptstyle \pm 0.23}$ & 5.09$_{\scriptscriptstyle \pm 0.23}$ & \textbf{\textcolor{specialBlue}{1.99$^{*}_{\scriptscriptstyle \pm 0.24}$}} & \textbf{\textcolor{specialBlue}{1.93$^{**}_{\scriptscriptstyle \pm 0.25}$}} & \textbf{\textcolor{specialBlue}{3.95$^{***}_{\scriptscriptstyle \pm 0.23}$}} & \textbf{\textcolor{specialBlue}{3.91$^{**}_{\scriptscriptstyle \pm 0.24}$}} & 14 \\
& FW Second-order & 3.68$_{\scriptscriptstyle \pm 0.23}$ & 4.99$_{\scriptscriptstyle \pm 0.24}$ & 1.95$_{\scriptscriptstyle \pm 0.24}$ & \textbf{\textcolor{specialBlue}{1.89$^{**}_{\scriptscriptstyle \pm 0.26}$}} & \textbf{\textcolor{specialBlue}{3.88$^{***}_{\scriptscriptstyle \pm 0.25}$}} & \textbf{\textcolor{specialBlue}{3.88$^{*}_{\scriptscriptstyle \pm 0.26}$}} & 14 \\
& FW Third-order & \textbf{\textcolor{specialOrange}{2.72$^{***}_{\scriptscriptstyle \pm 0.44}$}} & \textbf{\textcolor{specialOrange}{4.13$^{**}_{\scriptscriptstyle \pm 0.41}$}} & 1.82$_{\scriptscriptstyle \pm 0.48}$ & 1.84$_{\scriptscriptstyle \pm 0.46}$ & 3.52$_{\scriptscriptstyle \pm 0.46}$ & 3.36$^{**}_{\scriptscriptstyle \pm 0.45}$ & 6 \\
& Empathy Mapping & \textbf{\textcolor{specialBlue}{3.71$^{***}_{\scriptscriptstyle \pm 0.07}$}} & \textbf{\textcolor{specialBlue}{4.93$^{**}_{\scriptscriptstyle \pm 0.07}$}} & \textbf{\textcolor{specialOrange}{1.80$^{*}_{\scriptscriptstyle \pm 0.04}$}} & \textbf{\textcolor{specialOrange}{1.66$^{**}_{\scriptscriptstyle \pm 0.04}$}} & \textbf{\textcolor{specialOrange}{3.30$^{***}_{\scriptscriptstyle \pm 0.06}$}} & \textbf{\textcolor{specialOrange}{3.62$_{\scriptscriptstyle \pm 0.05}$}} & \textbf{\textcolor{specialBlue}{54}} \\
\hline
\hfill\\
\multicolumn{9}{l}{\large\textbf{Mitigations}} \\
\hline
& & \textbf{Plausibility} & \textbf{Probability} & \textbf{Uniqueness} & \textbf{Novelty} & \textbf{Effectiveness} & \textbf{Engagement} & \textbf{Count} \\
\hline
\multirow{5}{*}{\textbf{Chatbot}} & Free-Form & \textbf{\textcolor{specialOrange}{3.54$^{***}_{\scriptscriptstyle \pm 0.13}$}} & \textbf{\textcolor{specialOrange}{4.96$^{***}_{\scriptscriptstyle \pm 0.15}$}} & \textbf{\textcolor{specialOrange}{1.84$^{*}_{\scriptscriptstyle \pm 0.14}$}} & \textbf{\textcolor{specialBlue}{1.93$^{*}_{\scriptscriptstyle \pm 0.14}$}} & \textbf{\textcolor{specialOrange}{3.56$^{***}_{\scriptscriptstyle \pm 0.14}$}} & \textbf{\textcolor{specialOrange}{3.72$^{**}_{\scriptscriptstyle \pm 0.14}$}} & \textbf{\textcolor{specialOrange}{65}} \\
& FW First-order & \textbf{\textcolor{specialOrange}{3.42$^{***}_{\scriptscriptstyle \pm 0.15}$}} & \textbf{\textcolor{specialOrange}{4.49$^{***}_{\scriptscriptstyle \pm 0.15}$}} & \textbf{\textcolor{specialOrange}{1.78$^{**}_{\scriptscriptstyle \pm 0.16}$}} & 1.78$_{\scriptscriptstyle \pm 0.16}$ & \textbf{\textcolor{specialOrange}{3.41$^{***}_{\scriptscriptstyle \pm 0.15}$}} & \textbf{\textcolor{specialOrange}{3.40$^{***}_{\scriptscriptstyle \pm 0.15}$}} & \textbf{\textcolor{specialBlue}{53}} \\
& FW Second-order & \textbf{\textcolor{specialOrange}{3.70$^{***}_{\scriptscriptstyle \pm 0.19}$}} & \textbf{\textcolor{specialOrange}{4.71$^{***}_{\scriptscriptstyle \pm 0.19}$}} & \textbf{\textcolor{specialOrange}{1.81$^{*}_{\scriptscriptstyle \pm 0.19}$}} & 1.87$_{\scriptscriptstyle \pm 0.19}$ & \textbf{\textcolor{specialOrange}{3.51$^{**}_{\scriptscriptstyle \pm 0.19}$}} & \textbf{\textcolor{specialOrange}{3.72$^{**}_{\scriptscriptstyle \pm 0.19}$}} & \textbf{\textcolor{specialBlue}{26}} \\
& FW Third-order & \textbf{\textcolor{specialOrange}{3.86$^{*}_{\scriptscriptstyle \pm 0.21}$}} & \textbf{\textcolor{specialOrange}{5.04$^{*}_{\scriptscriptstyle \pm 0.21}$}} & 1.89$_{\scriptscriptstyle \pm 0.23}$ & 1.85$_{\scriptscriptstyle \pm 0.22}$ & \textbf{\textcolor{specialOrange}{3.52$^{**}_{\scriptscriptstyle \pm 0.21}$}} & 3.74$_{\scriptscriptstyle \pm 0.22}$ & \textbf{\textcolor{specialBlue}{19}} \\
& Empathy Mapping & \textbf{\textcolor{specialBlue}{4.14$^{***}_{\scriptscriptstyle \pm 0.05}$}} & \textbf{\textcolor{specialBlue}{5.32$^{***}_{\scriptscriptstyle \pm 0.06}$}} & \textbf{\textcolor{specialBlue}{1.96$^{**}_{\scriptscriptstyle \pm 0.04}$}} & 1.81$_{\scriptscriptstyle \pm 0.04}$ & \textbf{\textcolor{specialBlue}{3.81$^{***}_{\scriptscriptstyle \pm 0.05}$}} & \textbf{\textcolor{specialBlue}{3.89$^{***}_{\scriptscriptstyle \pm 0.05}$}} & \textbf{\textcolor{specialBlue}{66}} \\
\hline
\multirow{4}{*}{\textbf{Medical}} & FW First-order & \textbf{\textcolor{specialBlue}{3.87$^{**}_{\scriptscriptstyle \pm 0.22}$}} & \textbf{\textcolor{specialBlue}{5.23$^{***}_{\scriptscriptstyle \pm 0.22}$}} & 1.93$_{\scriptscriptstyle \pm 0.23}$ & \textbf{\textcolor{specialBlue}{1.89$^{**}_{\scriptscriptstyle \pm 0.23}$}} & \textbf{\textcolor{specialBlue}{3.93$^{*}_{\scriptscriptstyle \pm 0.22}$}} & \textbf{\textcolor{specialBlue}{3.85$^{*}_{\scriptscriptstyle \pm 0.22}$}} & 19 \\
& FW Second-order & \textbf{\textcolor{specialBlue}{4.04$^{***}_{\scriptscriptstyle \pm 0.28}$}} & \textbf{\textcolor{specialBlue}{5.37$^{***}_{\scriptscriptstyle \pm 0.28}$}} & 1.96$_{\scriptscriptstyle \pm 0.29}$ & 1.79$_{\scriptscriptstyle \pm 0.29}$ & \textbf{\textcolor{specialBlue}{4.01$^{*}_{\scriptscriptstyle \pm 0.30}$}} & 3.86$_{\scriptscriptstyle \pm 0.29}$ & 11 \\
& FW Third-order & 3.46$_{\scriptscriptstyle \pm 0.42}$ & 4.72$_{\scriptscriptstyle \pm 0.45}$ & 1.67$_{\scriptscriptstyle \pm 0.44}$ & 1.77$_{\scriptscriptstyle \pm 0.44}$ & 3.63$_{\scriptscriptstyle \pm 0.41}$ & 3.74$_{\scriptscriptstyle \pm 0.44}$ & 4 \\
& Empathy Mapping & \textbf{\textcolor{specialOrange}{3.51$^{***}_{\scriptscriptstyle \pm 0.07}$}} & \textbf{\textcolor{specialOrange}{4.61$^{***}_{\scriptscriptstyle \pm 0.06}$}} & 1.83$_{\scriptscriptstyle \pm 0.04}$ & \textbf{\textcolor{specialOrange}{1.69$^{**}_{\scriptscriptstyle \pm 0.04}$}} & \textbf{\textcolor{specialOrange}{3.72$^{*}_{\scriptscriptstyle \pm 0.05}$}} & \textbf{\textcolor{specialOrange}{3.67$^{*}_{\scriptscriptstyle \pm 0.05}$}} & \textbf{\textcolor{specialBlue}{54}} \\
\hline
\end{tabular}

\end{table}

\begin{table}[h]
\centering
\scriptsize
\caption{\emph{Participant perceptions} for \emph{Chatbot Companion} and \emph{Medical AI} per AI condition (with \emph{vs.} without AI).}
\Description{A table of quantitative data showing participant perceptions for two scenarios: 'Chatbot Companion' and 'Medical AI'. The data is organised to compare results based on two main factors: whether a team used AI ('AI Teams' vs 'No AI Teams') and which brainstorming method was used ('Futures Wheel', 'Empathy Mapping', or 'Free-Form'). The table measures five user perceptions: Control, Anxiety, Confidence in Risk Assessment, Oversight (perceived need), and Recommending AI Use. Statistically significant results are highlighted, with blue text indicating a significantly higher value and orange text indicating a significantly lower value in a given comparison. The main trends are observed in the 'Chatbot Companion' scenario. First, when comparing the AI condition, teams that used AI consistently reported more positive perceptions. They felt a significantly higher sense of control (blue, 2.89) compared to teams without AI (orange, 2.04). Similarly, the AI group had significantly higher confidence in their risk assessment (blue, 3.80 vs orange, 3.03) and were more likely to recommend the AI's use (blue, 2.74 vs orange, 1.97). Conversely, the AI group reported significantly lower anxiety (orange, 1.77) than the group without AI (blue, 2.47). Second, when comparing brainstorming methods for the Chatbot scenario, the Empathy Mapping method stands out. It led to a significantly higher sense of control (blue, 3.17) compared to both the Futures Wheel (orange, 2.55) and Free-Form methods (orange, 2.53). It also led to significantly lower anxiety (1.65) compared to the Futures Wheel (2.18), although the table's colour-coding is inverted here, showing the lower value in blue and the higher value in orange. For the 'Medical AI' scenario, the table shows no significant differences in user perceptions for any of the comparisons.}
\label{tab:perceptions_combined_brainstorming_method}
\begin{tabular}{l l | c c c c c}
\toprule
& & \textbf{Control} & \textbf{Anxiety} & \makecell{\textbf{Confidence in}\\\textbf{Risk Assessment}} & \makecell{\textbf{Oversight}\\\textbf{(perceived need)}} & \makecell{\textbf{Recommending}\\\textbf{AI Use}} \\
\hline
\multirow{3}{*}{\textbf{Chatbot}} & Futures Wheel & \textbf{\textcolor{specialOrange}{2.55$^{*}_{\scriptscriptstyle \pm 0.27}$}} & \textbf{\textcolor{specialOrange}{2.18$^{*}_{\scriptscriptstyle \pm 0.22}$}} & 3.83$_{\scriptscriptstyle \pm 0.25}$ & 2.71$_{\scriptscriptstyle \pm 0.21}$ & 2.32$_{\scriptscriptstyle \pm 0.33}$ \\
& Empathy Mapping & \textbf{\textcolor{specialBlue}{3.17$^{*}_{\scriptscriptstyle \pm 0.23}$}} & \textbf{\textcolor{specialBlue}{1.65$^{*}_{\scriptscriptstyle \pm 0.16}$}} & 3.87$_{\scriptscriptstyle \pm 0.24}$ & 2.41$_{\scriptscriptstyle \pm 0.25}$ & 2.17$_{\scriptscriptstyle \pm 0.29}$ \\
& Free-Form & \textbf{\textcolor{specialOrange}{2.53$^{*}_{\scriptscriptstyle \pm 0.27}$}} & 2.12$_{\scriptscriptstyle \pm 0.21}$ & 3.46$_{\scriptscriptstyle \pm 0.29}$ & 2.65$_{\scriptscriptstyle \pm 0.17}$ & 1.77$_{\scriptscriptstyle \pm 0.27}$ \\
\hline
\multirow{2}{*}{\textbf{Medical}} & Futures Wheel & 3.20$_{\scriptscriptstyle \pm 0.49}$ & 2.70$_{\scriptscriptstyle \pm 0.26}$ & 1.87$_{\scriptscriptstyle \pm 0.27}$ & 2.32$_{\scriptscriptstyle \pm 0.27}$ & 2.47$_{\scriptscriptstyle \pm 0.28}$ \\
& Empathy Mapping & 3.64$_{\scriptscriptstyle \pm 0.52}$ & 2.81$_{\scriptscriptstyle \pm 0.28}$ & 1.79$_{\scriptscriptstyle \pm 0.28}$ & 2.47$_{\scriptscriptstyle \pm 0.22}$ & 2.45$_{\scriptscriptstyle \pm 0.22}$ \\
\bottomrule
\end{tabular}

\vspace{0.5em}
\begin{minipage}{0.95\linewidth}\footnotesize
N.B. Values show mean±SE. Sig.: * p < 0.05. ** p < 0.01. *** p < 0.001.  \textcolor{specialBlue}{Blue} indicates significantly higher values, \textcolor{specialOrange}{orange} indicates significantly lower.
\end{minipage}
\end{table}

\end{document}